\documentclass[12pt]{amsart}
\usepackage{amsfonts,amssymb,amscd,amsmath,epsfig}
\usepackage{latexsym}
\usepackage{mathrsfs}
\usepackage{tocvsec2} 
\usepackage{graphicx} 
\usepackage{bmpsize} 

\usepackage{titlesec}
\titleformat{\section}[runin]{}{\bf\thesection.}{0.5em}{}[.]

\usepackage[
hyperindex=true,pagebackref=true,bookmarks=true,
colorlinks=true,linkcolor=blue,citecolor=red]{hyperref}
\def\~{{\rm --}}

\begin{document}

\renewcommand{\tilde}{\widetilde}
\renewcommand{\hat}{\widehat}

\newcommand{\BR}{{\mathbb R}}
\newcommand{\BQ}{{\mathbb Q}}
\newcommand{\BC}{{\mathbb C}}
\newcommand{\BP}{{\mathbb P}}
\newcommand{\BZ}{{\mathbb Z}}
\newcommand{\BN}{{\mathbb N}}
\newcommand{\BS}{{\mathbb S}}

\newcommand{\cH}{{\mathcal H}}
\newcommand{\cA}{{\mathcal A}}
\newcommand{\cB}{{\mathcal B}}
\newcommand{\ccF}{{\mathfrak F}}
\newcommand{\cD}{{\mathcal D}}
\newcommand{\cL}{{\mathcal L}}
\newcommand{\cF}{{\mathcal F}}
\newcommand{\cP}{{\mathcal P}}
\newcommand{\cX}{{\mathcal X}}
\newcommand{\cY}{{\mathcal Y}}
\newcommand{\cS}{{\mathcal S}}
\newcommand{\cSol}{\hbox{$\mathcal Sol$}}
\newcommand{\cT}{\hbox{$\mathcal T$}}

\newcommand{\Z}{{\mathbb Z}}
\newcommand{\Q}{{\mathbb Q}}
\newcommand{\N}{{\mathbb N}}
\newcommand{\C}{{\mathbb C}}
\newcommand{\R}{{\mathbb R}}
\newcommand{\X}{{\mathbb X}}
\newcommand{\Y}{{\mathbb Y}}

\newcommand{\CH}{{\mathcal H}}
\newcommand{\CA}{{\mathcal A}}

\def\HH{\mbox{${\mathcal H}$\kern-5.2pt${\mathcal H}$}}

\newcommand{\binomial}[2]{\genfrac{(}{)}{0pt}{}{ #1 }{ #2 }}
\newcommand{\qbinomial}[2]{\genfrac{[}{]}{0pt}{}{ #1 }{ #2 }_q }
\newcommand{\qbinom}[3]{\genfrac{[}{]}{0pt}{}{ #1 }{ #2 }_{ #3 } }


\def\der{\partial}
\def\tensor{\otimes}
\def\gam{\gamma} \def\Gam{\Gamma}
\def\del{\delta} \def\Del{\Delta}
\def\kap{\kappa}
\def\lam{\lambda} \def\Lam{\Lambda}
\def\Comp{{\mathbb C}}
\def\sM{{\mathcal M}}

\newtheorem{theorem}{Theorem}[section]
\newtheorem{maintheorem}[theorem]{Main Theorem}
\newtheorem{proposition}[theorem]{Proposition}
\newtheorem{definition}[theorem]{Definition}
\newtheorem{lemma}[theorem]{Lemma}
\newtheorem{corollary}[theorem]{Corollary}
\newtheorem{notation}[theorem]{Notation}
\newtheorem{remark}[theorem]{Remark}
\newtheorem{example}[theorem]{Example}

\newtheorem{theorem }{Theorem}[section]
\newtheorem{maintheorem }[theorem]{Main Theorem}
\newtheorem{proposition }[theorem]{Proposition}
\newtheorem{definition }[theorem]{Definition}
\newtheorem{lemma }[theorem]{Lemma}
\newtheorem{corollary }[theorem]{Corollary}
\newtheorem{notation }[theorem]{Notation}
\newtheorem{remark }[theorem]{Remark}
\newtheorem{example }[theorem]{Example}

\newtheorem{ maintheorem }[theorem]{Main Theorem}
\newtheorem{ theorem}{Theorem}[section]
\newtheorem{ proposition}[theorem]{Proposition}
\newtheorem{ definition}[theorem]{Definition}
\newtheorem{ lemma}[theorem]{Lemma}
\newtheorem{ corollary}[theorem]{Corollary}
\newtheorem{ notation}[theorem]{Notation}
\newtheorem{ remark}[theorem]{Remark}
\newtheorem{ example}[theorem]{Example}

\newtheorem{thm}{Theorem}[section]
\newtheorem{prop}[thm]{Proposition}
\newtheorem{lem}[thm]{Lemma}
\newtheorem{cor}[thm]{Corollary}
\newtheorem{conj}[thm]{Conjecture}
\newtheorem{con}[thm]{Conjecture}
\newtheorem{dfn}[thm]{Definition}
\newtheorem{df}[thm]{Definition}
 \newcommand{\rem}{{\bf Comment.\ }}
 \newcommand{\rmk}{{\bf Comment.\ }}
 \newcommand{\exmp}{{\bf Example.\ }}
 \newcommand{\ex}{{\bf Example.\ }}
 \newcommand{\prob}{{\bf Problem.\ }}

\newtheorem{note}{Note} 
\renewcommand{\thenote}{}
\newtheorem*{acka}{Acknowledgments}
\newtheorem{ack}{Acknowledgments}
\renewcommand{\theack}{}
\renewcommand{\appendixname}{\bf Appendix}
\renewcommand{\proof}{{\em Proof.\ }}

\hyphenation{
ap-pen-dix as-ymp-tot-ic at-trib-uted at-trib-ut-able
Bry-li-n-sky com-mu-ta-tion de-ge-ne-rate
de-riv-a-tive dis-trib-ute equi-vari-ant ex-tra-or-di-nary  
geo-met-ric griev-ance griev-ous grad-ed ho-lo-no-my ho-mo-thetic
in-fin-ite-ly in-fin-i-tes-i-mal Ha-rish Cha-n-dra mul-ti-plic-able 
non-euclid-ean non-iso-mor-phic non-smooth par-a-digm 
par-a-bol-ic pa-rab-o-loid pa-ram-e-trize phe-nom-e-non 
post-script pseu-do-dif-fer-en-tial pseu-do-fi-nite 
qua-drat-ics quad-ra-ture Han-kel rec-tan-gle semi-def-i-nite 
set-up wide-spread Euler-ian Feb-ru-ary Gauss-ian Grothen-dieck 
Hamil-ton-ian Her-mi-t-ian her-mi-t-ian Jan-u-ary 
Japan-ese Ka-shi-wa-ra Kor-te-weg Le-gendre No-vem-ber Rie-mann-ian 
Sep-tem-ber Za-mo-lo-d-chi-kov Kni-zh-nik quan-tum Op-dam
Mac-do-nald Ca-lo-ge-ro Su-ther-land Mo-ser 
Ol-sha-net-sky  Pe-re-lo-mov in-de-pen-dent ope-ra-tors 
cy-clo-to-mic ra-tio-nal de-gen-er-a-tion 
in-ter-est-ing de-for-ma-tions de-for-ma-tion pro-ce-dure 
fol-lows ope-ra-tors  pre-serve suf-fices ap-proach 
for-mu-las con-sider its com-ple-tion cor-re-spond-ing 
au-to-mor-phism be-cause pro-por-tional fi-nal-ly let-ting 
equi-v-a-lence ge-n-er-al-ized Mac-do-nald iden-ti-ties 
cor-re-s-pond sub-dia-grams par-ti-tion na-t-u-ral-ly 
or-dered stan-dard de-for-ma-tion ar-gu-ment com-bined 
sphe-r-i-cal rep-re-sen-ta-tions tri-go-no-me-t-ric
ge-n-er-al-ly speak-ing pri-m-it-ive ir-re-du-cible 
sum-ma-tion  rep-re-sen-ta-tives pro-por-ti-o-na-li-ty
ultra-sphe-ri-cal Ro-gers}

\def\ffor{\quad\hbox{ for }\quad}
\def\wwhen{\quad\hbox{ when }\quad}
\def\wwhere{\quad\hbox{ where }\quad}
\def\aand{\quad\hbox{ and }\quad}
\def\for{\  \hbox{ for } \ }
\def\iif{ \ \hbox{ if } \ }
\def\when{ \ \hbox{ when } \ }
\def\where{\  \hbox{ where } \ }
\def\and{\  \hbox{ and } \ }
\def\and{\  \hbox{ and } \ }
\def\oor{\  \hbox{ or } \ }
\def\proof{{\em Proof. \  }}

\def\equal{\stackrel{\,\mathbf{def}}{= \kern-3pt =}}

\def\la{\lambda}
\def\La{\Lambda}
\def\om{\omega}
\def\Om{\Omega}
\def\Th{\Theta}
\def\th{\theta}
\def\al{\alpha}
\def\be{\beta}
\def\ga{\gamma}
\def\ep{\epsilon}
\def\up{\upsilon}
\def\Up{\Upsilon}
\def\de{\delta}
\def\De{\Delta}
\def\ka{\kappa}
\def\kapp{\hbox{\bf \ae}}
\def\si{\sigma}
\def\Si{\Sigma}
\def\Ga{\Gamma}
\def\ze{\zeta}
\def\io{\iota}
\def\bio{b^\iota}
\def\aio{a^\iota}
\def\twio{\tilde{w}^\iota}
\def\hwio{\hat{w}^\iota}
\def\gio{\g^\iota}
\def\Bio{B^\iota}

\def\del{\delta}
\def\pa{\partial}
\def\vp{\varphi}
\def\ve{\varepsilon}
\def\inf{\infty}

\def\vph{\varphi}
\def\vps{\varpsi}
\def\vPh{\varPhi}
\def\vep{\varepsilon}
\def\vpi{{\varpi}}
\def\vth{{\vartheta}}
\def\vsi{{\varsigma}}
\def\vrh{{\varrho}}

\def\bph{\bar{\phi}}
\def\bsi{\bar{\si}}
\def\bvp{\bar{\varphi}}

\newcommand{\bS}{{\mathbf S}}
\newcommand{\bH}{{\mathbf H}}
\newcommand{\bF}{{\mathbf F}}
\newcommand{\bE}{{\mathbf E}}

\def\tal{\tilde{\alpha}}
\def\tbe{\tilde{\beta}}
\def\tde{\tilde{\delta}}
\def\tpi{\tilde{\pi}}
\def\txi{\tilde{\xi}}
\def\tPi{\tilde{\Pi}}
\def\tPhi{\tilde{\Phi}}
\def\tV{\tilde{V}}
\def\tJ{\tilde{J}}
\def\tla{\tilde{\lambda}}
\def\tga{\tilde{\gamma}}
\def\tGa{\tilde{\Gamma}}
\def\tvs{\tilde{{\varsigma}}}
\def\tu{\tilde{u}}
\def\tU{\tilde{U}}
\def\tw{\widetilde w}
\def\tW{\widetilde W}
\def\tB{\tilde B}
\def\tv{\tilde v}
\def\tV{\tilde V}
\def\tz{\tilde z}
\def\tb{\tilde b}
\def\ta{\tilde a}
\def\tih{\tilde h}
\def\trh{\tilde {\rho}}
\def\tx{\tilde x}
\def\tf{\tilde f}
\def\tg{\tilde g}
\def\tG{\tilde G}
\def\tk{\tilde k}
\def\tl{\tilde l}
\def\tL{\tilde L}
\def\tD{\tilde D}
\def\tR{\tilde R}
\def\tP{\tilde P}
\def\tH{\tilde H}
\def\tp{\tilde p}

\def\hH{\hat{H}}
\def\hh{\hat{h}}
\def\hR{\hat{R}}
\def\hY{\hat{Y}}
\def\hX{\hat{X}}
\def\hP{\hat{P}}
\def\hT{\hat{T}}
\def\hV{\hat{V}}
\def\hG{\hat{G}}
\def\hF{\hat{F}}
\def\hw{\widehat{w}}
\def\hW{\widehat{W}}
\def\hu{\hat{u}}
\def\hs{\hat{s}}
\def\hv{\hat{v}}
\def\hb{\hat{b}}
\def\hB{\widehat{B}}
\def\hze{\hat{\zeta}}
\def\hsi{\hat{\sigma}}
\def\hrh{\hat{\rho}}
\def\hth{\hat{\theta}}
\def\hy{\hat{y}}
\def\hx{\hat{x}}
\def\hz{\hat{z}}
\def\hg{\hat{g}}
\def\he{\hat{e}}
\def\hE{\widehat{E}}

\def\B{\mathbf{B}}
\def\I{\mathbf{I}}
\def\P{\mathbf{P}}
\def\G{\mathbf{G}}
\def\S{\mathbf{S}}
\def\F{\mathbf{F}}
\def\one{\mathbf{1}}
\def\Sn{\mathbf{S}_n}
\def\0{\mathbf{0}}
\def\H{\mathbf{H}}
\def\V{\mathbf{V}}

\def\f{\mathcal{F}}
\def\çF{\mathcal{F}}
\def\o{\mathcal{O}}
\def\t{\mathcal{T}}
\def\r{\mathcal{R}}
\def\l{\mathcal{L}}
\def\m{\mathcal{M}}
\def\k{\mathcal{K}}
\def\n{\mathcal{N}}
\def\d{\mathcal{D}}
\def\p{\mathcal{P}}
\def\cP{\mathcal{P}}
\def\a{\mathcal{A}}
\def\h{\mathcal{H}}
\def\c{\mathcal{C}}
\def\y{\mathcal{Y}}
\def\e{\mathcal{E}}
\def\v{\mathcal{V}}
\def\z{\mathcal{Z}}
\def\x{\mathcal{X}}
\def\s{\mathcal{S}}
\def\g{\mathcal{G}}
\def\u{\mathcal{U}}
\def\w{\mathcal{W}}
\def\i{\mathcal{I}}
\def\j{\mathcal{J}}
\def\b{\mathcal{B}}

\def\lan{\langle}
\def\llb{(\!(}
\def\ran{\rangle}
\def\rrb{)\!)}
 \def\dim{{\hbox{\rm dim}}_{\mathbb C}\,}
\def\lng{\hbox{\rm{\tiny lng}}}
\def\sht{\hbox{\rm{\tiny sht}}}
\def\sph{\hbox{\rm{\tiny sph}}}
\def\inv{\hbox{\rm{\tiny inv}}}

\def\br#1{\langle #1 \rangle}

\def\rank{\hbox{rank}}
\def\gl{\mathfrak{gl}_N}

\newcommand{\Aut}{\operatorname{Aut}}
\newcommand{\Hom}{\operatorname{Hom}}
\newcommand{\End}{\operatorname{End}}
\newcommand{\Ind}{\operatorname{Ind}}
\newcommand{\ad}{\operatorname{ad}}
\newcommand{\pr}{\operatorname{pr}}
\newcommand{\aweyl}{\tilde{\mathbb S}_n}
\newcommand{\hec}{{\mathcal H}^t_n}
\newcommand{\Func}{{\mathcal F}({\mathbb C}^n,{\mathcal H}^t_n)}
\newcommand{\tr}{\operatorname{tr}}
\newcommand{\Out}{\operatorname{Out}}
\newcommand{\Rad}{\operatorname{Rad}}
\newcommand{\Spec}{\operatorname{Spec}}
\newcommand{\id}{\operatorname{id}}
\newcommand{\Int}{\operatorname{Int}}
\newcommand{\ct} {\operatorname{ct}}

\newcommand{\rat}{{\mathbb Q}}
\newcommand{\real}{{\mathbb R}}
\newcommand{\cplx}{{\mathbb C}}
\newcommand{\zint}{{\mathbb Z}}

\newcommand{\sq}{\phantom{1}\hfill$\qed$}
\newcommand{\Rea}{\Re}
\newcommand{\Ima}{\Im}

\newcommand{\st}{\bowtie}
\newcommand{\modd}{\mbox{\,mod\,}}
\newcommand{\lr}{\langle}
\newcommand{\rr}{\rangle}
\newcommand{\eps}{\varepsilon}
\newcommand{\phk}{\phi^{(k)}}
\newcommand{\psk}{\psi^{(k)}}
\newcommand{\Res}{\mbox{Res}\;}
\newcommand{\sgn}{\mbox{sgn}}
\newcommand{\mn} {\left\{ \begin{array}{c}m\\
n\end{array}\right\}}

\def\sX{\mathscr{X}}
\def\sH{\mathscr{H}}
\def\sY{\mathscr{Y}}
\def\TT{\mathfrak{T}}
\def\JJ{\mathfrak{J}}
\def\HH{\mathfrak{H}}
\def\FF{\mathfrak{F}}
\def\GG{\mathfrak{G}}
\def\CC{\mathfrak{C}}
\def\LL{\mathfrak{L}}

\def\BB{\mathfrak{B}}
\def\AA{\mathfrak{A}}
\def\ZZ{\mathfrak{Z}}
\def\HH{\hbox{${\mathcal H}$\kern-5.2pt${\mathcal H}$}}
\def\HHH{\hbox{${\mathbb H}$\kern-4.2pt${\mathbb H}$}}
\def\tHH{\widetilde{\HH\ }}

\font\smm=msbm10 at 12pt 
\def\symbol#1{\hbox{\smm #1}}
\def\lsmash{{\symbol n}}
\def\rsmash{{\symbol o}}
\def\#{\sharp}

\font\tenbf=cmbx10
\font\tenrm=cmr10
\font\tenit=cmti10
\font\ninebf=cmbx9
\font\ninerm=cmr9
\font\nineit=cmti9
\font\eightbf=cmbx8
\font\eightrm=cmr8
\font\eightit=cmti8
\font\sevenrm=cmr7
\font\sevenbf=cmbx7


\title [Momentum managing epidemic spread]
{Momentum managing epidemic spread\\  and 
Bessel functions}
\author[Ivan Cherednik]{Ivan Cherednik, UNC Chapel Hill $^\dag$}

\begin{abstract}
Starting with the power law for the total number of detected
infections, 
we propose differential equations describing 
the effect of momentum epidemic management.
Our 2-phase formula matches very well the curves
of the total numbers of the
Covid-19 infections in many countries; the first phase
is described by Bessel functions. It provides 
projections for the saturation, assuming that the management
is steady. We discuss 
Austria, Brazil, Germany, Japan, India, Israel, Italy,  
the Netherlands, 
Sweden, Switzerland, UK, and the USA, including some 
analysis of the second waves.
\end{abstract}


\address[I. Cherednik]{Department of Mathematics, UNC
Chapel Hill, North Carolina 27599, USA\\
chered@email.unc.edu}


\thanks{$^\dag$ \today.
\ \ \ Partially supported by NSF grant
DMS--1901796 and the Simons Foundation}

\vskip -0.0cm
\maketitle
\smallskip


 \def\sht{\raisebox{0.4ex}{\hbox{\rm{\tiny sht}}}}
 \def\bysame{{\bf --- }}
 \def\~{{\bf --}}
 \def\rr{{\mathsf r}}
 \def\cc{{\mathsf c}}
 \def\mm{{\mathsf m}}
 \def\pp{{\mathsf p}}
 \def\ll{{\mathsf l}}
 \def\aa{{\mathsf a}}
 \def\bb{{\mathsf b}}
 \def\NS{\hbox{\tiny\sf ns}}
 \def\ssum{\hbox{\small$\sum$}}
\newcommand{\comment}[1]{}
\renewcommand{\tilde}{\widetilde}
\renewcommand{\hat}{\widehat}
\renewcommand{\V}{\mathbb{V}}
\renewcommand{\F}{\mathbb{F}}
\newcommand{\dagx}{\hbox{\tiny\mathversion{bold}$\dag$}}
\newcommand{\ddagx}{\hbox{\tiny\mathversion{bold}$\ddag$}}
\newtheorem{conjecture}[theorem]{Conjecture}
\newcommand*\toeq{
\raisebox{-0.15 em}{\,\ensuremath{
\xrightarrow{\raisebox{-0.3 em}{\ensuremath{\sim}}}}\,}
}

\vskip -0.0cm
\maketitle
\vskip -0.0cm
\noindent

\renewcommand{\natural}{\wr}


\vbadness=3000
\hbadness=3000

\baselineskip 12pt


\section{\bf Our approach and findings}\label{sec1}
A system of differential equations is proposed describing the 
effects of {\it momentum management 
of epidemics\,},  which is in this paper a set of reactive
measures mostly based on the latest total numbers of 
detected infections.
The {\it hard measures\,} are the key; the most important are the 
detection and isolation of infected people and closing the places 
where the spread is the most likely. If their intensity 
and consistency are high enough, our model
provides projections for the saturation of the epidemic spread,
followed by the {\it second phase}, which
is essentially a period of modest constant numbers of new 
infections. 
The {\em two-phase formula\,}, the main output of
this paper, was tested well for the spread of 
{\em Covid-19} in many countries.
The 1st phase is described by Bessel-type functions with 
surprisingly high 
accuracy. The exactness of the corresponding formulas for the
2nd phase is even more surprising taking into consideration 
many factors influencing the management of {\em Covid-19} in 
the later stages. The figures from Sections \ref{sec:twoph},
\ref{sec:auto} demonstrate our 
"2-phase solution" for Japan, Israel, Italy, Germany, the
Netherlands and UK. The latter and Sweden were the 
latest in Europe to reach the 2nd phase. The USA was 
close to it, but it was still in phase 1, when it
entered the second wave of {\it Covid-19}. The {\it second waves}
match out theory equally well. 
\vskip 0.2cm


 
The second phase is when "hard measures" are
relaxed or even abandoned. Self-isolation,
wearing the protective masks and 
social distancing become the key. Such and similar 
{\it soft measures\,}  reduce the transmission rate, but the 
1st phase is the key for reaching the saturation. 
\vskip 0.2cm

Our formula for the total number of cases for the 2nd phase is 
$C\,t^{c/2}\cos(d\log(t))$:  $t$ 
is the time, $d$ reflects the intensity of "soft" measures,
and $c$ is the {\it initial transmission rate\,}, which is as follows. 
The initial growth of the total number of cases is
$\sim t^{c}$, where $c$ is mostly 
from $2.2$ to $2.8$ for {\it Covid-19},
but reached $4.5-5.5$ for Brazil and India. 
The number of new daily infections becomes modest
during phase 2.
Also, asymptomatic (mild) cases begin to dominate,
which contributes to diminishing the spread too.
\vskip 0.2cm
\vfil

{\sf Focus on risk-management.}
We attribute the 
similarity of the curves of the {\it
total numbers of detected cases\,} in many
countries to the uniformity of the measures 
employed and to the ways people react to the threat,
more specifically, react to the 
growing numbers of infections.

The total number of cases is generally more reliable
and stable than other characteristics of {\it Covid-19},
though it depends.
It is not really important in our approach
that they mostly reflect {\em symptomatic
cases\,} and are frequently underreported.
{\it As far as they influence the decisions of the authorities
in charge and our own behavior, they can be used.\,}
Our focus on the epidemic
management resulted in algebraic-type formulas for the
curves of total cases, which explain very well the 
surprising uniformity
of such curves in so many different places.
 
\vskip 0.2cm

We identify 3 basic types of governmental  management.
The countries in the first group 
are determined to reach "double-triple digit numbers"
of new daily infections, which is our $(A)$-mode.
The second group is when the reduction  
of hard measures begins upon the 
first signs of the stabilization of the daily numbers,
even if they are high; this is a premature switch to the
 $(AB)$\~mode or $(B)$-mode for us.
The third group of countries is where
"hard" measures are not employed systematically, which
can be due to a variety of reasons, including  insufficient
medical capacities or political decisions. 

Some measures are of course
always in place: medical help,
self-isolation of those who think
that they can be infected, various self-imposed limits, 
and so on. We focus on {\it active momentum management}. 
\vskip 0.2cm

{\sf Testing our theory.}
Our research was organized as follows. The theory and
its applications to mode $(A)$, the most aggressive one,
were essentially completed around April 15 of 2020. Not too 
many countries reached
the "saturation" by then, but the middle stages matched
this new theory almost with an accuracy of physics laws.
The challenge was to understand the later stages. 
\vskip 0.1cm

The parameter $c$, the initial transmission rate,
can be obtained during the early stages; finding
$a$,  the intensity of hard measures,
generally requires the period till the "turning point".
Assuming that the measures are steady, this can be
sufficient for forecasting the spread, but not always. 
For instance,
$a,c$ coincided for the UK and the Netherlands. 
The latter reached the end of 
phase one after about 45 days
(counting from March 13), matching well our Bessel-type
formula. This appeared
significantly slower in UK, which we attribute to
the relaxation of the "hard" measures there after the 
"turning point".
It is even more visible in the USA, where the middle stage 
{\it perfectly\,} matched our formula
with the smallest $c$ we ever observed: $2.2$. Then the measures 
were reduced, phase 2 was not reached, and eventually
the USA entered wave 2.  
\vskip 0.2cm

At the end of April, we saw some signs of the switch from
mode $(A)$ to mode $(AB)$ in the USA and UK.
This is a transitional mode, where 
the main measures are "hard", but the response to the current 
number of infections is as in mode $(B)$, not really 
"aggressive". The corresponding $(AB)$-curves, called
$w$-curves in the paper, were calculated for the USA, UK 
on May 5. The expected dates of "phase-1 saturations"
under the $(AB)$\~mode were 
correspondingly May 30 and June 10 for these countries.
This worked well for UK, but did not materialize for the USA 
due to further significant reduction of the hard measures 
approximately in the middle of May.
The economic and societal impact of "hard measures"
is of course huge. However long periods of high daily numbers of
new infections obviously present significant risks.
     
\vskip 0.2cm
\vfil

Concerning the USA,
a period of essentially constant, but high,
numbers of new daily cases lasted for some time,
mathematically, similar to that in Sweden. 
However, when processing {\em automatically\,} all 50 states
individually, we found at the end of May,
that about 22 states were already in phase 2, which 
appeared sufficient to expect the general saturation
at 09/19. This projection remained quite stable till
the end of June, when the changes with the policies in 
{\it almost all} 50 states toward "opening" resulted in a 
very different scenario. 

The number of the states in phase 2 quickly dropped from 22 
only to 8: Colorado, Connecticut, Maine, Massachusetts, New Hampshire,
New Jersey, New York, Rhode Island as of July 8.
Then  USA entered the {\it second wave\,} at about
June 15, with the
starting number of total detected cases about $2M$. 
\vskip 0.1cm

{\sf The second wave.}
Upon subtracting initial numbers, the match
with the corresponding Bessel-type solution
appeared very good for the 2nd wave,
with the transmission rates $c\,$ 
comparable with those for the 1st wave. This was
expected in our theory. 
Our $c$ is some combination of the
transmission strength of the virus and the "regular" number of
contacts in the considered area. A preliminary analysis of
about 10 countries, shows that it can somewhat increase.
The parameter $a$, the intensity of the 
"hard" measures, certainly diminished significantly;
$1/\sqrt{a}$ is essentially proportional to the time till
the saturation. This is not unexpected: the second wave of
extensive 
lockdowns seems not too likely.

According to our automated program, the spread of {\it Covid-19}
remained mild in Western Europe till the middle of July, though
the following countries had clear second waves
as of July 8:
{\footnotesize Albania, Bosnia and Herzegovina, Bulgaria, Croatia,
Czech Republic, Greece, Kosovo, Luxembourg, Macedonia, Montenegro,
Romania, Serbia, Slovakia, Slovenia}. Also,
Sweden, Poland, 
Portugal and some other countries did not reach phase 2 at
that date. Then the number of new cases increased
at the end of July.

\vskip 0.2cm

We analyze in the paper 
the current second waves in Israel and the USA.
Among other factors, many schools
were open in Israel in June and quite a few summer activities
for children were held in July-August, 
which presumably contributed to the high magnitude of the 
second wave there. This is more complex in the USA, especially
because of the "unfinished" 1st wave. 

Nevertheless, it appeared that the curves of the total number 
of cases are very similar to each other in these countries, 
upon some natural rescaling. This uniformity and good match
with our Bessel-type solutions provide of course a confirmation 
of our approach. 
It is worth mentioning here that it is not
impossible that the measures can be not that "hard" for the
second wave to have the same effect, though the 
{\it detection-isolation-tracing\,} remains of course the key.  
\vfil

Our general theory remains  essentially unchanged 
from its first posted variant (April 13). However only now
all its main features are confirmed to
occur in reality, including the second  Bessel-type solutions, 
and the significance of modes $(B)-(AB)$. Let us comment on it.

The second, non-dominating,  Bessel-type solutions 
explained some "bulges" of the curves of total cases, 
during the early and middle stages in many countries.
Generally, {\it all} solutions of ODE, PDE must occur
when used for modeling.
Mode $(B)$ and the  
$\log(t)$\~saturation  
appeared  the key in  the second phase.
The $(AB)$-mode is not used much in this paper, but
creating "forecast cones" is of obvious importance.  
Last but not least, the fact that the Bessel-type formulas
describe well the {\em second waves\,} is 
a confirmation of our methods.

\vskip 0.2cm
\vfil

{\sf Classical theory.}
The solutions of
the basic classical equation for the number of infections
of communicable diseases is with the exponential growth 
of its solutions, which describes only 
initial stages of epidemics.
The growth is no greater than some power functions in time
during the middle stages, so these equations must be changed.

The equally 
classical {\it logistic\,} equations for the spread, 
as well as the SIR and SID models, assume that the number
of infections is comparable with the whole population,
which was not 
really the case with "major epidemics" we faced 
during the last 100 years.
This is mostly due to better disease control 
worldwide. The {\it herd immunity\,} is of course
of fundamental 
importance, but
many  epidemics were significantly reduced  
or terminated (well) before it had full effect. 

Thus, we must firstly address 
the {\it power-type growth of epidemics\,} except for  
short initial short periods of exponential growth (if any).
This can be clearly seen with
{\it Covid-19\,}, including sufficiently  long periods 
of essentially linear growth of the total number of infections.
And the saturation happens well before the number
of infected people becomes comparable with the whole population.

We mostly associate this  "polynomiality" with some assumptions on
the distribution of infected people, a kind of
local herd immunity: infected people do not transmit the
virus if {\it surrounded\,} by those infected or recovered.  
However some
sociological and biological factors contribute too; see e.g.  
\cite{CLL, ChAI, ChMed}.
\vskip 0.2cm

Whatever the theoretical foundations,
the {\it power law of epidemics\,} must be the starting 
point of any analysis if we want our mathematical models to be 
up to date, the challenges with {\it Covid-19\,}  included.
Our approach is  entirely based on the differential equations
that provide the power growth. The key was to extend them 
to the ones that could be used to model the 
saturation, which we see with {\it Covid-19} and other epidemics.
Our formulas
are not only accurate. They are simple and depend 
on very few parameters: essentially, the initial 
transmission rate $c$, and the intensities $a,d$ of the measures 
for phases 1,2. 
\vskip 0.2cm
\vfil

{\sf Behavioral aspects.}
There is a strong connection of our approach with
behavioral science, including {\it behavioral finance}. 
Our differential equations are actually from 
\cite{ChAI} devoted to {\it momentum risk-taking\,}, with
momentum investing as the main application.
The aggressive management of type $(A)$ from Section \ref{sec:Two}
is an almost direct counterpart of {\em profit taking\,} 
from Section 2.6 of this paper.
The measures of type $(B)$ are parallel to the investing regimes
discussed in Section 2.4.  The key link
to financial mathematics is that the 
{\em price function} from \cite{ChAI} is a 
counterpart of the {\em protection function\,} in the
present paper.
\vfil

Paper \cite{ChAI} can be considered as some step
toward {\it general purpose artificial 
intelligence}, the most difficult and ambitious 
among various {\it AI}\~related research directions.
 {\it Momentum risk-taking\,}
is a very universal concept. We even argue in
\cite{ChAI}  that 
the mathematical mechanisms we propose
can be present in {\em neural processes\,} 
in our brain; see Section 1.4 there.
The ways we manage risks real-time
are of course related to behavioral aspects of epidemics.


\section{\bf Two kinds of management}\label{sec:Two}
There is a long history and many
aspects of  mathematical modeling the epidemic spread;
see e.g. \cite{He} for a review. The present paper seems
the first one where the momentum management of epidemics 
is considered the {\it cornerstone\,}. To be more exact,
our $c$, the {\it initial transmission coefficient}, 
reflects the transmission strength of the virus and the 
"normal" number of contacts in the infected areas,
so it is "given". However,
the second parameter, the intensity $a$ of the measures,
is entirely about the management.
The basic management modes are as follows:
\vskip 0.1cm

{\em (A)}\ aggressive enforcement of the hard measures, where
their intensity is directly linked  to the current {\it absolute\,} 
number of infections;

{\em (B)}\ an approach when the {\em average\,} number of  
infections to date is the trigger and the employed 
measures are of more palliative nature.  
\vskip 0.1cm
\vfil

{\sf Hard and soft measures.}
To clarify, the main actions of type $(A)$ are testing, 
detection, and prompt isolation of infected people 
and those of high risk to be infected, as well as
closing places where the spread is the most likely.
Wearing protective masks, social distancing,
recommended self-isolation, restrictions on the size of events,
travel restrictions  are typical for $(B)$. 
The following is important to us.

Mathematically, the modes $(A)$ are $(B)$ are based on
different ways to respond to 
the total number of detected  infections: 
the {\it absolute\,} current number 
of infections is the trigger for $(A)$,
whereas the {\it average\,} number of infections to date
is the key under $(B)$.
 
The $(A)$\~type approach provides the fastest possible
"hard" response to the changes with the number of infections. 
With $(B)$,  we postpone with our actions until 
the averages reach proper levels, and the measures we 
implement are "softer". Mathematically, the averages
are better protected against stochastic fluctuations, 
but $(B)$ alone significantly delays the termination of the 
epidemic, as we will show within our model.
There is an 
analogy with {\it investing\,}, especially with
{\it profit taking}: an investor either directly
uses the {\it price targets\,} or prefers to rely on 
the so-called {\it technical analysis\,}, based on  
charting averages. 
\vskip 0.2cm

Let us emphasize that 
by "response" and "actions", we mean not only those by the authorities
in charge of the epidemic. Our own ways of 
reacting to epidemic figures are equally important. We can
"monitor"  the {\it total\,} number of infections 
and act accordingly, or mostly "consider" this number divided
by the time from day 1, some substitute for 
the number of new daily infections. For instance, when 
the number of new infections is constant (even high), this
can be acceptable for some. On the other hand, this means 
some steady growth of the total number of infections, which
can be troubling if these numbers  are already high.
Epidemiology  has strong roots in 
behavioral science, psychology,
sociology, mass and collective behavior; see e.g. \cite{St}.  
\vskip 0.2cm

{\sf Saturation.}
Both modes, $(A)$ and $(B)$, are momentum,  
responding to the latest data. As we will see, 
both 
provide essentially the same kind of growth of
the total number of infections in the beginning:
$\ \sim t^{c}\,$ in terms of time $t$ and
for the initial transmission rate $c$. 
The main difference is that the solutions of
the differential equations of type $(A)$ are
{\it quasi-periodic\,}: asymptotically  periodic
functions in $t$ multiplied 
by power functions. This automatically
grants sufficiently fast "saturation", which is
 reaching the first
maximum. Then they cannot be used for modeling the
{\it total\,} number of infections: it always increases. 
For $(B)$, we have $\log(t)$-quasi-periodicity, 
but it reveals itself only in the late stages;
it can be really seen with {\it Covid-19}
during {\it phase two\,}.
   
The periodicity of our solutions is  
not related to the
periodicity of the epidemic models based on seasonal 
factors, various 
delays and other mechanisms of this nature;
see e.g. \cite{HL}. The periodicity and saturation
we consider are entirely 
due to active {\it momentum\,} management.  
\vskip 0.2cm

The "forced" saturation of this kind can be unstable.
Reducing the measures too much on the first 
signs of improvement is likely to result in further {\it waves\,} 
of the epidemic, which can be very intensive. This
was expected theoretically and now it becomes a serious concern.
\vskip 0.2cm

\section{\bf Power law of epidemics}
Here and below we will assume that the number of people
perceptive to the virus is unlimited, i.e. we do not take into 
consideration in this paper any kind of saturation when the
number of infected people is comparable with the whole
population. Accordingly, {\it herd immunity\,} and similar
factors are not considered.
Also, we disregard the average duration of the disease 
and the durations of the quarantine periods. 

The {\it total 
number of detected infections\,} is what we 
are going to model. This is commonly used; the corresponding
data are widely available and seem the most reliable.   
\vskip 0.1cm
  
For any choice of units, days are the most common, let
$U_n$ be the number of infected individuals at the moments 
$n=0,1,2\ldots$. The simplest equation of the epidemic
spread and its
solution are:
\begin{align}\label{Un}
U_n\!-\!U_{n-1}\!=\!\si \,U_{n-1} \for n=1,2,\ldots, \ \,\and  
U_n= C(1\!+\!\si)^n
\end{align}
for some constant $C$ and  the {\it intensity\,} $\si$.
Here $\si$ is
the number of infections transmitted by an average
infected individual during 1 time-unit, 
assuming that the "pool" of 
non-infected perceptive people is unlimited.
The problem here is
that the  exponential growth of $U_n$
can be practically present only during 
the early stages of epidemics, especially with the epidemics we
faced during the last 100 years.

An important factor ignored in (\ref{Un}) is that
$U_n-U_{n-1}$
is actually proportional to  $U_{n-1}\!-\!U_{n-p}$, where 
$p$ is the period when the infected people transmit the virus in 
the most intensive way. Switching to $p$ as the time-unit, we
arrive at $U_n-U_{n-1}=\rho(U_{n-1}-U_{n-2})$, where $\rho$
is essentially the {\it basic reproduction rate\,} $R_0$.
One has: $U_n=C_1\rho^n+C_0$ for some constants $C_{0,1}$;
so the growth is still exponential for $\rho>1$.  
\vfil

We are going to replace $U_{n-1}-U_{n-2}$ by 
$U_{n-1}/(n-1)$ with some coefficient of
proportionality. This can be obviously done
if $\,U_n\,$ grows essentially linearly.
Generally,  the reasons for such 
a significant change must be and really are of fundamental nature. 
\vskip 0.2cm

{\sf Some psychological aspects.}
Presumably, the linear growth of $U_n$ is not really 
sufficient  to force us, people, to change our behavior,
even with epidemics. 
However, we certainly begin reducing our contacts and 
consider other protective measures
if the {\it trend} seems faster than linear. This is
not only with epidemics. The question is what we mean by 
"trend" and how we measure it, at 
conscious and subconscious levels.

The ratio $U_{n-1}/(n-1)$ seems almost perfect
to represent it. Generally, $(U_{n-1}\!-\!U_{n-2})$
gives this, but it may be not what we really use
in this and similar decision-making situations. 
First, such differences or the corresponding derivatives 
are poorly protected 
against the random fluctuations of $U_n$.
Second, momentum decision-making is very intuitive
and sometimes entirely subconscious.  
"Storing and processing"
the prior $U$-numbers  in our brain is more involved than
"keeping" $U_{n-1}/(n\!-\!1)$. 
Our brain can perfectly process the input like
"yesterday number was $U_{n-1}$ and this took $(n\!-\!1)$ days",
though the  mechanisms are complicated. The concept of time
is quite sophisticated. 

Also, let us mention here that the actions of infected people or
those who think they can be infected become
less chaotic over time and they generally receive better medical 
help in the later stages, 
which reduces the transmission of the virus.

\vskip 0.2cm

\vfil
{\sf Some biological aspects.}
The {\it viral fitness} is an obvious component of  $\si$.
Its diminishing  over time can be expected, but this
is involved. This can happen because of  
the virus replication errors, especially typical 
for {\it RNA\,} viruses, which are of highly variable and 
adaptable nature. The {\it RNA\,} viruses, 
{\it Covid-19\,} included, replicate with fidelity 
that is close to error catastrophe. See e.g. \cite{CJLP,Co}
for some review, perspectives and interesting predictions. 
Such matters are
well beyond this paper, but one biological aspect must
be mentioned: {\it asymptomatic cases}. 
\vskip 0.1cm

The viruses mutate at very high rates. They can "soften" 
over time to better coexist with the hosts, though fast and 
efficient spread is of course the "prime objective" of any virus. 
For instance, quarantine measures may 
"force" the virus to stay longer in a host; mild strains may
have advantages. 
Such softening  certainly results in an increase of 
asymptomatic cases.
Since we model the available (posted)
numbers $U_n$, which  mostly reflect the symptomatic 
cases, the coefficient $\si$ automatically diminishes 
when the percentage of the asymptomatic cases increases. 
Anyway, softening the virus over time 
contributes to diminishing $\si$. 
We will provide more direct
"geometric" reasons for the switch from $\si$ to $\frac{\si}{n-1}$
below: some kind of "local herd immunity". 
\vskip 0.2cm

{\sf Master equation revisited.}
We arrive at the following equation for the {\it middle 
stages of epidemics\,},  resulting in the power growth 
(asymptotically) of 
its solutions:
\begin{align}\label{Unpower}
U_n-U_{n-1}= \frac{\si}{n-1}\,U_{n-1},\ n=2,3,\ldots,\ \,
U_n \approx Cn^{\si}
\end{align}
for some constant $C$. When $\si=1$, $U_n$ becomes exactly
$C n$. 

Actually, it is not that important in this paper what are
the exact reasons for dividing $\si$ by 
the time here, though we suggest some underlying 
principles. What is the key for us is that the switch from 
(\ref{Un}) to (\ref{Unpower}) is a {\em mathematical necessity};
the exponential growth is unsustainable. 
\vskip 0.1cm

The power growth
can be unsustainable long-term too, but such
growth is really present in epidemics, 
{\em Covid-19\,} included, especially during the middle
stages. Needless to say that power laws are 
fundamental everywhere in natural sciences. We outlined
some behavioral and biological aspects above, but we
think that the most plausible explanation of the applicability
of (\ref{Unpower}) is "geometric".
\vskip 0.2cm

Equation (\ref{Unpower}) and its differential 
counterpart serve several seemingly different "situations".
One example is {\it news propogation\,} 
from \cite{ChAI}; this is connected with the spread
of epidemics. Let us discuss briefly two other
examples where the same equation is applicable; they are 
from \cite{ChAI} too. 
\vskip 0.2cm

{\sf Tree growth.} Equation (\ref{Unpower}) reasonably
describes
the height $U_n$ of a tree in year $n$
at least for $\si\approx 1$. The division of $U_{n-1}$
by $(n-1)$
can be interpreted as follow. The radius $r$ of the root system
can be assumed proportional to the tree radius.
Therefore the root system, which is basically flat, must
provide nutrition for the whole tree which is 3-dimensional.
We obtain that the growth of a tree during one year is 
essentially proportional to $r^2/r^3=1/r$ times its prior
height, $U_{n-1}$. The final assumption is that $r$ is 
proportional to $n$, which essentially means that
the distances between consecutive 
{\it tree rings\,} are constant. 

In the beginning and at the
end of the life cycle of a tree, this can be different. The
volume of the tree can be rather $r^2$ than $r^3$ in the very
beginning, so the growth can be faster than polynomial.
On the other hand, the nutrition the root system provides becomes 
proportional to $r$ rather than to $r^2$ in the later stages.
The corresponding term $\frac{U_{n-1}}{(n-1)^2}$
readily results in the saturation of the tree size, which is 
well-known and used practically in {\it bonsai}.  
\vskip 0.2cm

{\sf Neural activities.}
We expected in \cite{ChAI} that (\ref{Unpower}) can serve
some basic processes in our brain. The assumptions
here are of geometric nature.
Let the number of neurons used for a particular task be 
$U_n$ at the moment $n$.
The neural architecture of our brain is obviously about  
connections (axons) rather than about physical distances 
between neurons in the brain. Some {\it auxiliary\,}
$N$-dimensional
space is needed to present the corresponding "graph" as 
an image in $\R^N$ where the distances provide the numbers 
of connections. Then its 
{\it frontier\,} (border) is the main source of the 
expansion of this brain activity. 
\vskip 0.2cm

Assuming that the neurons involved in this task 
are uniformly
distributed in the {\it image} of radius $r$, we obtain that
$U_n\!-\!U_{n-1}$ is essentially proportional to 
$\frac{U_{n-1}}{r}$, where
$U_n\sim r^{\si}$ for $\si\le N$. Finally, we assume that
the radius $r$ grows linearly in time.

Here $\si\le N$ can be actually any number due to the shape of
the image in $\R^N$.
We arrive at $U_n-U_{n-1}=\si\,\frac{U_{n-1}}{n-1}$ for
proper $\si$. This model is of course a gross
simplification; in contrast to trees and epidemics,
we do not know much about the real processes
in our brain. 
\vskip 0.2cm
\vfil

{\sf Back to epidemics.}
Following the last example, we represent the population
of some area perceptive to the virus as an image 
in some auxiliary $\R^N$; the  numbers of connections/contacts 
between people then becomes essentially proportional to
the {\it geometric\,} distances between the corresponding points.
The infected individuals  fully "surrounded" by infected
 people or those already recovered do not 
transmit the 
disease. Geometrically, they are represented by the
points {\it inside\,}
the image. Its boundary matters the most: these people 
are better positioned to transmit the virus. In a sense, the
{\it herd immunity\,} is reached {\it inside\,} this image.
As above, we assume
that the points representing people are distributed uniformly
in the image and its radius is basically proportional to
the time. We arrive at (\ref{Unpower}) with $\si\ge 2$;
if the contacts are mostly between the neighbors then expect
$\si\approx 2$, though the {\it intensity\,} of the contacts 
and the strength of the virus matter too of course. 

The uniformity assumption we made is quite standard in physics,
say in theory of gases or statistical physics. 
As everywhere in natural sciences,
the simplicity and universality of
the resulting equation are of course important considerations.
Note that in contrast
to $\si$ from (\ref{Un}), the constant $\si$ in
(\ref{Unpower}) and in its differential version
are {\it dimensionless}, i.e. this equation
remains the same if we change the time units. 
This is significant: this constant serves as an exponent. 
\vfil

Recall that we discuss here the epidemic spread without any active
management. We see that the polynomial growth of the spread can
be "deduced" essentially from one {\it postulate\,} of
geometric nature: {\it the 
"frontier" contributes the most to the epidemic spread}. 


\vskip 0.2cm

\section{\bf Hard measures}\label{sec:Pro}
In the realm of differential equations, our starting
equation (\ref{Unpower}) 
becomes as follows:
\begin{align}\label{Ut}
\frac{dU(t)}{dt}\ =\  c\,\frac{U(t)}{t}, 
\hbox{\ \,where $t$ is time\,}
\hbox{\, and $c$ replaces $\si$}.
\end{align}
We apply Taylor formula to $U_{n}-U_{n-1}$, so 
$c$ is "essentially" $\si$; we will use mostly $c$
from now on.

We will model now the impact of aggressive protective 
measures. The key is the introduction of some
{\it protection function\,} $P(t)$,
the total output of the measures.

It is some counterpart of  the 
{\it price function\,} in \cite{ChAI}. Essentially, we have
two types of measures: $(a)\,$ isolating infected
people, and $(b)\,$ general diminishing the number of contacts. 
The are "hard" and  "soft" measures, considered above, corresponding
to modes  $(A)$ and $(B)$; see below and Section \ref{sec:Two}. 
Let us begin with the "hard" ones.   
\vskip 0.2cm

{\sf Protection function.} 
The main "hard" measures are  {\it testing\,} and 
{\it detection} of infected individuals, followed by their
isolation and "tracing".  They are hard by any standards, especially
if there are many cases and the treatment
is unknown, as with {\it Covid-19}. 

Generally speaking,  $P(t)$ provides the total numerical 
"output" of all employed  measures. Its derivative gives their 
{\it productivity} by definition.
For "detection \& isolation", we are going to
use the following natural definition of $P$:
\vskip 0.2cm

\centerline{\it  
$P(t)=\frac{1}{c}\,($the total number of prevented infections 
from $0$ to
$t)$.}
\vskip 0.2cm

The primary measure here is {\it testing\,}; the number $T(t)$
of tests till $t$ is what we can really implement and control. 
The {\it detection\,} of infected people is its main purpose,
but the number of tests is obviously not directly related to the 
number of detections, i.e. to the number of {\it positive 
tests\,}, and to the number of resulting {\it isolations}.  

The efficiency of testing is complicated to be  measured
directly. However, the number of positive tests
can be mostly assumed  a stable fraction of the total
number of tests, especially during the periods of
stable growth of the number of total cases. 
When $U(t)=\ga T(t)$ for some $\ga$, finding $\ga$
experimentally is not a problem. Accordingly,
the system of differential equations will be only
for $U$ and $P$.


We assume now that the  effect of 
isolating infected people is approximately linear.
More exactly,
an infected person isolated at $t_\bullet<t$
will {\it not\,} transmit 
the disease to $(t-t_\bullet)$ people till the
moment $t$ (now).
This kind of "linearization" is very common when composing 
differential equations; it 
does not mean of course that the solutions of our equations will 
grow linearly. 
\vskip 0.1cm

Thus, if the isolations occur at the moments $\{t_i\}$,
then these infected individuals
will {\it not\,} infect $P(t)=\frac{1}{c}\sum_i (t-t_i)$
people till $t$ (now). If the latter group of people
were not protected,
they would infect $\de P(t)$ people from  $t$ to $t+\de$, 
assuming the transmission with the coefficient $c$ for them.
So $P(t)$ must be subtracted from $dU(t)/dt$. 

Finally, $c\,dP(t)/dt$ is the number of isolated people
by construction. We assume it to be
 $\al\,U(t)$, a constant fraction of  all 
infected people, isolated or not.
This is another linearization.  The coefficient $a\equal \al/c$
is the {\it intensity} of the isolation process.
We arrive at the system:
\begin{align}\label{UPa}
dU(t)/dt= c\,U(t)/t - P(t),\  dP(t)/dt= a\,U(t). 
\end{align} 

The quarantine period, 
mostly 14 days for {\it Covid-19}, the duration of the disease
and similar factors are disregarded here. However, they
are partially incorporated in  (\ref{Ut}) through $c$.
Interestingly, such and similar simplifications 
appeared quite relevant and provided high accuracy of 
our modeling. We also completely disregard the stochastic nature
of epidemics in this paper. Random processes are supposed
to be used in a more systematic theory.
\vskip 0.1cm

With closing factories, schools and other places of high risk,
not all people there would be potentially 
infected if they continued to operate. The {\it exact\,} 
effect of closing (and then reopening) 
factories and other places where the fast spread of the disease 
can be expected is almost impossible  to estimate. 
However statistically, this preventive measure is actually no different from
the isolation of infected individuals.

\vskip 0.2cm
Other $P$-functions can be used here, but this
particular one has a very important feature: {\em it
does not depend on the moments of time when the
infected individuals were detected}.  If $P(t)$ depends on
$\{t_i\}$, the required mathematical tools can be much
more involved. 
Since $P(t)$ is actually for us to define, it makes sense 
to "postulate" its main mathematical properties needed to
compose the corresponding differential equations;
this will be done in $(i),(ii)$ below. 
\vskip 0.2cm

\vskip 0.2cm

We emphasize that the  measures of type $(A)$ impact $U(t)$ in
complex ways: isolating one individual prevents a 
{\it ramified\,} sequence of
transmissions.  The "soft" measures of type $(B)$, to be considered
next, are simpler: they result in a kind of 
reduction of the $c$\~coefficient.
\vskip 0.2cm

\section{\bf Soft measures} \label{sec:soft}
These measures are different from "hard" ones.  
Wearing protective masks is a key  
preventive measure of this kind; social distancing, 
considered mathematically, is of similar type. Now:
\vskip 0.1cm

{\it $P(t)=$ the number of {\it infected\,}
people who began wearing the masks before $t$ multiplied 
by the efficiency of the mask and the $c$\~coefficient.}
\vskip 0.1cm

Using the masks for
the {\it whole\,} population, infected or not, "simply" changes $c$\,;
no new differential equations are necessary.  Let $V(t)$ be the
number of infected people wearing the masks. Then $V(t)\le U(t)$
and

\begin{align}\label{UPb}
\frac{dU(t)}{dt}= c\,\frac{U(t)\!-\!V(t)}{t} + 
c'\,\frac{V(t)}{t} =
c\,\frac{U(t)}{t} - (c-c')\,\frac{V(t)}{t}
\end{align}  
for $c-c'=c(1-c'/c)=c\kappa$, where $\kappa$
is the mask efficiency. For instance, $\kappa=1$ if $c'=0$,
i.e. if the mask is fully efficient for infected people
who wear it.
The same consideration works for 
{\it social distancing\,}, with $\kappa$ being 
the efficiency of the corresponding distance.  
\vskip 0.1cm

An instructional example is when we assume that the
fraction of those wearing the masks among all infected
people is fixed. If $V(t)=\nu U(t)$ for some $0\le \nu\le 1$, 
then we have:
$$
dU(t)/dt= c\,(U(t)-\nu U(t))/t + c'\nu\,U(t)/t =
(c-\nu(c-c'))\,U(t)/t.
$$
I.e. this measure results in fact in a recalculation of  
the $c$\~coefficient under the proportionality assumption.
 
Generally speaking, the output of "soft"
measures is heavily based on probabilities. However, we
only need the following: 
the greater {\it intensity\,} of any
measure the greater the reduction of new infections. So we
"allow" only one way to control the efficiency of a measure:
by changing its {\it intensity\,}.
The exact mechanisms of its impact are not really needed 
to know in this approach. Let us formalize this.

\vskip 0.2cm
{\sf General approach.} 
From now on, the  {\it intensity\,} of a measure or 
several of them will be the main control parameter.
As we already discussed, the greater
the intensity, the greater the number of isolated infected
individuals in $(A)$ or the number of infected people 
who use the masks in $(B)$. 
If this dependence is of linear type, which can be expected,
the corresponding coefficient of proportionality is
sufficient to know. Respectively, 
the exact definition of the protection function 
$P(t)$ is not really needed to compose the corresponding 
differential equations.  What we really need is as follows:

\vskip 0.2cm
($\boldsymbol{i}$)\, the usage of $P$ reduces $dU(t)/dt$, possibly
with some coefficient of proportionality, by  $P(t)/t\,$ for 
mode $(B)$, 
the average of $P$ taken from $t=0$, or directly by $P(t)$ under
the most aggressive mode $(A)$;

($\boldsymbol{ii}$)\, the derivative $dP(t)/dt$, the productivity,
 is proportional
to $U(t)/t$ under $(B)$, the average number of infections from
$t=0$, or directly to $U(t)$ in the most aggressive variant,
which is $(A)$ considered above.
\vskip 0.3cm

Item $(i)$ has been already discussed. Let us clarify  $(ii)$, 
which provides  the {\it productivity of
the measures\,} $dP(t)/dt$ in terms of $U(t)$. 

The proportionality to
$U(t)$ for $(A)$ and $U(t)/t$ for $(B)$
 are the most natural
choices.
Indeed, the effect of the current number of infections on
our actions can be either direct or via some averages.
If the averages are used, $U(t)/t$ is quite reasonable,
as we argued above,
Mathematically, 
a relatively fast reaching the saturation requires mode $(A)$
or the transitional mode $(AB)$,
which is defined
as follows. 
\vskip 0.2cm

We employ "hard" measures as in $(A)$, however follow
less aggressive "management formula" for $dP(t)/dt$
from $(B)$. I.e. this is really some {\it transitional\,} mode. 
As we will see, the epidemic will end under $(AB)$, but the time to 
the "saturation" will be longer than under $(A)$. If only $(B)$
is used, this is uncertain.

\section{\bf Type (B) management}
Let $t=0$ be the starting point of the management;
so we can assume that $P(0)=0$. The following
normalization is somewhat convenient:
 $u(t)\equal U(t)/U(0)$ and $p(t)\equal
P(t)/U(0)$, i.e. $u(0)=1$ and $p(0)=0$. Recall that
$U(t)$ and $P(t)$ are the number of infections at the moment
$t$ and the corresponding value of the $P$-function.
Here $P$ is naturally for the sum of all measures 
in the considered mode, the sum of their $P$-functions. 

\vskip 0.2cm
Obviously  mode $(A)$ is significantly
more aggressive than $(B)$. 
Philosophically, the smaller interference in  
natural processes, which is under $(B)$,
the better: the effect will last longer. 
But with epidemics, we cannot afford waiting too long.  
\vskip 0.2cm

{\sf Type $(B)$ equations.}
As it was stated in Section \ref{sec:Pro},
we couple $(i)$ of type $(B)$, which is 
relation (\ref{UPb}), with $(ii)$ for the same mode. 
I.e. the derivative of $u(t)$ will be "adjusted" by 
$-p(t)/t$ and, correspondingly,
the rate of change of $p$ will be taken 
proportional to $u(t)/t$.  
This means that the {\it impact of $u(t)$ to $p(t)$ and vice versa
goes through the averages\,}; i.e. the response to $u(t)$
is not "immediate" as in $(A)$.

Note that the
exact  $-p(t)/t$ (later, $-p(t)$) in the equation 
for $du(t)/dt$, i.e. without a coefficient of
proportionality,  is a matter of  normalization. 
This coefficient  can and will
be "moved" to the second equation.
\vskip 0.1cm

The equations under $(B)$ become as follows:
\begin{align}\label{1}
&\frac{du(t)}{dt}\, =\, c\, \frac{u(t)}{t}- \frac{p(t)}{t},\\
&\frac{dp(t)}{dt}\, =\, \frac{a}{ t}\, u(t).\label{2}
\end{align}

Here $c$ is the initial transmission coefficient, and $a$ is the intensity of
the protection measure(s). This system can be readily integrated.
Substituting $u(t)=t^r$,
the roots of the characteristic equation are 
$r_{1,2}=c/2\pm \sqrt{D}$, where   $D=c^2/4-a.$
Accordingly, when $D\neq 0, t>0$:
\begin{align}\label{6}
&u(t)=C_1 t^{r_1}+C_2 t^{r_2}\, \hbox{\ \, if\ \,} D>0 
\ \, \hbox{for constants\,}\ \, C_1, C_2\,,\, \and\\
&u(t)=t^{\frac{c}{2}}(C_1 \sin(\sqrt{-\!D}\log(t))\!+\!
C_2 \cos(\sqrt{-\!D}\log(t)))\, \hbox{\ if\ } D<0,
\label{7}
\end{align}  
where the constants are adjusted to ensure the
initial conditions. Note that a proper 
branch of $\log$ 
must be chosen for $t$ near zero.

So for $a\!<\!c^2/4$, the initial $t^c$ (for $a=0$) will be 
reduced up to $t^{r_1}$ with  $c/2\!<\!r_1\!<\!c$. 
When $a>c^2/4$, the power growth is always of type 
$t^{c/2}$, and the period of $\sin$ and $\cos$ with respect 
to $\log(t)$ is $2\pi/(\sqrt{a\!-\!c^2/4})$. The corresponding
"saturation" is defined as the first $t$ such that $du(t)/dt=0$;
here  $t^{c/2}$ obviously contributes.
 
\vskip 0.2cm

{\sf Positivity of $du/dt$.} The control parameter $a$,
the intensity, is actually not arbitrary. To see this
let us invoke  $v(t)=V(t)/U(t_0)$, where $V(t)$ is the
number of infected people wearing the protective masks.
It was used in the definition of $P$:\ $P(t)=\kappa c V(t)$,
where $\kappa$ is the efficiency of the mask (from $0$ to $1$).
We then have:
\begin{align}\label{1v}
&\frac{du(t)}{dt}\, =\, c\, \frac{u(t)}{t}- \kappa c\frac{v(t)}{t},
\ \ \frac{dv(t)}{dt}\, =\, \frac{a}{\kappa c}\, \frac{u(t)}{t}.
\end{align}
Here $v(t)\le u(t)$ by the definition. This inequality
generally provides no restriction on the derivatives. However,
if $v(t)$ is essentially proportional to $u(t)$, 
there are some consequences. Let us assume that
$v(t)\sim \ga u(t)$. Then we obtain the relation:
$a=c^2\kappa\ga(1-\kappa\ga)$. Accordingly, the
maximal value of $a$ is $a_{max}=\frac{c^2}{4}$ in this case, which
is at $\kappa=\frac{1}{2\ga}$ provided that 
$1\!\ge\! \ga\!\ge\! \frac{1}{2}$.
This gives  $D=0$ for $\ga=1$, i.e. this is the case
when all wear the masks
of efficiency $1/2$; then $u(t)=Ct^{c/2}$, which is
obvious from to the definition of $\kappa$. 
\vskip 0.1cm

Generally, 
$du(t)/dt\ge c(1-\kappa)u(t)/t$, since $v(t)<u(t)$, and
$d\tilde{u}(t)/dt\ge 0$ for $\tilde{u}(t)=
u(t)t^{c(\kappa-1)}$. Thus the solutions $u(t)$
from (\ref{6})\&(\ref{7}) qualitatively 
grow faster than 
$t^{c(1-\kappa)}$ due to (\ref{6}).
\vskip 0.2cm

Using the 
$log(t)$\~periodicity (if present),  the
{\em $(B)$\~type saturation points} is the first maximum 
of $u(t)$.  We must stop
using $u(t)$ after this, because it will begin to decrease.
This $(B)$-saturation  
appeared an important factor during the 
{\em second phase\,} of our "$2$-phase solution".
If the {\em first phase}, which is under $(A)$, 
reaches its saturation and results in
a relatively small numbers of new daily infections,  
then mode $(B)$ will governs "the rest". The countries
then almost automatically switch to "soft" measures.
See Section \ref{sec:twoph}.

\vskip 0.2cm

\section{\bf Type (A) management} \label{sec:typeA}
The most "aggressive" model of momentum
management is of type $(A)$, described by system in 
(\ref{UPa}).   
We replace the average $p(t)/t$ in (\ref{1})
by $p(t)$, and $u(t)/t$ by $u(t)$ in (\ref{2}). Actually,
the first change affects the solutions greater than the second. 
One has:
\begin{align}\label{14}
& \frac{du(t)}{ dt}\ =\ c\,\frac{u(t)}{t}-
p(t),\\ \label{15}
&\frac{dp(t)}{ dt}\ =\ a\,u(t).
\end{align}

Solving  system (\ref{14})\&(\ref{15})
goes as follows:
\begin{align}\label{ptut}
&t^2\frac{d^2p}{dt^2}-c t \frac{dp}{dt}
+et^2p\ \, =\ \,0\ \,=\ \, 
 t^2\frac{d^2u}{dt^2}-c t \frac{du}{dt}
+et^2u+cu,\\
&u\!=\!A_1u^1\!+\!A_2u^2,\  
u^{1,2}(t)\!=\!t^{\frac{1+c}{2}} 
J_{\al_{1,2}}(\sqrt{a}t) \hbox{\, for\, }
 \al_{1,2}\!=\!\pm\frac{c\!-\!1}{2}. 
\label{ptut1}
\end{align} 
Here the parameters $a,c$ are assumed generic,
$A_{1,2}$ 
are undermined constants, and we
use the {\it Bessel functions\,} of the first kind:
$$ J_\al(x)=\sum_{m=0}^\infty 
\frac{ (-1)^m (x/2)^{2m+\al}}{m! \Ga(m+\al+1)}.
$$
See \cite{Wa} (Ch.3, S 3.1). We will also need the
asymptotic formula from S 7.21
there:
$$ J_\al(x)\sim \sqrt{\frac{2}{\pi x}}\cos(x-\frac{\pi\al}{2}
-\frac{\pi}{4}) \for x>\!> \al^2-1/4. 
$$
It gives that $u^{1,2}(t)$ are approximately
$\c\, t^{c/2}\, \cos(\sqrt{a} t-\phi_{1,2})$ for
some constant $\c$ and
$\phi_{1,2}=\pm\frac{c-1}{2}\pi+\frac{\pi}{4}$. 
\vskip 0.2cm

{\sf Quasi-periodicity.}
We conclude that
for sufficiently big $t$, the function $u(t)$ is basically:

\begin{align}\label{pas}
u(t)\approx 
t^{c/2}\bigl( A\sin (\sqrt{a}t+\pi c/2)+B\cos(\sqrt{a}t-\pi c/2)
\bigr),
\end{align}
for some constants $A,B$. So it is {\it quasi-periodic\,},
which means that the periodicity is up to a power function and
only asymptotically; the asymptotic  $t$-period is
$\frac{2\pi}{\sqrt{a}}$. 


In our setting, $u(t)$ must always grow, so $du(t)/dt\ge 0$.
The technical end of phase one is when $u(t)$ reaches its 
first maximum; $\frac{\pi}{2\sqrt{a}}$ is a reasonable {\it estimate},
but not too exact. For instance,
$u(t)=t^{c/2+1/2}J_{(c-1)/2}(\sqrt{a}t)$ for $c=2.2, a=1/5$
reaches its first maximum at about $t=4.85$, not at
$\frac{\pi}{2\sqrt{a}}=3.51$; see Figure \ref{fig:Bessel}.
\vskip 0.2cm

{\sf Transitional mode.}
It can be applicable to model the epidemic spread when
hard measures are certainly used, but are
applied "cautiously". 
This mode is transitional
between $(A)$ and $(B)$: one replaces (\ref{15})
by $dp(t)/dt=au(t)/t$, but couple it with 
the equation for $du(t)/dt$
from mode $(B)$, which is (\ref{1}):

\begin{align}\label{14a}
& \frac{dw(t)}{ dt}\ =\ c\,\frac{w(t)}{t}-
p(t),\ \ 
\frac{dp(t)}{ dt}\ =\ b\,\frac{w(t)}{t},
\end{align}
where we replaced $u(t), a$ by $w(t), b$; the 
$c$ remains unchanged.
It was called
{\it transitional $(AB)$\~mode\,}
at the end of Section \ref{sec:Pro}. Actually, this
mode is the part of the theory practically tested the least.

It is aggressive enough to provide the saturation,
and better protected  against fluctuations than $(A)$. 
This makes sense
practically; let us see what this gives theoretically.

The system remains still integrable in 
terms of Bessel functions; see formulas (2.20), (2.21) in 
\cite{ChAI}. The leading fundamental solution is
\begin{align}\label{Bes2}
w^1(t)=t^{(c+1)/2}J_{c-1}(2\sqrt{bt}), \where c>1.
\end{align}
The second one is $w^2(t)=t^{(c+1)/2}J_{1-c}(2\sqrt{bt})$.
One has:
\begin{align}\label{uwGa}
&u^1(t)\approx t^{c}\,\frac{(\sqrt{a}/2)^{(c-1)/2}} {\Ga((c-1)/2)},
\ w^1(t)\approx t^c\,\frac{(\sqrt{b}/2)^{c-1}} {\Ga(c-1)},
\end{align}
when $t$ is sufficiently close to zero. Practically,
they are "almost" proportional with the coefficient calculated 
from (\ref{uwGa}) in sufficiently large intervals of $t$,
not only for $t\sim 0$. This is important
for forecasting. 

The quasi-periodicity will be now
with respect to 
$t^{1/2}$ and  the corresponding asymptotic 
period will be $\pi/\sqrt{a}$. I.e. the process of
reaching the saturation is "slower"
than for $(A)$, but still significantly faster than for $(B)$, where 
the "mathematical periodicity" (if any!) is in terms of 
$\log(t)$ in (\ref{7}).
This perfectly matches the qualitative description of $(AB)$
as a transition from $(B)$ to $(A)$.
\vskip 0.2cm

Actually we have a family $(AB)_\mu$ for  $-1\le \mu<1$
of such modes, described
by the system
\begin{align}\label{14mu}
& \frac{dw(t)}{ dt}\ =\ c\,\frac{w(t)}{t}-
p(t)/t^\mu,\ \ 
\frac{dp(t)}{ dt}\ =\ b\,\frac{w(t)}{t}.
\end{align}
Here the usage of the term $p(t)/t^\mu$ means that the impact 
of one isolation of an infected individuals grows non-linearly
over time, which actually makes sense. Assuming that $c>2$, the 
dominant solution is:

\begin{align}\label{Besnu}
w(t)=t^{\frac{c+1-\mu}{2}}J_{\frac{c}{1-\mu}-1}
\,\bigl(\frac{2\sqrt{b}}{1-\mu}t^{(1-\mu)/2}\,\bigr), \,\,
 w\sim t^c \for t\approx 0.
\end{align}
The second one is for $1\!-\!\frac{c}{1-\mu}$;
we follow \cite{ChAI}. When $\mu=-1$, the argument of
$J$ is
 const$\cdot t\,$, and we arrive at some counterpart of $u(t)$.  
\vskip 0.2cm

{\sf Connection to 
statistical framework}. The solution $t^r$  for $r=c$ of
our starting equation
(\ref{Unpower})
is the  square root of the {\it variance\,}
$Var(B^H)$ of the {\it fractional Brownian motion\,}
 $B_H(t)$  ({\it fBM} for short)
for the {\it Hurst exponent\,} $H=r$. It becomes 
$r=r_{1,2}$ for the solutions from (\ref{6}).
Alternatively, the parameter $r$ can be obtained from
the self-similarity property 
of {\it fBM}:  $B_r(ts)\sim t^r B_H(s)$. 
{\it Qualitatively\,} 
the connection with our approach 
is that the expected (percent) growth of the epidemic spread  is 
essentially proportional to the 
standard deviation of the corresponding stochastic process.
The important conclusion is
that the {\it volatility of the spread is directly 
related to $r$},
which of course can be expected qualitatively.
   
One can try to introduce generalized
{\it fBM}  for the {\it full\,} solutions from
(\ref{7}), or for our $u(t),w(t)$ 
in terms of Bessel functions. See e.g. \cite{Che} for the theory
of mixed fractional Brownian motion and its applications 
in financial mathematics, which is related to our modeling here.
\vskip 0.2cm

A more systematic way to link our 
ODE to SDE, Stochastic Differential Equations, 
 is via the Kolmogorov-type
equations for the {\it transition probability 
density\,}; see e.g. equation (1.7) from \cite{Kat} for
the definition of {\it Bessel processes}. 
This is beyond the present paper.
Obviously the detailed analysis of our ODE and their applications
for the spread of epidemics, what we are doing in this paper,
is necessary before we can go to the level of random processes. 

\section{\bf Using the {\em u}-curves}\label{sec:swiss}
In the range till April 14 the 
graphs of our solutions $u(t)$ 
matched surprisingly well the {\it total\,} number
of infections in the examples we provide. The
starting moments where when these numbers 
began to grow  "significantly"; these moments are approximately 
around March 16 for the USA and UK, March 13 for Switzerland,
and March 7 for Austria. Only $u^1$ is used in this section.

\vskip 0.2cm
Here and below,
{\tt https://ourworldindata.org/coronavirus} is the main source 
of the {\it Covid-19 data}; this site is
 updated at about 11:30 London time. We also use
{\tt "worldometers"}. From now on,
$x=\hbox{days}/10$. As for $y$, to improve the 
readability it will be always
the number of total (detected) cases divided by a proper 
power of $10$.
\vskip 0.2cm

\begin{figure*}[htbp]
\hskip -0.2in
\includegraphics[scale=0.5]{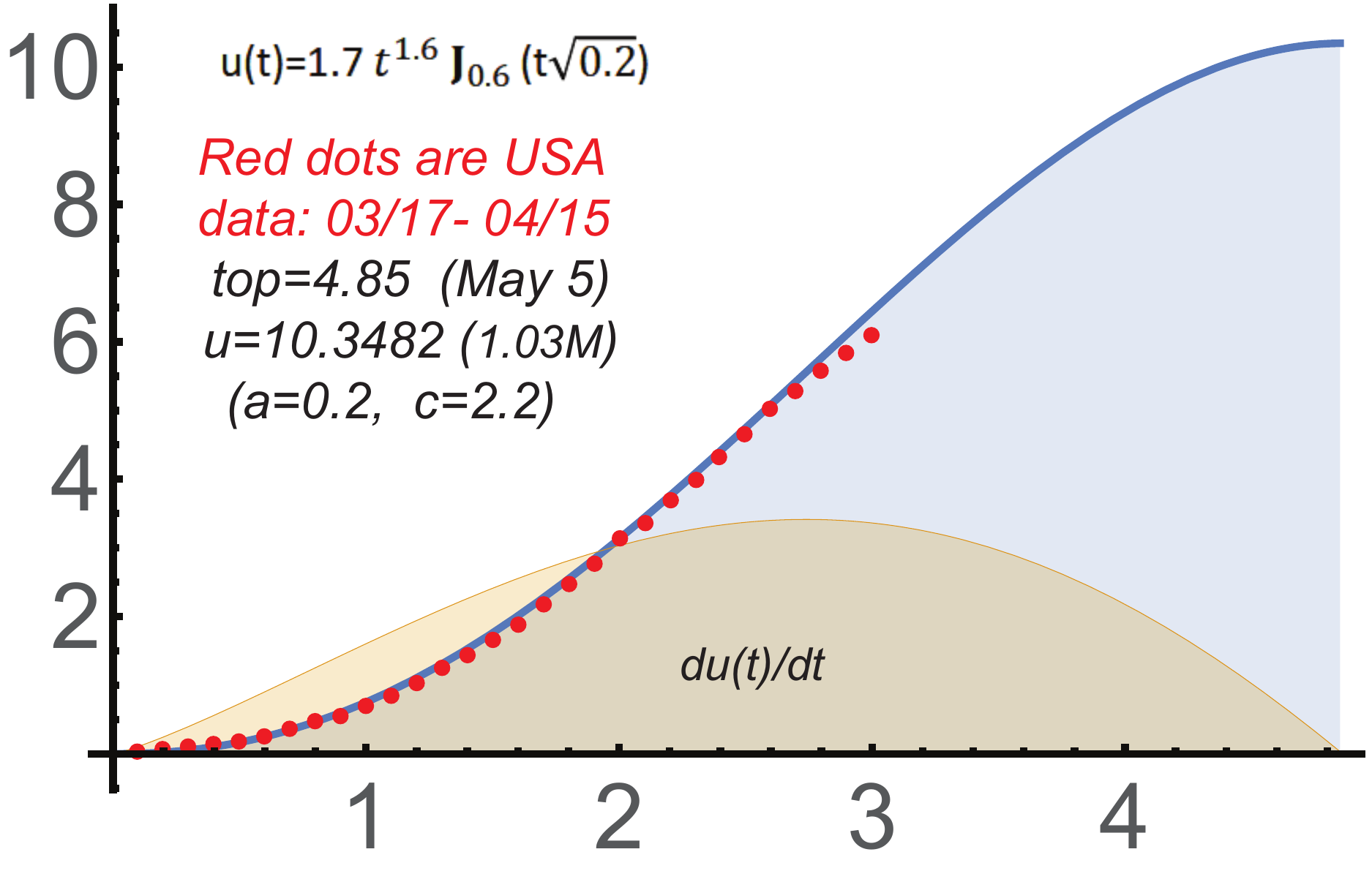}
\vskip -0.2in
\caption{$u(t)\!=\!1.7\,t^{(c+1)/2}J_{(c-1)/2}(\sqrt{a}t)$ for 
$c\!=\!2.2,a\!=\!0.2$}
\label{fig:Bessel}
\end{figure*}

\begin{figure*}[htbp]
\hskip -0.2in
\includegraphics[scale=0.5]{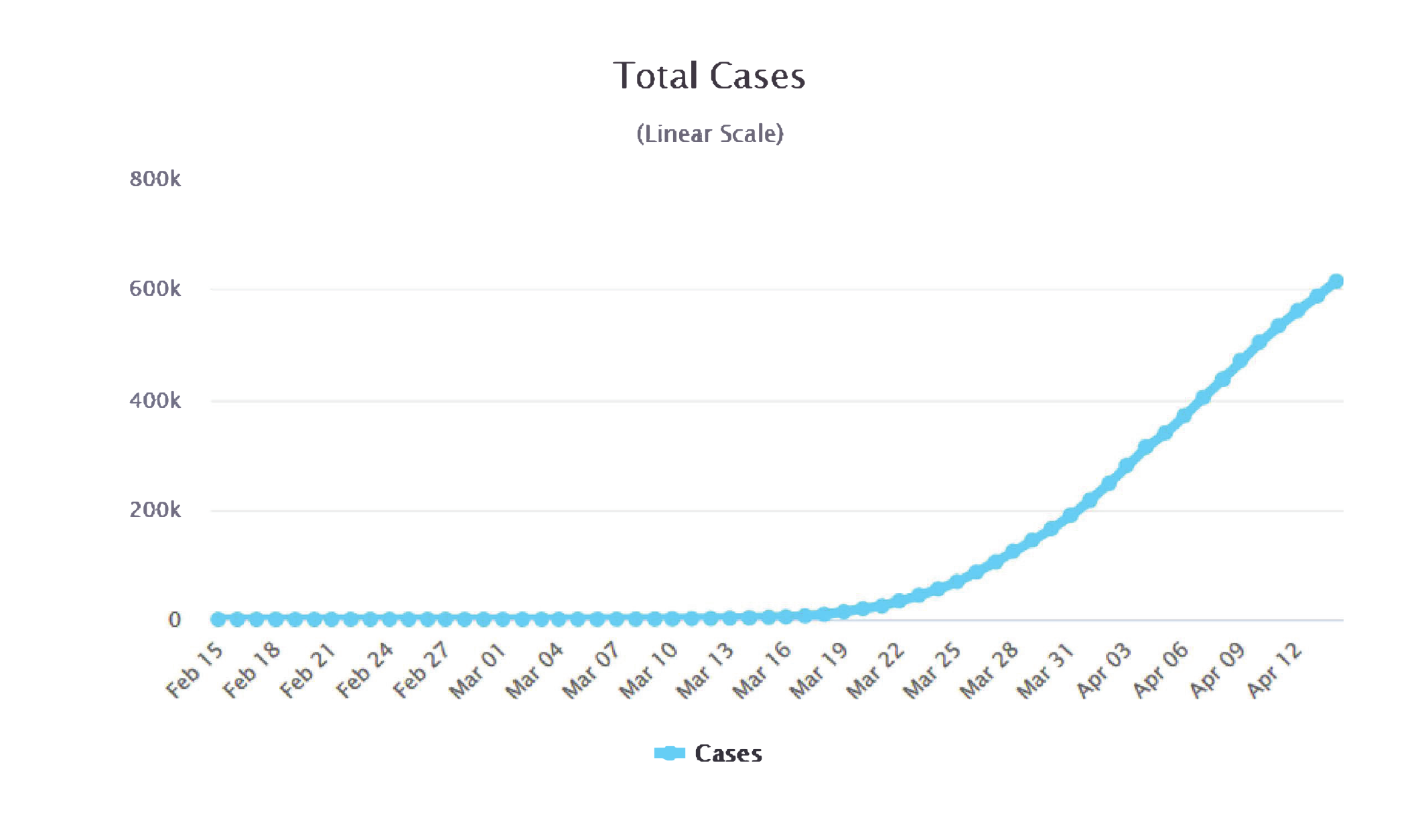}
\vskip -0.3in
\caption{Covid-19 in the United States}
\label{fig:US}
\end{figure*}

{\sf The USA data.}   The 
scaling coefficient
$1.7$ in Figure 
\ref{fig:Bessel} is adjusted to match 
Figure \ref{fig:US} for the United States.
For the USA, we set 
$y=\hbox{infections}/100K$, and take
March 17 the beginning of the period
of "significant growth". Then $c=2.2$, $a=0.2$
appeared perfect in the considered range.
\vskip 0.2cm

Recall that $c$, the initial transmission rate, reflects
the virus transmission strength and the way 
people respond to the current
numbers of infections. These numbers of course
depend on the information provided by the
authorities in charge and mass
media. 


\vskip 0.2cm
The {\it dots\,} in all figures
show the corresponding actual {\it total\,}
 numbers of infections. They {\it perfectly\,}
match $u(t)=1.7\, t^{1.6}\,J_{0.6}(t\sqrt{0.2})$  in Figure 
\ref{fig:Bessel}. Accordingly, assuming the same intensity of hard 
measures, the projection for the USA was:
$t_{top}=4.85$ ($48.5$ days from 03/17 till the saturation at
May 5) with 
$u_{top}=10.3482$, i.e. with $1034820$ infections
(it was $609516$ at 04/15). This did not happen; see below.

\vskip 0.2cm
The saturation at $t=t_{top}$ is of course 
a technical one, not the end of the spread of
{\it Covid-19}, even if it is reached.
The data from quite a few other countries demonstrated that  
some modest linear-like growth of the total
number of cases can be expected after $t_{top}$, and
then the {\it second phase\,} begins.
The epidemic remains far from over.

With these reservations, our graphs and  many
others we considered for {\it Covid-19\,} demonstrate 
that Bessel functions describe very well the
{\it first phase\,}, the period 
of "aggressive management" of type $(A)$.

"Hard" measures are of course not the only  force during phase 1. 
They are  {\it always\,} combined with "soft" ones. 
There are other factors and different stages. 
The graph for  Switzerland is used to
demonstrate various stages during this phase in  
Figure \ref{fig:swiss} below.

Assuming that Bessel-type functions model well
the period of the
intensive growth, one can try to
"capture" the parameters $c,a$ 
before or near the turning point of the epidemic. If $a,c$ are 
known, the first local maximum of the 
corresponding Bessel function times $t^{c/2+1/2}$ 
gives an estimate for the possible
"saturation", the technical end of phase 1. 
This is of course under the steady $(A)$-type management,
which appeared the case in sufficiently many countries. 
\vskip 0.1cm

In the data we provide, $a$, the intensity of
$(A)$\~measures, is $0.2$ for the USA, UK, Italy and
the Netherlands. It is $0.3-0.35$ for Israel, Austria,
Japan, Germany; $0.1$ for Sweden.
The parameter  $c$, the initial transmission rate, is 
$2.2$ for the USA, $2.4$
for UK, Austria, Sweden, the Netherlands, $2.6$ for Israel,
Italy, Germany,
Japan; it reaches $4.5-5.5$ for Brazil and India (with $a<0.05$).
 Here $c$ can be mostly "captured" during the early stages;
$a$ became "reliable"  later, when the management the epidemic
reached some stability. 
\vskip 0.2cm

\begin{figure*}[htbp]
\hskip -0.2in
\includegraphics[scale=0.5]{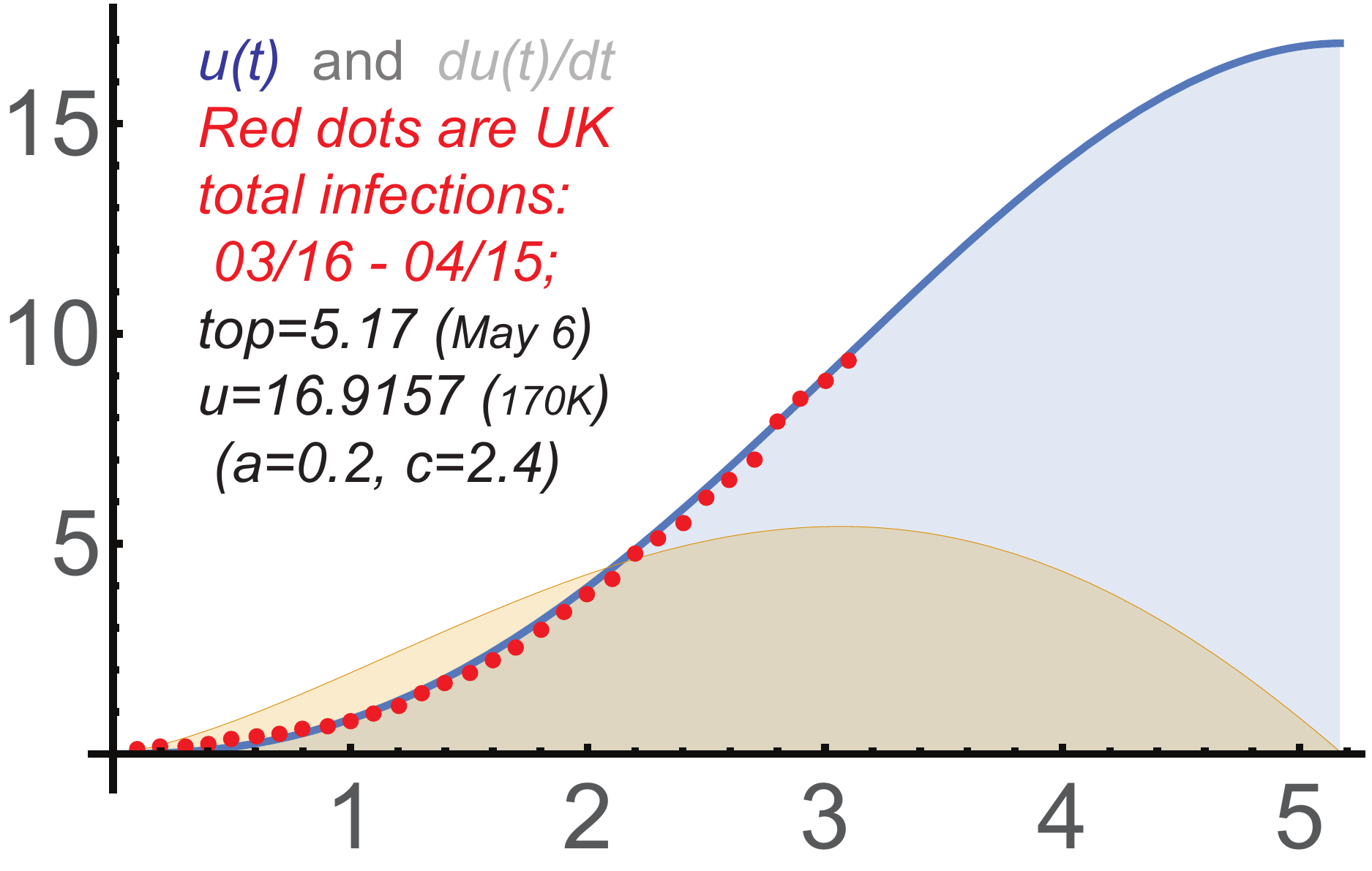}
\vskip -0.2in
\caption{$u(t)\!=\!2.2\,t^{(c+1)/2}J_{(c-1)/2}(\sqrt{a}t)$ for 
$c\!=\!2.4,a\!=\!0.2$}
\label{fig:Bessel-uk}
\end{figure*}

{\sf Covid-19 in UK.}
The coincidence of the {\it red dots\,}
with our $u(t)$  for the USA could be because this
was an average over 50 states.
Let us consider UK for the period from March 16 till 
April 15; see Figure \ref{fig:Bessel-uk}.

Then $c=2.4, a=0.2$; the scaling
coefficient is $2.2$. The total number of cases will be
now divided by $10K$, not by $100K$ as for the USA. The increase
of $c$ to $c=2.4$ qualitatively indicates 
that the "response" of the population to {\it Covid-19}
was a bit weaker in UK  than in the USA for
this period, and/or the regular number of contacts was
somewhat greater.  Though this is really some little change.

Recall that if the spread of disease is not 
{\it actively\,} managed,
then the growth $\approx t^{c}$ can be expected,
so $c$ reflects the strength of the virus and how we,
especially the infected people and those who think
that they are infected, react to the numbers of
infections {\it before\,} the active management begins.
Providing the numbers of infections, 
discussing them by the authorities and
in the media are some management too,
but {\it passive}. By "active", we mean the
modes $(A),(B)$ in this work.

The estimate for the "saturation moment" 
was $5.17$, i.e about $51$ days after March 16, somewhere
around May 6 with the corresponding number 
of infections about $170000$,
assuming that the "hard" measures would continue
as before April 15. The latter did not materialize.
\vskip 0.2cm

\begin{figure*}[htbp]
\hskip -0.2in
\includegraphics[scale=0.5]{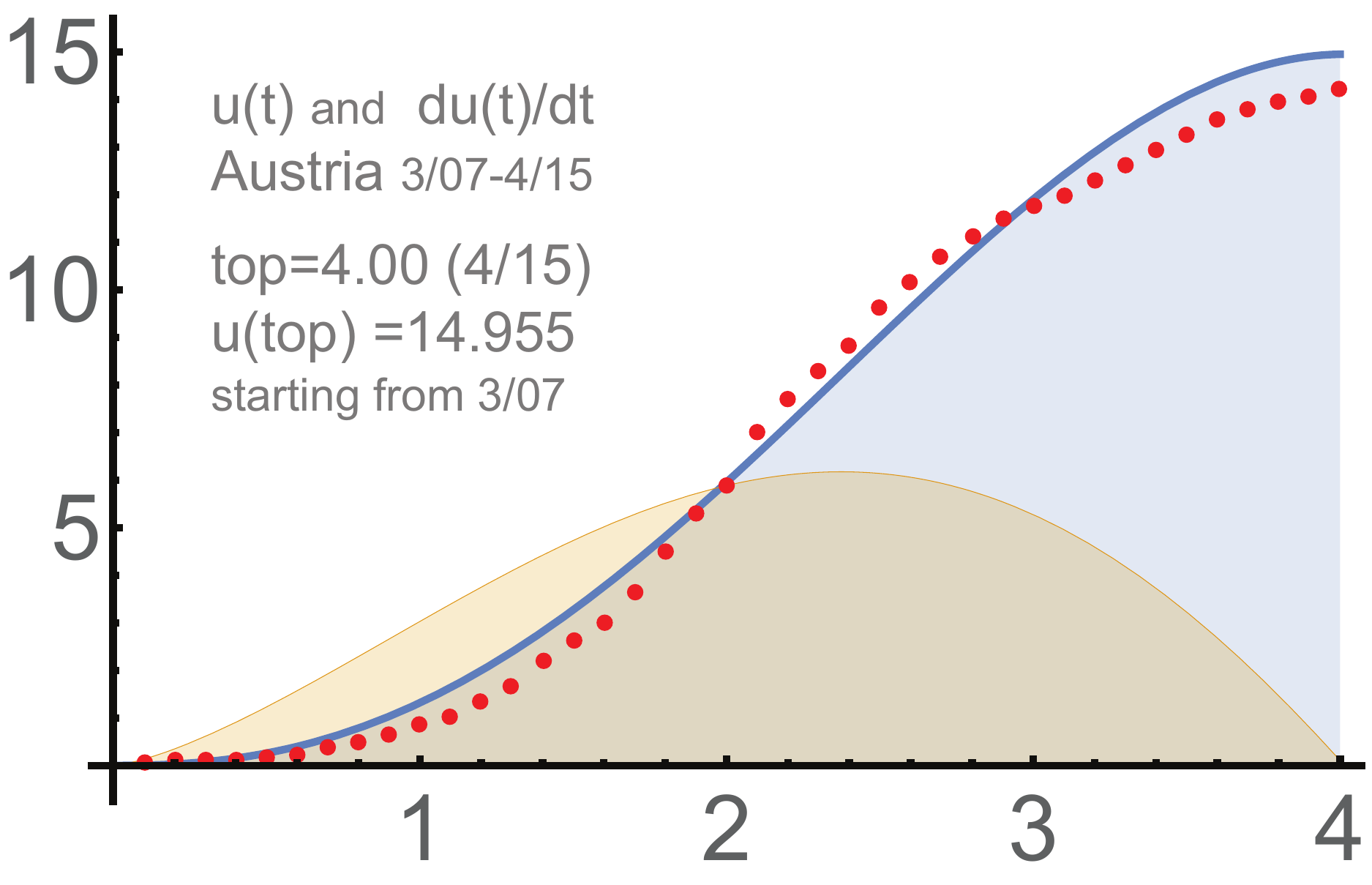}
\vskip -0.2in
\caption{$u(t)\!=\!3\,t^{(c+1)/2}J_{(c-1)/2}(\sqrt{a}t)$ for 
$c\!=\!2.4,a\!=\!1/3$}
\label{fig:Bessel-austria}
\end{figure*}

{\sf Austria: 3/07-4/15.}
This is one of the earliest examples
of a complete phase 1. 
There is some switch
to a linear growth around and after the "saturation", 
typical in all countries that reached the saturation.
Modeling this period by  Bessel function 
appeared quite doable, so the management was
steady and of type $(A)$. See Figure \ref{fig:Bessel-austria}.

Here we start with March 07; Austria began earlier with
the protection measures than UK and the USA. 
The parameters 
$a=0.33, c=2.4$ and the scaling coefficient $3$
appeared the best for Austria;  we divide now the number of 
infections by $1000$, i.e. $14$ in the graph 
corresponds to  $14000$ infections. 

\vskip 0.2cm

{\sf Switzerland: different stages.}  
We focus on the period March 13- April 12. 
As we have already discussed,  $(A)$  is supposed to dominate
closer to the turning point and then toward $t_{top}$. 
Mode $(B)$ is present too, but we mostly see in the
figures the result
of "hard" measures.

Recall that only the total number of the detected
infections is considered in this paper. The starting
moment is $t=0$. This is when the intensive growth 
of power type begins, and also the beginning of active 
management. It was 
around March 12 in Switzerland in Figure \ref{fig:swiss}.
The first results of the employed measures can be 
seen around March 20 : a transition of the initial 
(mostly) parabolic growth to the 
linear one. 

\vskip 0.2cm
Some "free" spread of quadratic-type  is present in practically
all countries in the early stages of {\it Covid-19}. 
Though in Brazil
and India, the initial growth was higher: 
$\sim t^{4.5}$ and $\sim t^{5.5}$. The usage of Bessel
functions continuously (and automatically)
diminishes the exponent here during phase 1.

This is different for mode $(B)$: the
transition from the "free"
growth $u(t)\sim t^c$ to $u(t)\sim t^{c/2}\cos(\cdots)$
is due to the increase of $a$ from $0$ to  
$a>c^2/4$; then  the exponent will stabilize even if
$a$ continues to increase. {\it Phase 2} is with $a>c^2/4$;
The role of $\cos(\cdots)$ and $\sin(\cdots)$
in (\ref{7}) becomes significant and 
provides the "final saturation": the technical end of
the {\it 2nd phase} (and the {\it 1st wave}). 

\vskip 0.2cm
Let us briefly comment here on changing the starting point from the 
absolute start of the epidemic to some $t_{\bullet}$. 
The equations
will remain the same, but $c/t$ must be replaced by
$c_\bullet/(t-t_\bullet)$ for the current $c_\bullet$.
There can be different stages of epidemics, so this can
be necessary.
Such a split of the total investment
period into intervals treated independently is very common in
stock markets. {\it However, we do not change the starting point
$t=0$ when switching from phase 1 to phase 2}. The corresponding
$(B)$-type solution is considered as started from the very 
beginning (though it is  used only around and after $t_{top}$ for 
mode $(A)$).
\vskip 0.2cm

 
\vspace* {-0.6cm}
\begin{figure*}[htbp]
\hskip -0.2in
\includegraphics[scale=0.5]{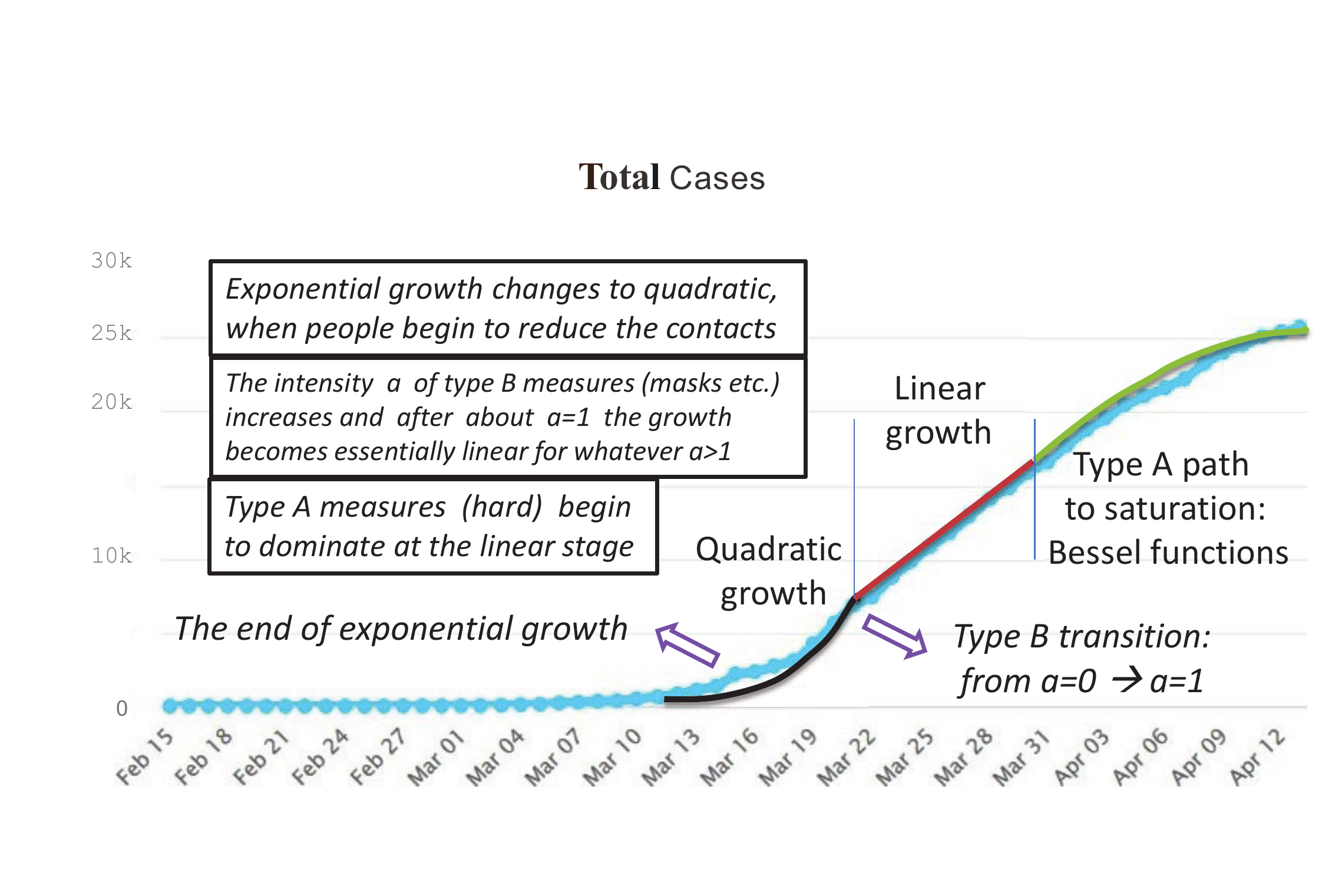}
\vskip -0.2in
\caption{Covid-19 in Switzerland}
\label{fig:swiss}
\end{figure*}
\vskip 0.2cm

\section{\bf Main findings}
The "power law of epidemics" presented as
a differential (initially, difference) equation is the starting
point of our approach to momentum management. This law
is different from other power laws for infectious diseases; 
compare e.g. with \cite{MH}. We deduce it from 
certain "geometric"
assumptions concerning the contacts of infected people,
a sort of principle of local herd immunity, and from some
behavioral and sociological analysis.  
The restrictions people impose themselves during the epidemic, mostly
the reduction of their contacts are related 
to the general theory of "momentum risk-taking"
from \cite{ChAI}.

The main problem here is to "couple" the {\it power law
of epidemics\,} with some mechanisms for their ending. 
These are major challenges,  biologically,
psychologically, sociologically and mathematically. 
We demonstrate that mathematically there is a path: 
Bessel functions serve as a natural way from
the power law of epidemics to the "saturation". 
Our differential equations
are based on various simplifications; however 
their match with the 
real data appeared almost perfect.
\vskip 0.2cm

The "megaproblem" of 
understanding and managing 
epidemics is ramified and interdisciplinary;
significant assumptions are inevitable in any models.
We model only the period when the total number of infections
is significantly smaller than the whole population. We also 
disregard the recovery and quarantines, though they are 
indirectly incorporated  in our differential equations via
$c$ and $a$.
\vskip 0.3cm

We fully aware of the statistical nature of the problem.
The usage of random processes, like {\it Bessel processes},
is reasonable here to address the randomness around epidemics;
see the end of Section \ref{sec:typeA}.
Some combination with the traditional SIR model can be 
possible too; see e.g. \cite{CLL}. Our equations have natural
"logistic" extensions.
\vskip 0.2cm
\vfil

{\sf Saturation via hard measures.}
The main outcome of our modeling and this paper 
is that the measures of "hard type", like detecting 
infected people, followed by their isolation,
and closing the places where the spread is 
likely, are the key for ending an epidemic. Such measures
are a must for at least the initial and middle stages
of the epidemic to ensure the saturation in our approach.
 
Moreover, they must be employed strictly 
proportionally to the current number of infections,
not its derivative of any kind, to be really efficient.
This is the most aggressive way to react, which
we call mode $(A)$ in the paper. This can be
seen in many countries, at least during the middle
stages of the epidemic.
\vskip 0.2cm
  
Then the point $t_{top}$ of the first maximum of the corresponding
Bessel function times $t^{c/2+1/2}$ is a good estimate for the 
duration of phase 1; it also gives some projection 
for the expected maximum  of the number of infections
during phase 1. This is assuming "hard measures"
and steady management of type $(A)$.
This formula worked very well for middle periods almost
everywhere and even for the {\it whole\,} period of the
extensive spread in the countries and areas 
that reached phase 2. It seems a real discovery.

If the saturation is successfully reached at $t_{top}$, then
the {\it second phase\,} begins, which is essentially the switch to
mode $(B)$. Mathematically, the curve of total cases significantly
changes around $t_{top}$. This switch can be clearly
seen in many countries: frequently,
a well-visible break of the derivative. 

The second phase is more challenging mathematically;
it is heavily influenced by economic, political and other factors.
So it is even more surprising, that it 
can be modeled with high accuracy using our system of differential
equations of type $(B)$. See our graphs for Israel, Italy, Germany,
Japan, and the Netherlands in Section \ref{sec:twoph}.

The quasi-periodicity of our $u$\~functions under 
momentum management of type $(A)$ and that under $(B)$ in
terms of $\log(t)$ is the key mathematical reason for reaching
the saturation in our approach. This is
not connected at all with the periodicity of epidemics
associated with seasonal factors, biological reasons, 
or various delays; see e.g. \cite{HL}. The saturation we
model entirely
results from the active management and general "response"
of the population to the threat. {\it Covid-19} is exceptional in
many ways, the scale and the intensity  of the
measures used to suppress it are quite unique.
The {\it second wave\,}, which can be
now observed in several countries, the USA, Israel, Japan,
several in Europe, and more of them, is a clear confirmation of 
the leading role of hard measures. It really looks like the main 
"control parameter" is their intensity. 
\vfil

With this reservation about the source
of our "periodicity", there is some analogy
with {\it Farr's law of epidemics}. Under mode $(A)$, this is
some reflection symmetry of $du(t)/dt$ for 
$u(t)=t^{c/2+1/2}J_{(c-1)/2}(\sqrt{a}t)$
in the range from $t=0$ to $t_{top}$, 
the first local maximum of $u(t)$. The portions of the 
corresponding graph before and after
the {\it turning point\,} are not exactly 
symmetric to each other, but are close enough to this.  
See e.g. Figure \ref{fig:Bessel}, where the turning point is at 
$\max\{du(t)/dt\}$. This holds for
$w(t)$, describing the $(AB)$\~mode, but the second "half"
becomes somewhat longer than the first. This is for 
the first phase.
\vfil 

\vskip 0.2cm
{\sf The second phase.}
Mode $(B)$ is the usage of the average 
$u(t)/t$ instead of $u(t)$ and 
relying mostly on  "soft" measures, like wearing protective 
masks and social distancing. Closing schools is certainly 
a hard measure by any standards, and a very important one. 
Some travel restrictions are "hard" too. Self-imposed
limitations are of course important too, and can be 
quite hard. They can significantly complement the governmental
management, as occurred many times in the history of epidemics.

When used alone,
soft measures are generally insufficient to "reach zero",
which follows mathematically from our model.
However, the usage of "hard measures" combined
with the response
based on $u(t)/t$ instead of $u(t)$ can work reasonably
quickly toward the saturation; this is mode $(AB)$. 
\vskip 0.1cm

The best way to address the 
late stages appeared our "two-phase solution". Since the
hard measures are reduced or
even abandoned at the end of phase 1 (or even before),
we switch from mode $(A)$ to $(B)$. This 
can be seen in quite a few countries that reached
the "saturation". There is a clear linear-type pattern around
the "technical saturation", which is $t_{top}$, the
point of maximum of our Bessel-type $u(t)$. 
Then the total number of detected infections closely 
follows
$\,C t^{c/2}\cos(d \log(t)\,$ for some 
$C$ and $d=\sqrt{-D}$ from
 (\ref{2}),(\ref{7}). Here $c$ is the exponent of 
the initial power growth of $u(t)$; it remains the same for
any modes, $(A),(B)$ or $(AB)$. As for $d$, it must be adjusted
numerically to match the shape of the curve of the
actual number of total
cases near and after the saturation;
$C$ is a (constant) rescaling coefficient.
\vskip 0.2cm
  
This formula describes the late stages with surprising accuracy
in almost all countries that entered phase 2; there are
already quite a few, for instance, almost all Western Europe
by now. The usage of
our {\it two-phase solution\,} is less relevant
for the countries that begin reducing or abandoning 
hard measures upon the first signs of the stabilization of
new daily cases or implement hard measures insufficiently. 
This includes some countries, as UK and the USA, with solid 
type $(A)$ response during the initial and middle stages. 
 See Section \ref{sec:forecones}; the USA
(as the whole) did not reach the end of phase 1 before switching
to the second wave.
 
If the period of linear growth becomes long, then $d$ in
$\cos(d\log(t))$ can be more difficult to find. Such long
periods of linear growth of the total number of
detected infections were typical for the countries that 
do not employ "hard" measures systematically, like Sweden.
These measures are present in some forms anywhere, but this
can be insufficient. 
The growth of the number of total cases
can be significantly faster than linear for 
sufficiently
long periods, which can be due to a variety of
factors, including insufficient
detection-isolation capacities and some general weakness of the 
health-care systems; Brazil and India are examples.

\vskip 0.2cm
\vfil

{\sf The risks of recurrence.}
The recurrence of epidemics is 
a clear challenge for microbiology, epidemiology
and population genetics;
very much depends on the type of virus.
The "natural" $c$\~coefficient
reflects both, the transmission strength of the virus and the 
"normal" intensity of our contacts. The second component
is likely to return back to normal after the epidemic is
considered  "almost finished". If the "microbiological 
component" remains essentially the same as it was before, 
the recurrence of the epidemic is almost inevitable after  
the protection measures are removed. The recurrence of the 
epidemic becomes a sort of "cost" of our aggressive interference 
in its natural course.
\vskip 0.2cm

The whole purpose of aggressive management, mode $(A)$
in our approach, is to keep daily new case  at
reasonably small levels.
Then {\it detection-isolation-tracing\,} becomes much
simpler and the pressure on medical facilities reduces.
 However the  "saturation" in one place 
is of course not the end
of the epidemic. New clusters of infections are
likely to emerge, and the epidemic suppressed 
in one area can still continue in other places. 
\vskip 0.2cm

These are real risks. The economic and other "costs" of 
hard measures are very high. It is understandably easy to 
"forget" that the turning point and the saturation
were achieved reasonably quickly 
mostly due to the hard 
measures imposed. The corresponding underlying 
mathematical processes have a strong 
tendency to become periodic, which can  
"play against us" as we approach the end of the epidemic.  
\vskip 0.2cm

Unless the {\it herd immunity\,} is reached or the
virus lost its strength, 
"restarting" the epidemic
with few infected individuals left after the previous
cycle is standard. Significant numbers of those 
recovered and the asymptomatic cases provide some
protection, but it can be insufficient.
For the next cycle(s),  our better preparedness can be 
expected, which can diminish $c$. However this is not
confirmed so far in the countries with 2nd waves;
there are obvious problems with restarting hard measure,
including keeping schools close.   

\vskip 0.2cm

Finally, let us
emphasize that our differential equations describe an
{\it optimal\,} momentum response aimed at diminishing the
new infections only under the following key
condition. We assume that people and the authorities in
charge {\it constantly\,} monitor the numbers 
of infections and momentarily respond to them in the 
"most aggressive" way.

This is actually similar to the ways stock markets 
work. Professional traders 
simply cannot afford {\it not\,} to react to any news and
any change of the stock prices, even if they seem
random, temporary  or insignificant. Indeed, any particular 
event or a change of the share-price can be 
a beginning of a new trend. Applying this
to managing epidemics, we are really supposed to closely follow
the data. The  "flexibility" here is only
the usage of some average numbers as the triggers, 
as a "protection" against random fluctuations. This is exactly 
mode $(A)$ versus "more defensive"  $(B)$ or $(AB)$.
\vskip 0.2cm

\begin{figure*}[htbp]
\hskip -0.2in
\includegraphics[scale=0.5]{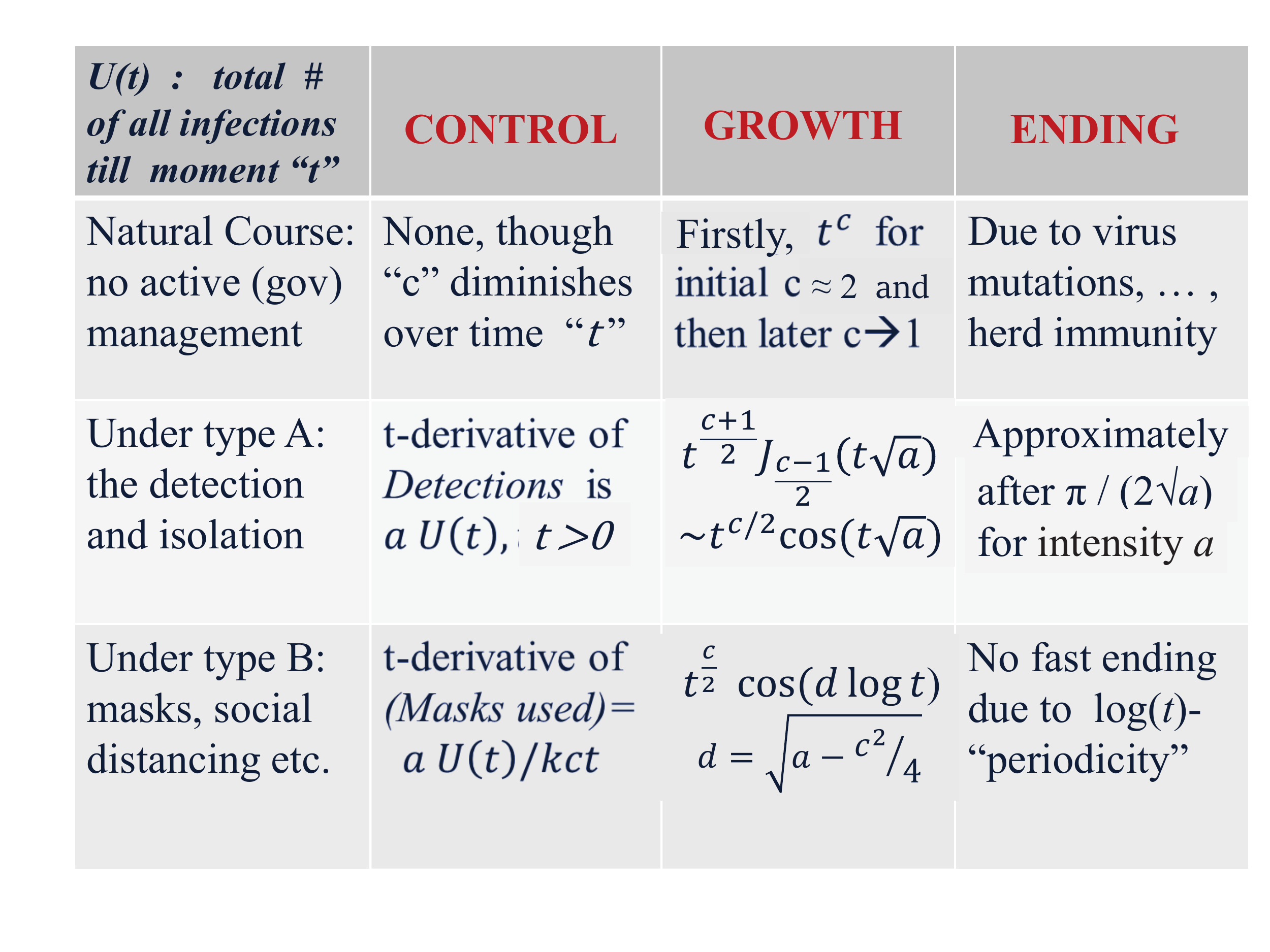}
\vskip -0.2in
\caption{Two types of momentum management}
\label{fig:2typ}
\end{figure*}

\vfil 
{\sf The table: $(A)$ vs. $(B)$.}
For convenience of the readers,
let us provide a basic table presenting the $(A)$-mode 
and the $(B)$-mode with some simplifications. This is
Figure \ref{fig:2typ}. 
Recall that $c$
is the initial transmission rate, $a$ the control parameter,
which is the intensity of the management, 
$\kappa$ the efficiency of protective
masks.

 
By "Detections", we mean the total number
of detected individuals till $t$, which is 
essentially proportional to the number of tests
performed. This is up to "us" to control.

\vfil
The response of type $(A)$ is basically as follows:
if the number of infections doubles, then
the {\it rate of increase\,} of tests must be
doubled, not just the number of tests. For $(B)$, the rate of
change of the measures is assumed proportional
to the rate of change of the infections. The $(AB)$ mode
uses this kind of response for hard measures. So under $(AB)$,
if the number of
infections doubles, then the number of tests must be
doubled too.
\vskip 0.2cm

This kind of response will
provide the quasi-periodicity with respect to $\sqrt{t}$, so
the saturation will take longer than under $(A)$, where the
quasi-periodicity is for $t$. However, the time till saturation
under $(AB)$ (roughly, $1/4$th of the period)
is generally "quite acceptable". See 
Section \ref{sec:forecones}  below and 
Sections \ref{sec:Pro}, \ref{sec:soft}.

Recall that the key hard measure is  
{\it testing \& detection \& isolation}, followed by
{\it tracing}. Also, 
we use {\it only\,} the total numbers 
of detected infections, without "subtracting" the closed cases.

Switching to $(B)$ in the table, by 
"Masks used", we mean the total number of infected people who began
using the masks till $t$. If the number of infections doubles,
than the total number of masks must be doubled too under $(B)$, 
{\it not the rate of change} as for $(A)$.

\vskip 0.2cm

\section{\bf Unusual patterns} \label{sec:forecones}
Our key finding is that the {\it total\,} numbers of 
detected infections for {\it Covid-19} can be described with 
high accuracy by
$u(t)=C\,t^{c/2+1/2}J_{c/2-1/2}(\sqrt{a}t)$ during 
the initial and middle stages of phase 1, which was 
practically in all countries we considered. 
This is the period of intensive growth of the spread when
the hard measures are coupled with the most
aggressive response to the changes of the
numbers of total cases. A typical example is
Figure \ref{fig:Bessel}.

It appeared that  
the whole phase 1 was covered well by such functions
if the  $(A)$\~mode was employed 
until the numbers of new detected infections dropped 
to really small levels. South Korea, Austria, 
Israel, Italy, Germany, Japan, the Netherlands and many
other countries (almost all Western Europe) did exactly this. 
Then $t_{top}$, the first zero 
of $du(t)/dt$, is a reasonable estimate
for the "technical end" of phase 1. 

To understand the usage of type $(A)$ curves
for forecasting, we determine the parameters $a,c,C$
for the initial periods, the 
{\it red dots\,}, and then analyze sufficiently long test
periods, the {\it black dots\,},  without changing $a,c,C$. 
A challenge here is the usage
of $a,c,C$ obtained in some  {\it early\,} stages for forecasting
the later ones. 

This seems doable, if the hard measures are employed,
and when the management is steady, which
was not (always) the case with Sweden, USA, UK and some other
countries. There were significant periods of essentially
linear growth of 
the curves of total (detected) cases in these countries.
In our charts,
the {\it black dots\,} 
remained essentially linear 
significantly longer than it was "allowed" by  the corresponding
Bessel-type $u(t)$, based on the parameters obtained
at earlier stages.
In the USA, the process appeared the most complicated from 
these 3 countries.
Brazil and India also must be considered exceptional
because  their $c$-coefficient are exceptionally high.
By now (August 25), they are still in the middle half 
of phase 1.

We note that the parameters $a,c$ were the same
for the Netherlands and UK, and not too different for those for
the USA. However, the Netherlands reached phase 2 following
the $u$-curve, as well as many countries in Western Europe.
So this was not about the values of the parameters,
but rather about significant changes with the management
of {\it Covid-19} in the USA and UK. 

One can expect the $(AB)$-mode to occur
after the mode $(A)$ is "relaxed". 
Actually, the moments of the change of the
management can be clearly seen in the graphs for 
the USA and UK. We found  
the corresponding $w(t)$ curves of type $(AB)$ 
matching the {\it red dots} and {\it black dots} till
these moments for these countries.
We also calculated the $w$-curves 
for Israel and Sweden, some  "control group";
$w(t)$ is not 
needed for Israel, which reached $t_{top}$ under our $u(t)$,
and the actual curve of total cases 
for Sweden was not supposed to match $u(t)$
or $w(t)$.  
\vskip 0.2cm

Recall that
mode $(AB)$ is when the hard measures are still present, but the 
response to the total number of infections
becomes softer, as under $(B)$. 
For instance, 
if the number of new cases is essentially a 
constant, even uncomfortably high, the testing-detection under
the $(AB)$ will be essentially constant too. This
is much more aggressive with $(A)$. 
Considering the $(AB)$ mode is not necessary for the 
countries that reached $t_{top}$  under $(A)$. For such countries,
the passage will be from $(A)$ directly to mode $(B)$,
to the {\it second phase}. 

\vskip 0.2cm
\vfil

The {\it blue dots\,} till 05/27 after the
{\it black dots\,} were added to monitor 
the usage of the $w(t)$\~functions. They did 
not stay within $w(t)$ for Sweden and the USA,
This was expected for Sweden, since
no aggressive approach was implemented there at that time. 
For the USA, there was another 
significant reduction of  hard measures.
UK  "managed" to stay within the 
{\it forecast cone\,}
between $u$ and $w$; its "expiration" was June 10. Then
it entered phase 2, and the curve of total cases  
matched well our "2-phase solution". Sweden eventually
reached phase 2 too, but this occurred significantly
later.
\vskip 0.2cm
  
When using $w(t)$,
the point $t^w_{top}$, the (first) maximum of $w(t)$,
is a reasonable upper bound for the "technical 
saturation". The prior $t^u_{top}=t_{top}$ then is a 
lower bound; we obtain some forecast cone. 
\vskip 0.1cm

Let $u(t)=Ct^{\frac{c+1}{2}}J_{(c-1)/2}(t\sqrt{a})$, \,
$w(t)=Dt^{\frac{c+1}{2}}J_{c-1}(2\sqrt{tb})$. Using
(\ref{uwGa}), the match $u(t)\approx w(t)$ near $t=0$ gives:
\begin{align}\label{CDGa}
\frac{C}{D}\approx \Bigl(\frac{2b}{\sqrt{a}}\Bigr)^{\!\!\frac{c-1}{2}}
\, \,\frac{\Ga((c+1)/2)}{\Ga(c)} \hbox{\ \, for the $\Ga$-function}.
\end{align}

\vskip 0.2cm
 
\begin{figure*}[htbp]
\hskip -0.2in
\includegraphics[scale=0.5]{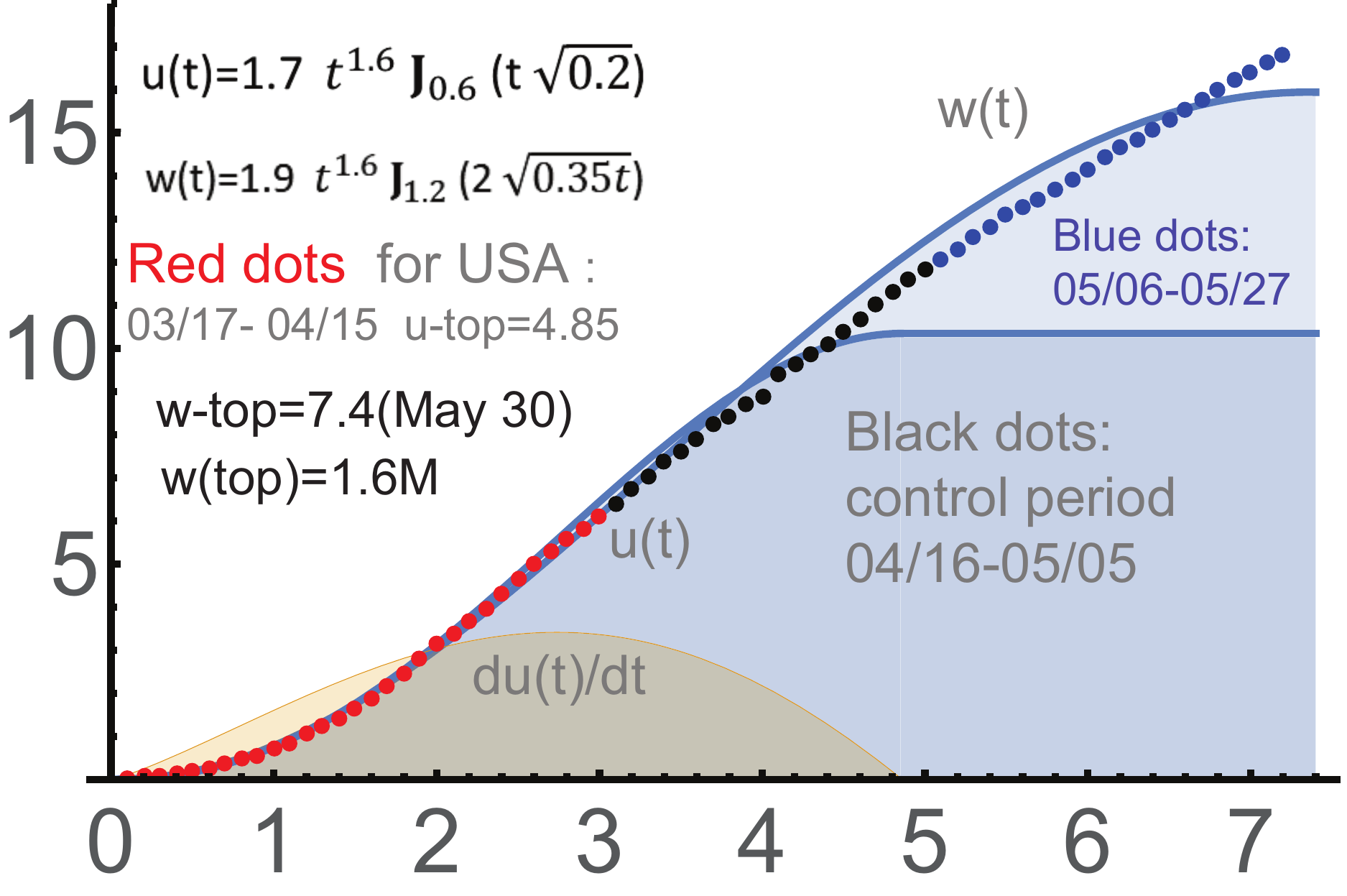}
\vskip -0.2in
\caption{
USA: $c\!=\!2.2, a\!=\!0.2, C\!=\!1.7;\, b\!=\!0.35, D\!=\!1.9$.}
\label{fig:Bessel-usa-new}
\end{figure*}

{\sf The cone for the USA.}
The  prior $t_{top}=t_{top}^u$ for $u(t)$ 
was May 5, with $a,c,C$ calculated on the bases of the data till
April 16, marked by {\it red dots}.
It was under the expectations that the "hard" measures 
would be applied as in $(A)$, i.e. as before April 16.
The {\it black dots\,} represent the control period.
The challenge was to understand how far $u(t)$ can be used
toward the "later stages". The match was very good for quite
a long period of time (the same with UK), but then the curve
went up, which we mostly attribute to the softening  the
management mode, namely, relaxing hard measures.
The "response" of the population of the USA to the threat 
noticeably softened too.  

 Whatever the reasons, the switch to $w(t)$, describing the
$(AB)$-mode, appeared necessary. The 
trend was initially toward $w(t)$ for the USA, but then the {\it
blue dots\,} "went through" $w(t)$.
Generally, the test dots
can be expected to stay within the {\it forecast cone\,}, subject
to standard reservations: the usage of hard measures and
steady management. 
The cone was the domain between
$u(t)$ extended by a constant $u(t_{top})$ for $t>t_{top}^u$ 
and the graph
of $w(u)$ till $t_{top}^w$, which was approximately May 30, 2020.

The parameters $b,D$ of $w(t)$ where calculated to ensure good match
with {\it red dots}; 
$c$, the initial transmission rate,
is {\it always\,} the same for $u(t)$ and $w(t)$.
The graphs of $u(t)$ and $w(t)$ are very close to 
each other in the range of {\it red dots}. 
The above relation for $C/D$ holds with the accuracy
about $20\%$. 

\vskip 0.1cm

The $w$\~saturation value was around 1.6M,
but mostly we monitor the  {\it trend\,}, the "derivative"
of the graph of black dots, which is supposed to
be close to the derivative of
$w(t)$ or $u(t)$.  See Figure \ref{fig:Bessel-usa-new}.
Some spikes with number of cases are inevitable; it is acceptable
if the dots continue to be mainly "parallel" to $u(t)$ or 
$w(t)$ (or in between). 

\begin{figure*}[htbp]
\hskip -0.2in
\includegraphics[scale=0.5]{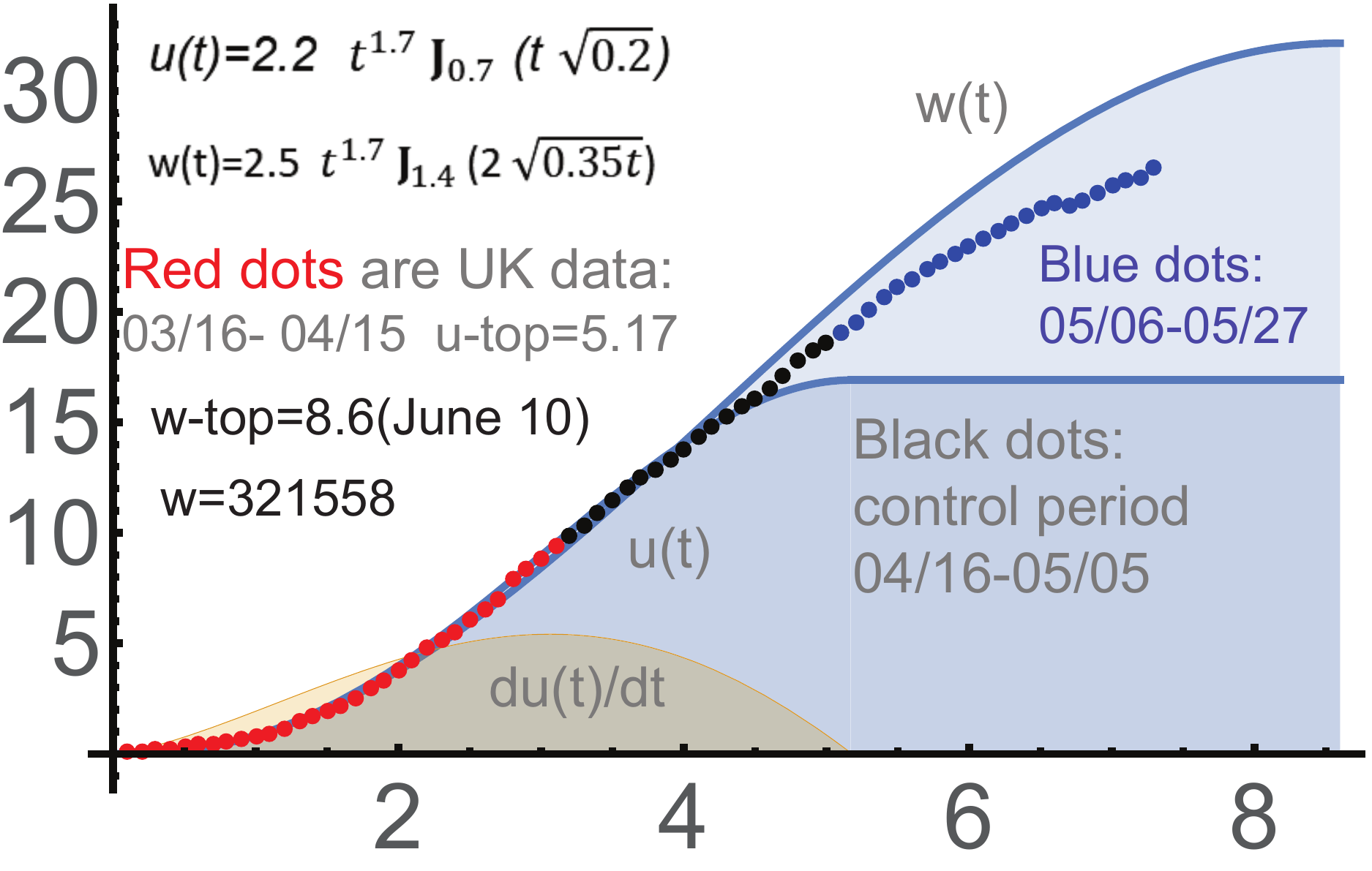}
\vskip -0.2in
\caption{
UK: $c\!=\!2.4, a\!=\!0.2, C\!=\!2.2;\, b\!=\!0.35, D\!=\!2.5$.}
\label{fig:Bessel-uk-new}
\end{figure*}

\vskip 0.2cm
{\sf UK till June 10.} 
\comment{
Let us first demonstrate how close
$u(t)$ and $w(t)$ can be during the period of {\it red dots\,}
and even after it. The match is practically ideal, though we
adjusted $b,D$ in a very simple way. As we have already
noted, fluctuations of $a,C,b,D$ can lead to some 
sizable changes over time. In these examples, they are integers
divided by $10$, so we do only "basic" adjustments of the
parameters. 

\begin{figure*}[htbp]
\hskip -0.2in
\includegraphics[scale=0.5]{bessel-uk-uw-eps-converted-to.pdf}
\vskip -0.2in
\caption{
UK: how close $w(t)$ is to $u(t)$.}
\label{fig:Bessel-uk-uw}
\end{figure*}
}
The  graphs of $u(t),w(t)$ with prior red-black
dots and the blue ones added after May 5 are in Figure 
\ref{fig:Bessel-uk-new}. The expected $t_{top}^w$ was around
June 10. The total number of detected infections was expected
330K or below. These "predictions" of course depended on many
unknown factors. However, some argument
in favor of the stability of our model is that
any spikes with the numbers
of infections are supposed to trigger additional actions of
the authorities in charge and increase our own protective
measures. Such "self-balancing" seems a rationale for 
relatively uniform patterns of the spread of {\it Covid-19}
in different countries.   
\vskip 0.2cm

\begin{figure*}[htbp]
\hskip -0.2in
\includegraphics[scale=0.5]{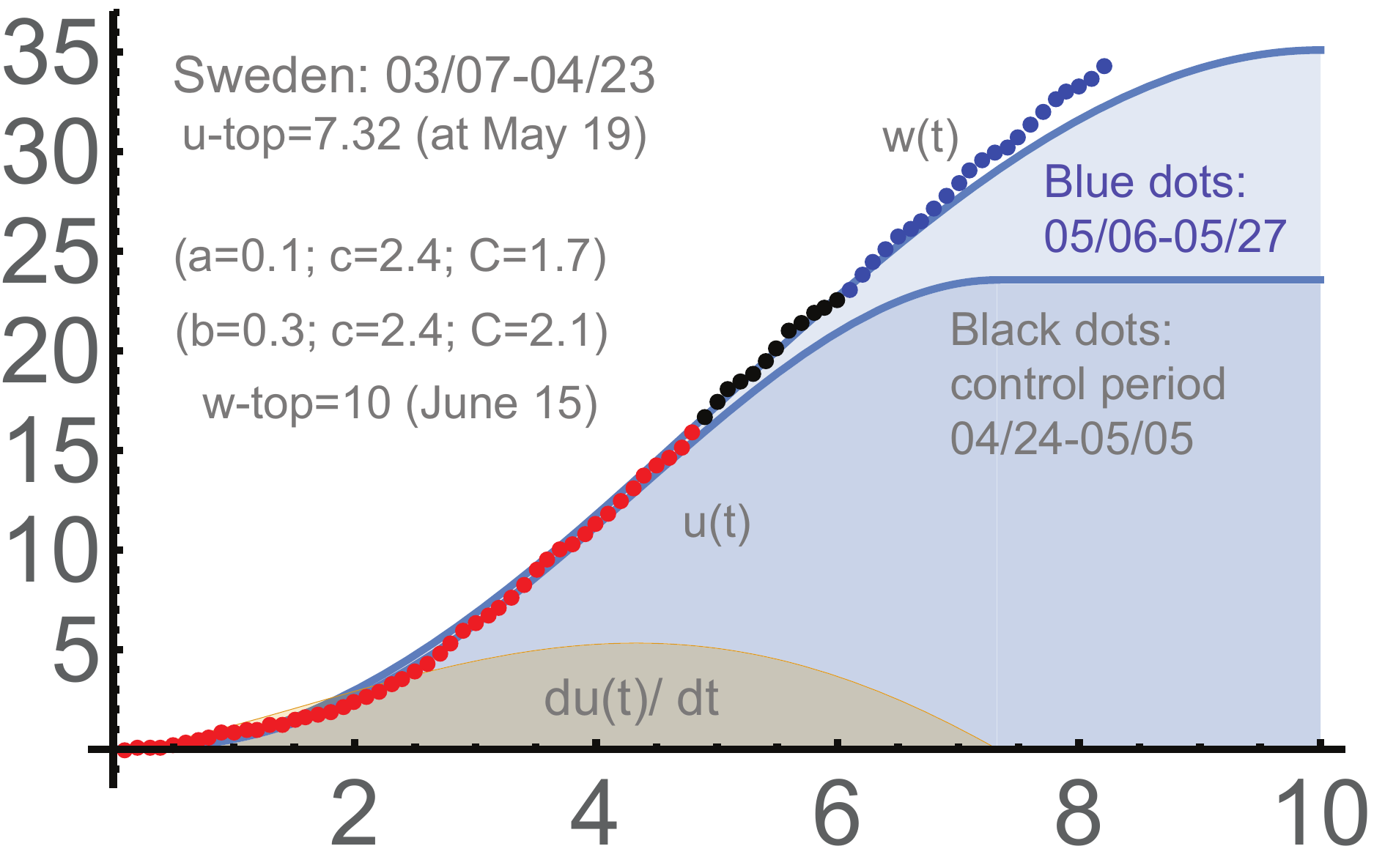}
\vskip -0.2in
\caption{
$w(t)\!=\!2.1 t^{(c+1)/2}J_{c-1}(2\sqrt{bt}); 
b\!=\!0.3, c\!=\!2.4$.}
\label{fig:Bessel-sweden-new}
\end{figure*}

\begin{figure*}[htbp]
\hskip -0.2in
\includegraphics[scale=0.5]{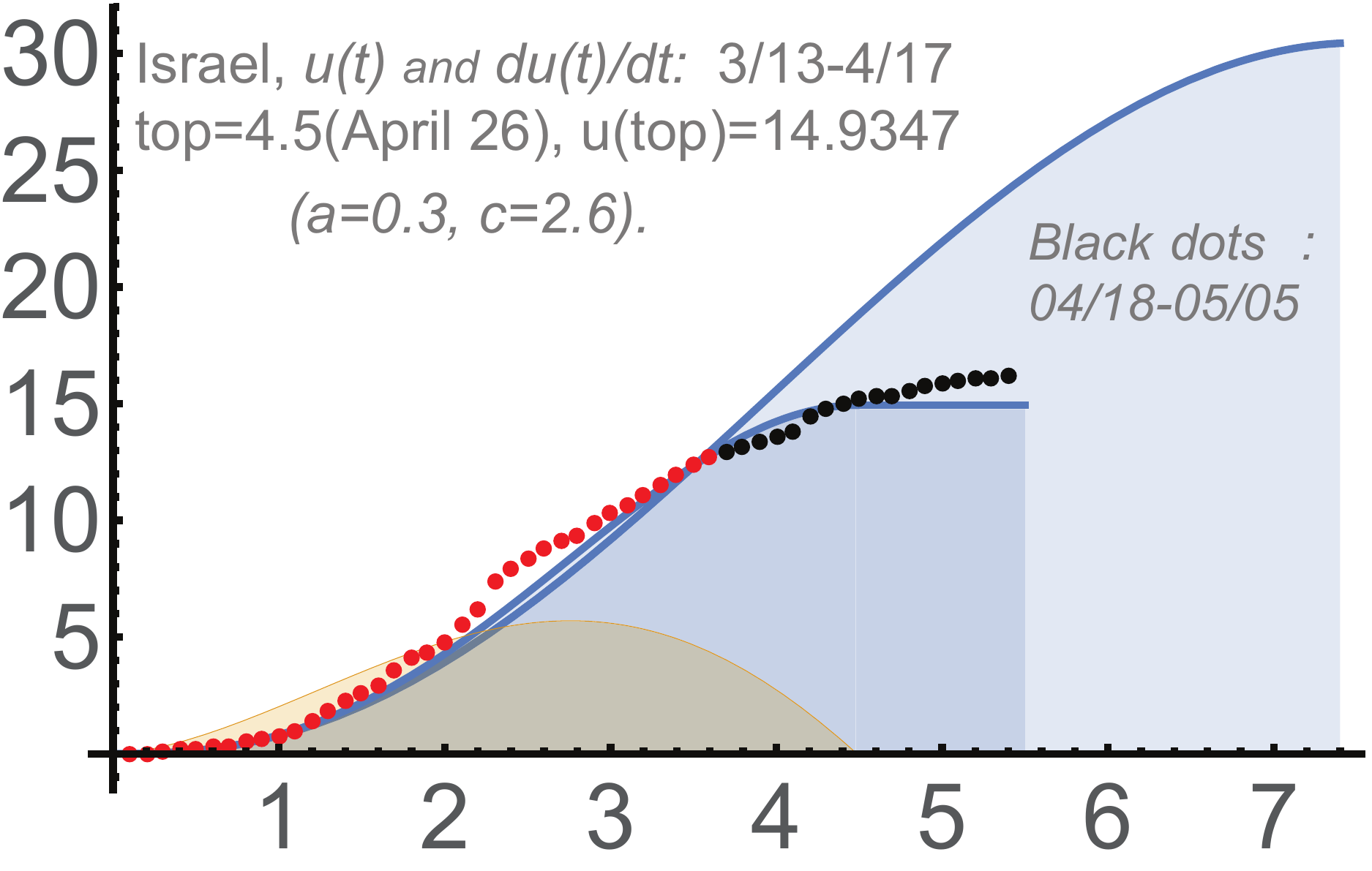}
\vskip -0.2in
\caption{$u(t)\!=\!2.2\,t^{(c+1)/2}J_{(c-1)/2}(\sqrt{a}t)$, 
$c\!=\!2.6,a\!=\!0.3$.}
\label{fig:Bessel-israel-new}
\end{figure*}

{\sf Sweden: anti-forecast.}
Sweden is actually expected {\it not\,} to follow
our $u(t)$ and $w(t)$; this country did not follow hard ways
with fighting {\it Covid-19}. We calculated
$w(t)$ to confirm this;  see Figure 
\ref{fig:Bessel-sweden-new}. Its expiration 
was approximately after 100 days ($x=10$); i.e. 
the $w$-saturation was about June 15. We did not expect this 
saturation to occur. Definitely some measure were in 
place in Sweden; for instance,
people readily contact the authorities with 
any symptoms of {\it Covid-19}. However this did not change
much the linear growth of the spread until recently,
when the country really did something to reduce
the spread (and the vacation period helped).

{\sf Israel: $u(t)$ worked.}
Here the usage of $u(t)$ was sufficient to model
the total number of detected cases till the "saturation",
which occurred almost exactly when $u(t)$ reached its maximum.
The black dots demonstrate well the "linear period" after
the saturation, which was essentially of the same type as
in South Korea, Austria and quite a few countries that went
through the saturation. Nevertheless, we provide the 
{\it forecast cone\,} calculated as above on the basis of $w(t)$.
In contrast to the USA, UK and, especially, Sweden, the black dots
are much closer to the flat line started at the $u$-top, which
was at $t=4.5$ (April 26).

\begin{figure*}[htbp]
\hskip -0.2in
\includegraphics[scale=0.5]{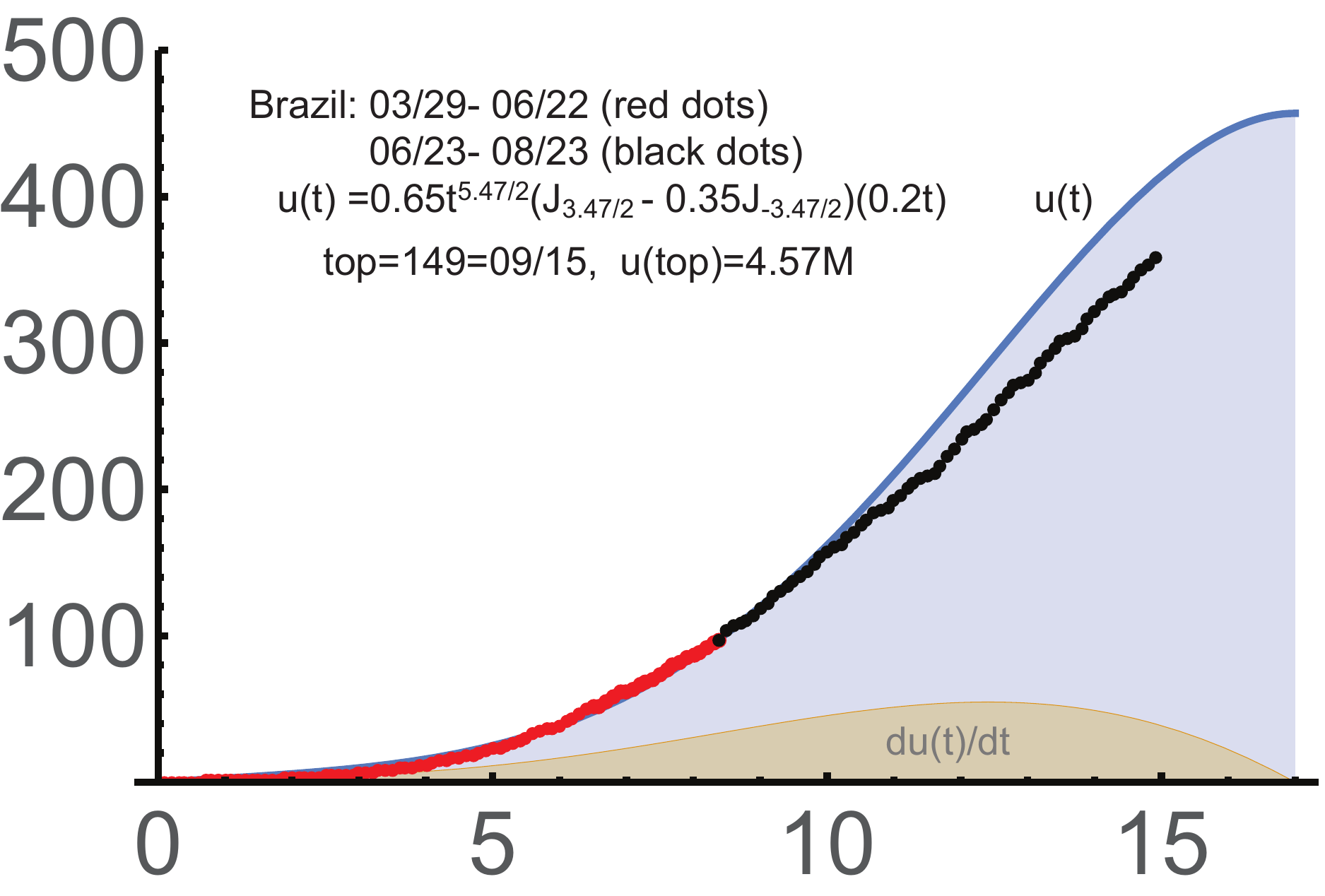}
\vskip -0.2in
\caption{
Brazil: $c\!=\!4.47, a\!=\!0.04, C\!=\!0.65$.}
\label{fig:brazil-fin}
\end{figure*}

\vskip 0.2cm
{\sf Brazil: 3/29-8/23.}
This is an ongoing process with high $c$ and low $a$. The initial
parameters were determined for the {\it red dots}:
from 3/29-6/22 with the starting number $3904$ of the total cases.
The black dots (the testing period) were till 8/23, 2020.
See Figure \ref{fig:brazil-fin}.

They parameters are $a=0.04, c=4.47, C=0.65$, where we 
set $y=$cases$/10K$.
In contrast to the previous examples, we use $u_1=Cu^1$ {\it and\,} 
the second (non-dominant) solution $u_2=Cu^2$: 
$$u(t)=u_1(t)\!-\!0.35 u_2=
C\,t^{c/2+0.5}\bigl(J_{c/2-0.5}(\sqrt{a} t)\!-\!
0.35 J_{0.5-c/2}(\sqrt{a} t)\bigr).
$$
The cone will not be provided, since the curve of real
detected cases is still within our $u$-curve. Note that
the parameter $a$, the intensity of "hard measures", is very
low, and $c$ is quite large.

\vskip 0.2cm
{\sf India.} The parameters are even more extreme for India:
$a=0.035, c=5.2$. This country is still in the period
of polynomial growth of the spread, as of August 25.
 The analysis of India is obviously important to 
understand the future of {\it Covid-19}. We
provide the corresponding $u$\~curve. Recall that it
is under the assumption that the "hard"  measures
are used in a steady manner, which was the case
in almost all Western Europe, China, South Korea, 
the USA (during some periods), and in several other
countries. Also, a sufficiently high level of 
health-care system and the uniformity of measures in the whole
country are necessary for the success of such measures.

Even when all these factors are present, the spread can be
very intensive. This occurred in the USA, where the $u$-curve
and $w$-curve served well only the middle stages.  
In  India, the parameters of the spread were
the most extreme among the countries we consider in this paper:
$$u(t)=C\,t^{c/2+0.5}\bigl(J_{c/2-0.5}(\sqrt{a} t)\!+\!
0.2 J_{0.5-c/2}(\sqrt{a} t)\bigr), c\!=5.2,\, a=0.035.
$$

\begin{figure*}[htbp]
\hskip -0.2in
\includegraphics[scale=0.5]{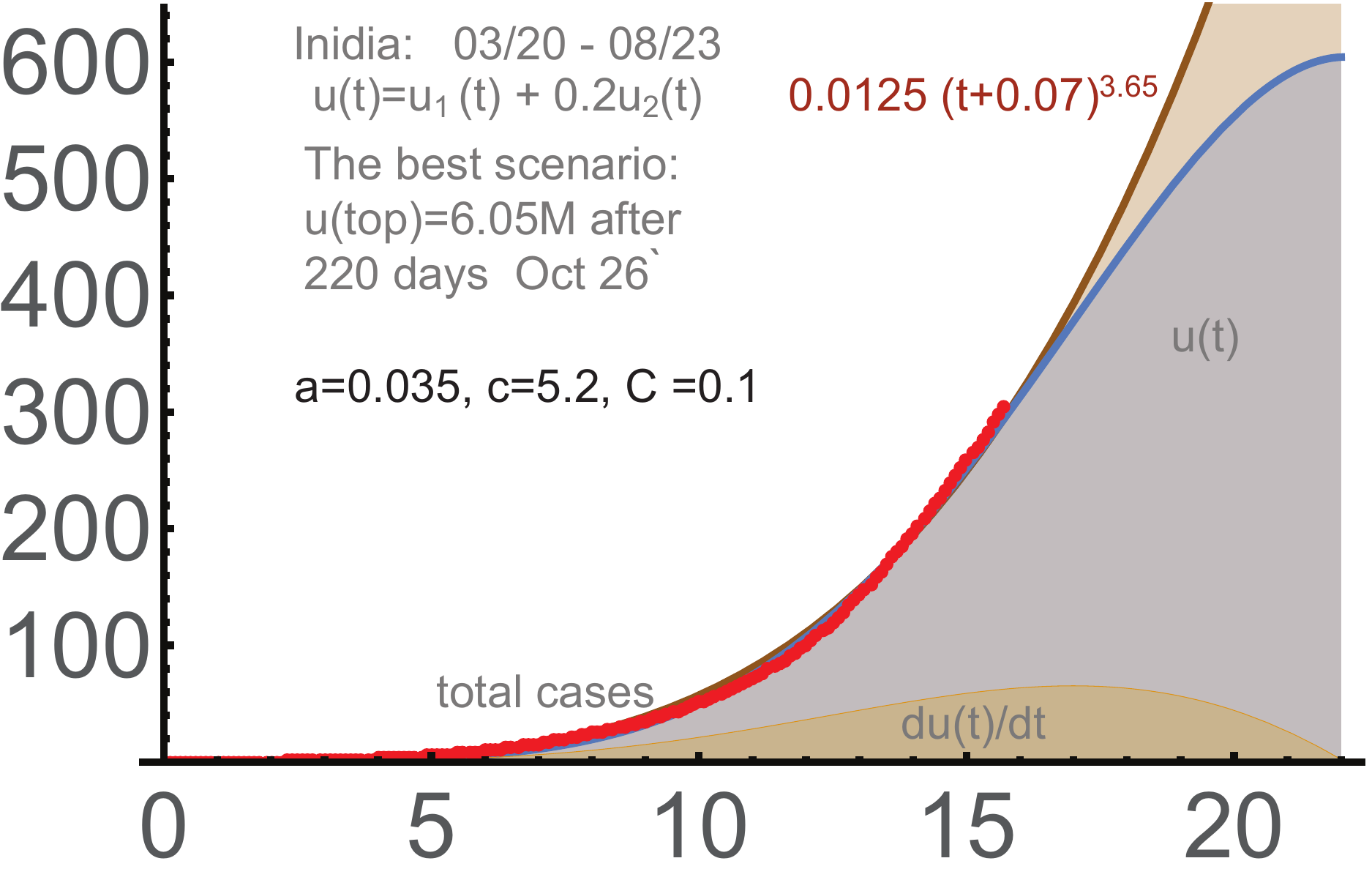}
\vskip -0.2in
\caption{
$India: 3/20-8/23, c\!=\!5.2, a\!=\!0.035, C\!=\!0.1$.}
\label{fig:india-fin}
\end{figure*}

We set $y=$cases$/10K$ in Fig. \ref{fig:india-fin}.
The period we used to determine
the parameters was 3/20-8/23 (no "black dots").
The starting number of detected 
cases was $191$, which was subtracted, 

Actually, $0.0125(t+0.07)^{3.65}$ gives a
good approximation for the main part
of the period; see Figure \ref{fig:india-fin},
where the graph of this function is shown as brown.
So this is really a period of (strong) 
polynomial growth, but {\it obviously\,} not of
any exponential growth. We never observed 
exponential growth of the number of detected
infections with {\it Covid-19} beyond
some very short initial periods.

\section{\bf Two-phase solution}\label{sec:twoph}
Let us describe our {\it two-phase\,}
model for the total number of infections. It is  
from the beginning of the intensive growth of cases
through the saturation $t_{top}$ of the $(A)$\~phase, 
and from $t_{top}$  till the "final saturation"  of the  
corresponding $(B)$\~phase. It works surprisingly well for the 
countries that managed to reach reasonably small numbers
of new daily infections during the first phase;
the "forecast cone" and the $(AB)$\~mode 
appeared unnecessary for them.

We determined below the parameters
$a,c,d$, and the scaling coefficient $C$ for the
period till May 22 (2020). So all dots are red 
unless for Israel, where we used $a,c,C$ above found in the middle of
April. The epidemic is of course far from over, but 
the first "wave"  ended in
these countries; so no test periods, "black dots", 
were considered.

 Recall that 
the parameters $a,c,C$ are mostly obtained to ensure the match
for the intervals till the 
"turning points";
$d$ can be determined only in the vicinity of $t_{top}$,
if it is reached. The
initial transmission rate $c$, must be the same for 
$(A)$ and $(B)$ according to our theory. 

The leading solution from (\ref{2})
is $u_{B}(t)=C_Bt^{c/2}\cos(d\log(Max(1,t)))$, where $c$ is the
one in $u(t)=C_A\,t^{(c+1)/2}J_{(c-1)/2}(\sqrt{a}t)$,
$d\!=\!\sqrt{a_B\!-\!c^2/4}$.  
The scaling coefficient $C_A, C_B$ are adjusted independently;
the intensity parameters $a=a_A$ and $a_B$ have
different meanings too.
In the later stages, we assume that $a_B>c^2/4$.
Here $Max(1,t)$ is due to some ambiguity
at $t=0$; anyway, we use $u_B(t)$ only for $t>1$. 
Here  $d$ equals very approximately   
to $\sqrt{a}$ for considered countries.

Importantly, we start $u_B(t)$ at $t=0$, though
it is used only after or around $t_{top}$. This results in the
following nice formula for the end\, $t_{end}$ of phase two: 
$t_{end}=\exp\bigl(\frac{1}{d}\,tan^{-1}(\frac{c}{2d})\bigr)$.

\vskip 0.2cm
{\sf Israel: 03/13-05/22.} We already closely considered the
first phase: $u(t)=2.2\,t^{(c+1)/2}J_{(c-1)/2}(\sqrt{a}t)$, 
till $t_{top}$ for $c=2.6,a=0.3$. For
the second phase, $d=0.6$ and 
$C_B=3.4$.

\vskip 0.2cm

\begin{figure*}[htbp]
\hskip -0.2in
\includegraphics[scale=0.5]{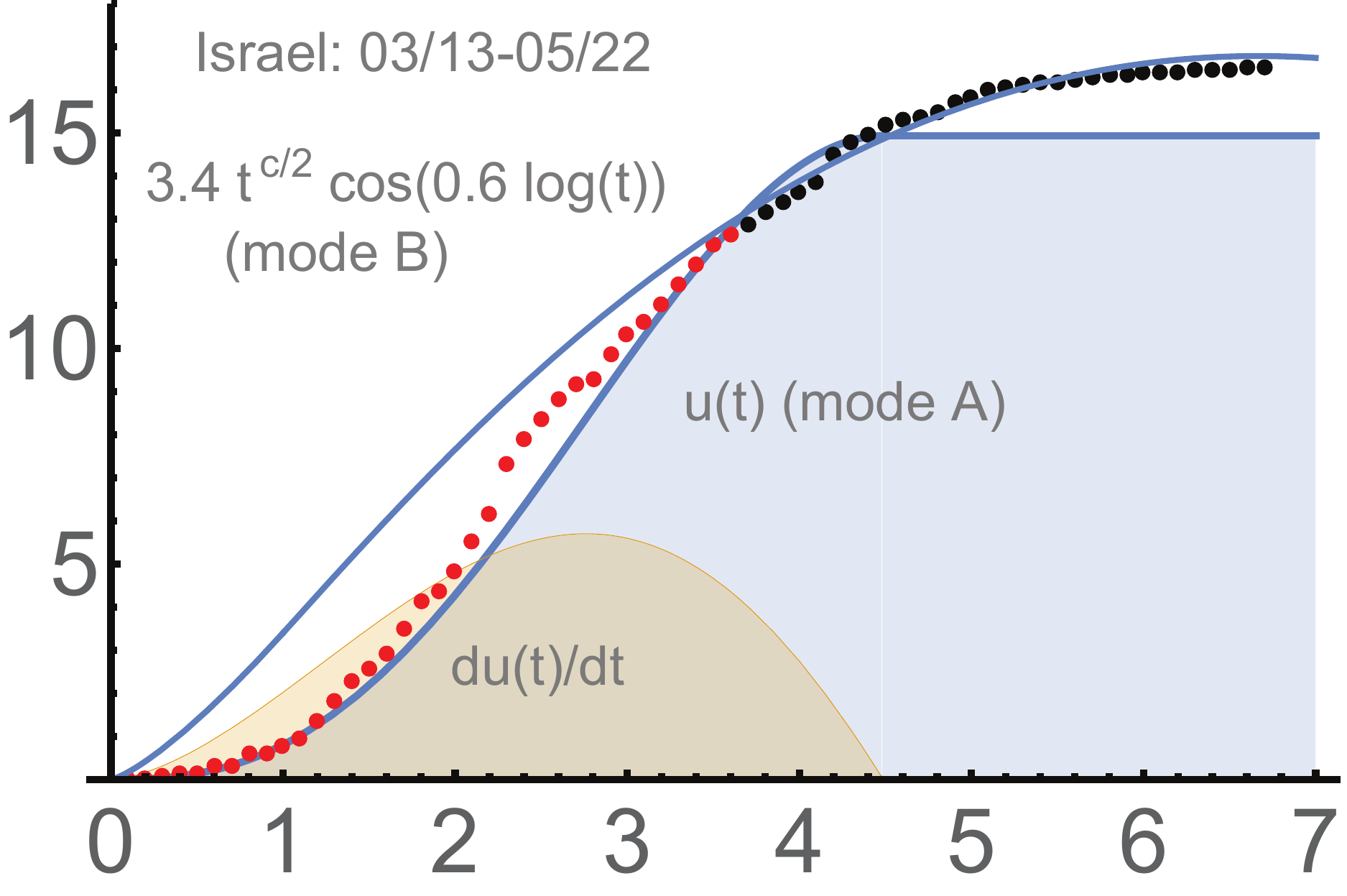}
\vskip -0.2in
\caption{
Israel: $c\!=\!2.6,a\!=\!0.3,d\!=\!0.6$.}
\label{fig:israel-2}
\end{figure*}

\begin{figure*}[htbp]
\hskip -0.2in
\includegraphics[scale=0.5]{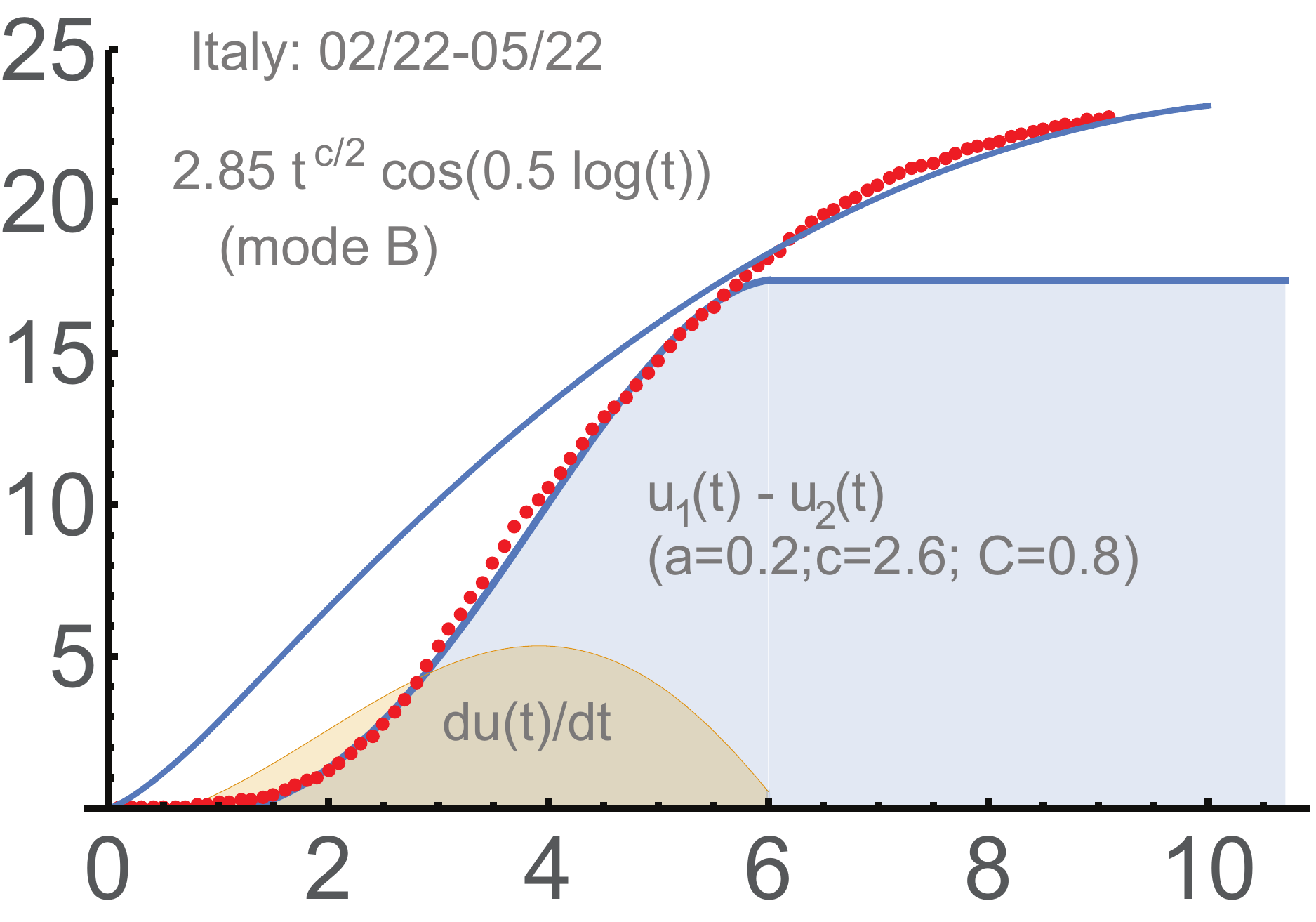}
\vskip -0.2in
\caption{
Italy: $c\!=\!2.6,a\!=\!0.2,d\!=\!0.5$.}
\label{fig:italy-2}
\end{figure*}

{\sf Italy: 2/22-5/22.} See Figure \ref{fig:italy-2}.
The starting point was $2/22$, when the total number
of infections was $17$; in this paper, we always subtract
this initial value when calculating our {\it dots}. One has:
\begin{align*}
&u_{1,2}(t)=0.8\,t^{(c+1)/2}J_{\pm \frac{c-1}{2}}
(\sqrt{a}t),\ \, u(t)=u_1(t)-u_2(t), \,\and \\
& 
u_{B}(t)=\!2.85\, t^{c/2}\cos(d\log(Max(1,t))),
c\!=\!2.6, a\!=\!0.2, 
d\!=\!0.5.
\end{align*}
We use here the second
solution $u^2(t)$ from (\ref{ptut1}). For $t\approx 0$,
it is approximately $\sim t$, i.e. smaller than $\sim t^c$ for 
the dominating solution $u^1$.
The top of the bulge occurs when $u^2(t)$ reaches its minimum;
which is approximately at  $t_{top}/2$, where
$t_{top}$ is that for $u_1$. 
Actually, we can see similar bulges for Austria, Israel 
and in the Netherlands too, 
but they are relatively short-lived and $u_1$
appeared sufficient. 
The linear combination $u_1(t)-u_2(t)$ was found numerically;
the coefficient of $u_2$  is not always $-1$.
Here and further in this section, $u_1=Cu^1$ and
$u_2=Cu^2$ for $u^{1,2}$ from (\ref{ptut1})
with the same rescaling coefficient $C$, found numerically.

\vskip 0.2cm

\begin{figure*}[htbp]
\hskip -0.2in
\includegraphics[scale=0.5]{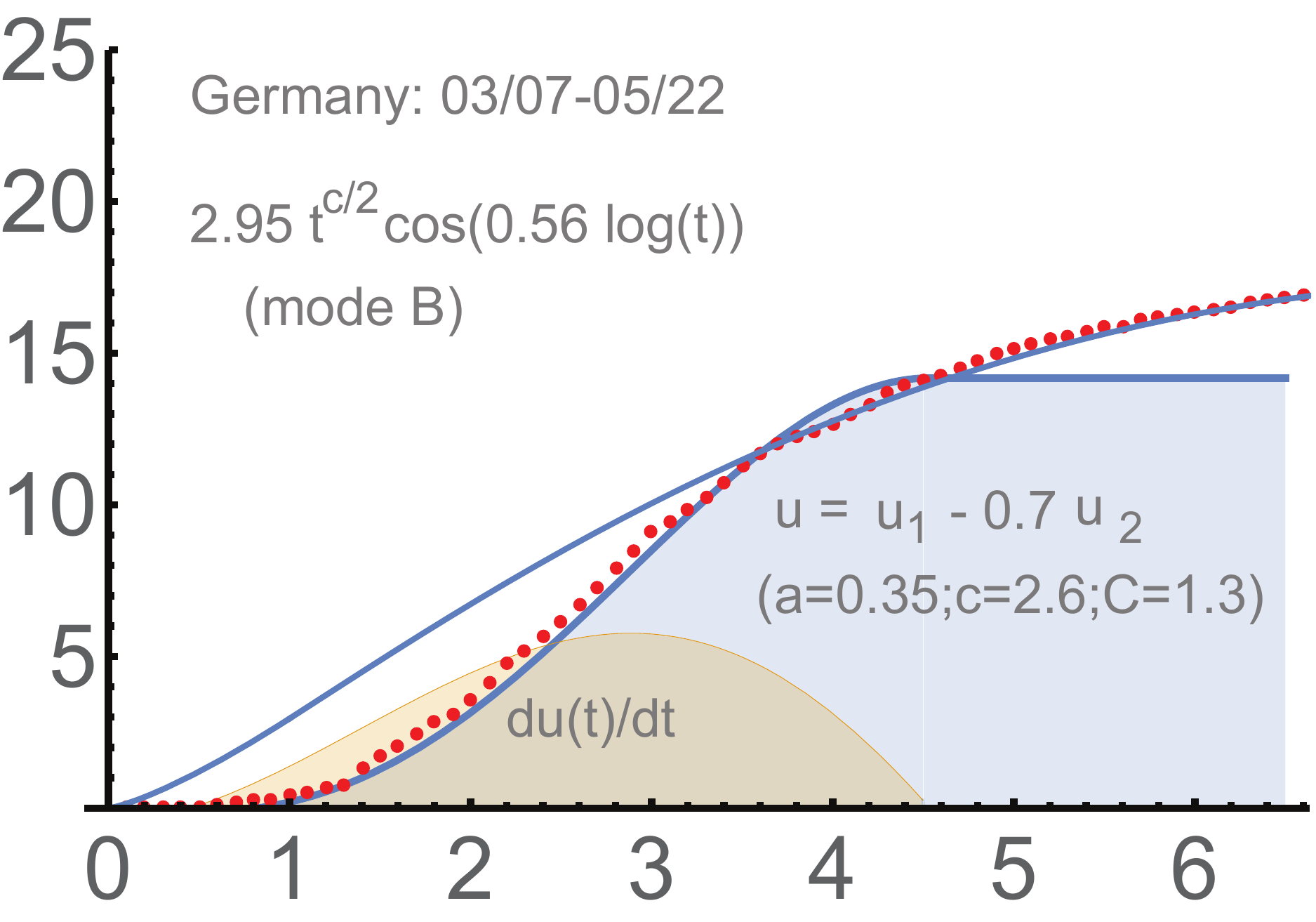}
\vskip -0.2in
\caption{
Germany: $c\!=\!2.6,a\!=\!0.35,d\!=\!0.56$.}
\label{fig:germ-2}
\end{figure*}

{\sf Germany: 3/07-5/22.} See Figure \ref{fig:germ-2}.
We began here with the initial number of total
infections $684$, which must be subtracted when calculating
the {\it red dots}. This was the moment when the curve
began to look stable, i.e. a systematic management began.
One has:
\begin{align*}
&u_{1,2}(t)=\!1.3\,t^{(c+1)/2}J_{\pm \frac{c-1}{2}}
(\sqrt{a}t),\  u(t)=u_1(t)-0.7 u_2(t), \hbox{\,\, \ and} \\
& 
u_{B}(t)=\!2.95\, t^{c/2}\cos(d\log(Max(1,t))),\, 
c\!=\!2.6, a\!=\!0.35, 
d\!=\!0.56.
\end{align*}

\begin{figure*}[htbp]
\hskip -0.2in
\includegraphics[scale=0.5]{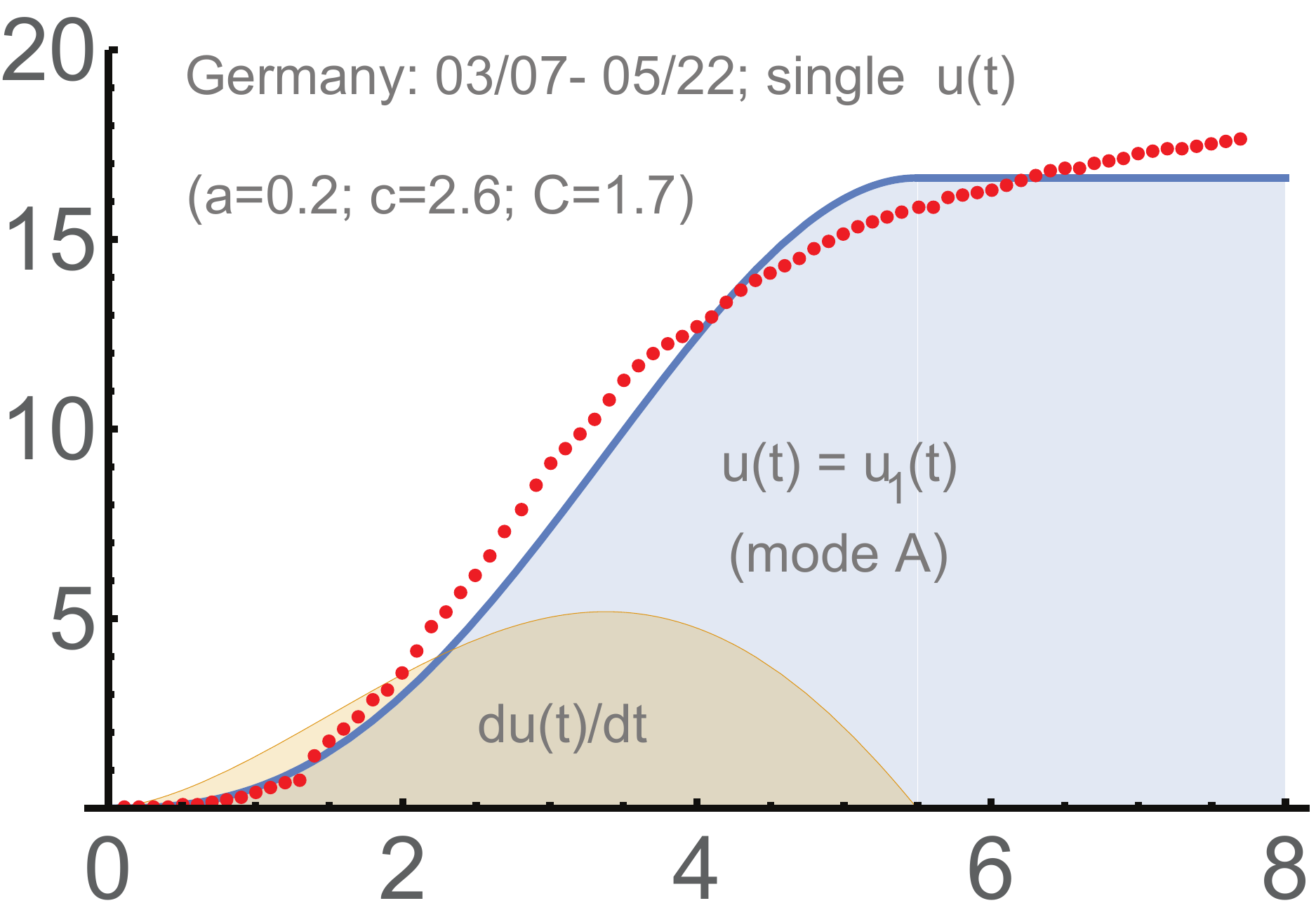}
\vskip -0.2in
\caption{
Germany: $u=u_1= 1.7 t^{1.8}J_{0.8}
(t\sqrt{0.2})$}.
\label{fig:germ1}
\end{figure*}

It makes sense to provide the $u$\~curve for the first phase
in Germany using {\it only\,} $u_1$;
Figure \ref{fig:germ1}. We disregard the bulge
in the middle of {\it phase one\,}, which is 
actually similarly to what
we did for Israel. The parameters are different:
$a=0.2,c=2.6,C=1.7$, i.e. now $u(t)= 1.7 t^{1.8}J_{0.8}
(t\sqrt{0.2})$.

For forecasting, $u_1(t)$ 
seems the most reasonable to use regardless of the bulges.
We "miss" the bulges, but then we mostly return to the curve
calculated with $u^2$, similar to what we observed above with
Israel. This remains essentially the same for Italy and Japan.
Let us give the corresponding
parameters for the usage of $u=u_1$ only: 
Italy ($c=2.5, a=0.1, C=1$),
Japan ($c=2.6, a=0.1, C=0.9$).

Practically, we always begin with using $u^1$ only
and obtain $a,c,C\,$ for the interval before or around
the turning point, when the bulge (if any) can be not
really visible. Then they must be adjusted constantly,
and $u^2$ can be added if necessary.
The initial $a,c,C$  are of course subject to quite a few 
uncertainties.

\begin{figure*}[htbp]
\hskip -0.2in
\includegraphics[scale=0.5]{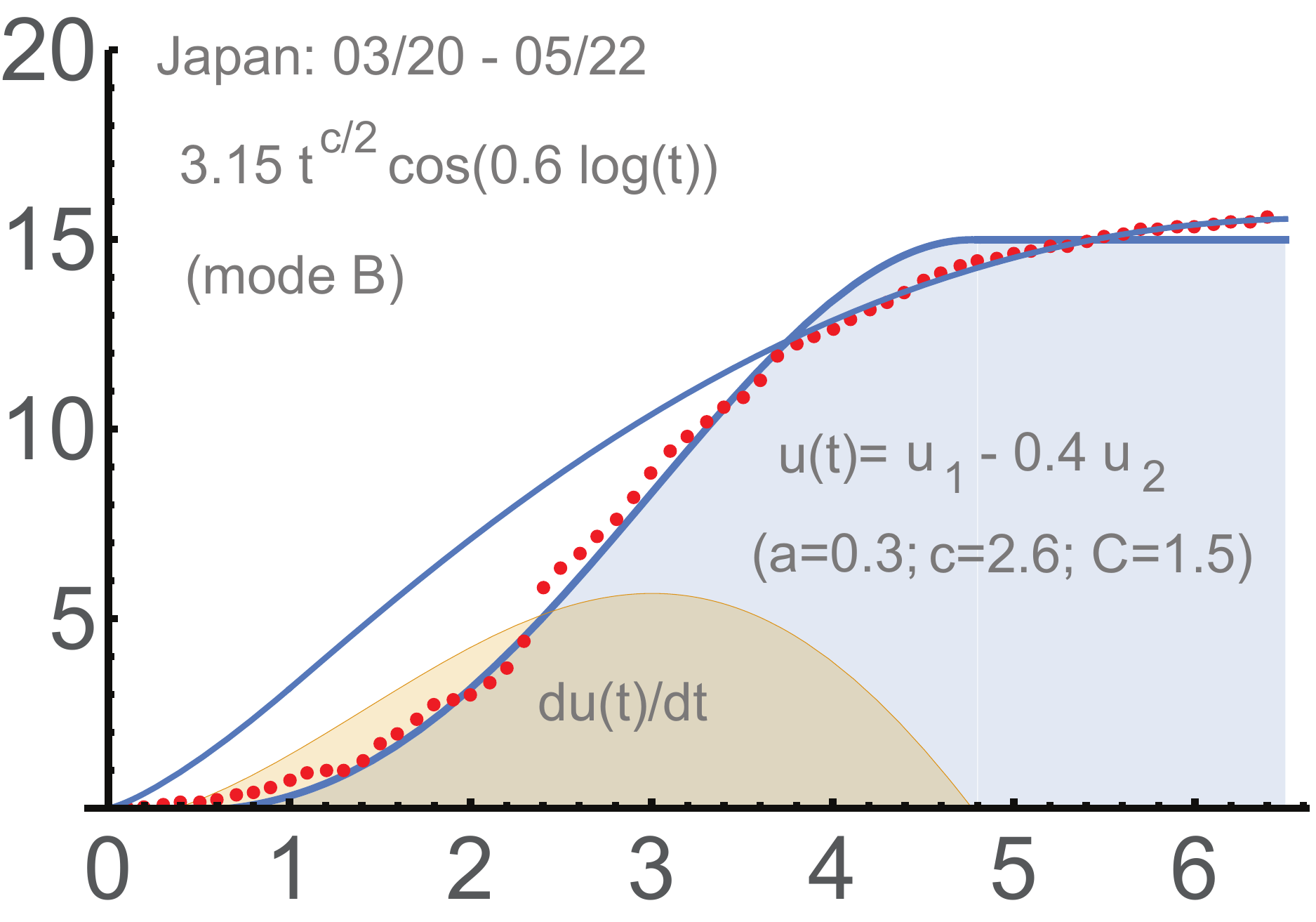}
\vskip -0.2in
\caption{
Japan: $u= 1.5 t^{1.8}(J_{0.8}-0.4J_{-0.8})
(t\sqrt{0.3})$.}
\label{fig:japan-2}
\end{figure*}

\vskip 0.2cm
{\sf Japan:  3/20- 5/22.} See Figure \ref{fig:japan-2}.
It was on top of some prior stage, so there were 
already $950$ (total) infections on
March 20. The curve is not too smooth,
but manageable well by our {\it 2-phase solution\,:}
\begin{align*}
&u_{1,2}(t)=\!1.5\,t^{(c+1)/2}J_{\pm \frac{c-1}{2}}
(\sqrt{a}t),\  u(t)=u_1(t)-0.4 u_2(t), \hbox{\, and} \\
& 
u_{B}(t)=\!3.15\, t^{c/2}\cos(d\log(Max(1,t))),\ \,
c\!=\!2.6, a\!=\!0.3, 
d\!=\!0.6.
\end{align*}

\begin{figure*}[htbp]
\hskip -0.2in
\includegraphics[scale=0.5]{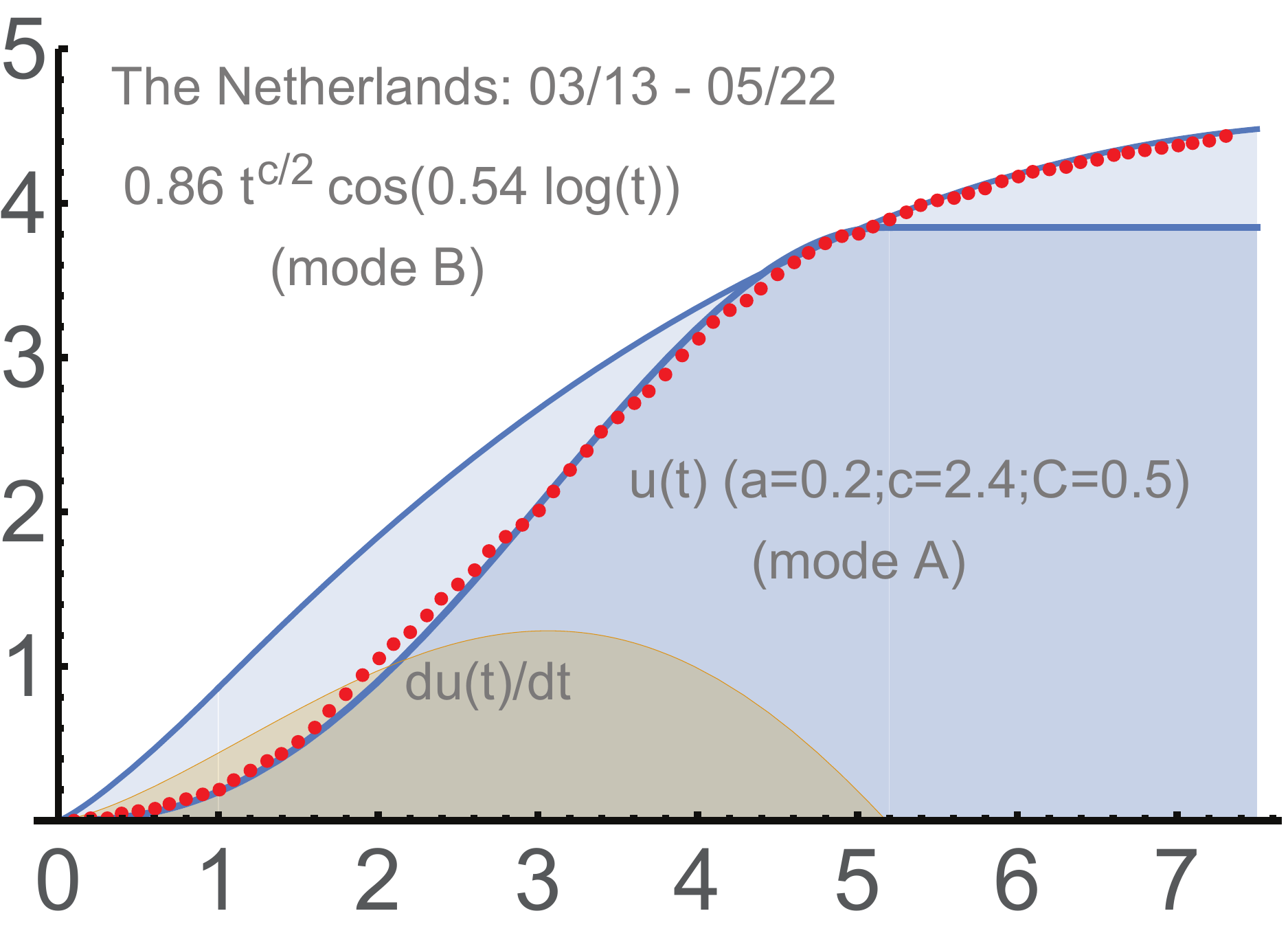}
\vskip -0.2in
\caption{
The Netherlands: $u= 0.5 t^{1.7}J_{0.7}(t\sqrt{0.2})$.}
\label{fig:nether-2}
\end{figure*}

\vskip 0.2cm
{\sf The Netherlands: 03/13-5/22}. Figure \ref{fig:nether-2}.
The response to {\it Covid-19\,}  was relatively late in
the Netherlands; the number of the total case was
$383$ on 3/13, the beginning of the intensive
spread from
our perspective. A small "bulge" in the middle of the
phase one can be seen here too,
but $u_1$ appeared sufficient: the country perfectly
reached the saturation at $t_{top}$ and then smoothly 
switched to phase two:
\begin{align*}
&u(t)\,=\, 0.5\,t^{(c+1)/2}\,J_{\frac{c-1}{2}}
(\sqrt{a}t),\ \ \, c\!=\!2.4,\  a\!=\!0.2, \\
& 
u_{B}(t)=\!0.86\, t^{c/2}\cos(d\log(Max(1,t))),
d\!=\!0.54.
\end{align*}


To finalize, the best ways to use our curves for
forecasting seem as follows:
\vskip 0.1cm 

\noindent
{\it (1):} determine $a,c, C\ $ when the spread looks
essentially linear;

\noindent
{\it (2):} update them constantly till the turning point
and beyond;

\noindent
{\it (3):} expect the "bulges" to appear and add the
$u^2(t)\,$ if needed;

\noindent
{\it (4):} try to adjust the intensity of the measures to 
match\  $u(t)$\ ;

\noindent 
{\it (5):} at the turning point, determine 
$b,D\ $ and the bound\ 
$w(t)$;

\noindent
{\it (6):} after the saturation at $t_{top}$, 
find $\ d$\, and switch to {\it phase 2\,},


Generally, the data must be  as uniform and "stable" as 
possible. Then 
underreporting the number of infections, the fact that these
numbers mostly reflect symptomatic cases, and inevitable 
fluctuations
with the data may not influence too much the applicability of the
$u,w$\~curves for phase 1, and $u_B$ for phase 2. 
"Testing-detection-isolation-tracing" must be of course continued 
non-stop; the recurrence of the epidemic is a real threat. 


\section{\bf Entering second wave}\label{sec:auto}
{\phantom{} }
The second waves are supposed to occur (well) after the
first ones. However this is not granted
for {\it Covid-19}, where the spread heavily depends
on active management.  Efforts  
are needed to finish the first wave
as quickly as possible to avoid building the second wave on
the top of the first. So there must be reliable tools
for forecasting during the second phase, which we actually
have. Their importance is clear from the example of the USA.

As the whole, the country remained in phase 1, when the
second wave began. However, there was a sufficiently
long period when  many states were actually in phase two.
It was similar to Western Europe in July, but all major
countries there were in phases 2 in contrast to 
the USA.  However this appeared
sufficient to predict the end of phase 2 in the USA
around 09/19, which forecast was very stable for sufficiently
long period. We used the following
concept of "interaction".

Generally, all 50 states, must
be considered separately; then the 
sum of the resulting curves is the forecast.
However, such a straightforward approach
does not take into consideration that 
improvements in one state positively impact
the other ones. This is similar to stock 
markets. For instance, the components of {\it SP500\,}
are supposed to be modeled individually, however 
their interaction must be taken into consideration. 

The simplest approach to the interaction is what
we actually used when taking the sums of two 
Bessel-type solutions, $u_1$ and  $u_2$. The second 
(non-dominating) 
can decrease
as far as the total $u(t)$, a proper linear combination
of $u^1$ and $u^2$, increases. 
Similarly, the {\it superposition\,} of $u$\~functions
for different states or countries can be defined
as their sum where "we allow"
the components, modeling  different states,
to decrease as far as the total goes up. Here 
the physics intuition is relevant, but we will
not specify.

\begin{figure*}[htbp]
\hskip -0.2in
\includegraphics[scale=0.5]{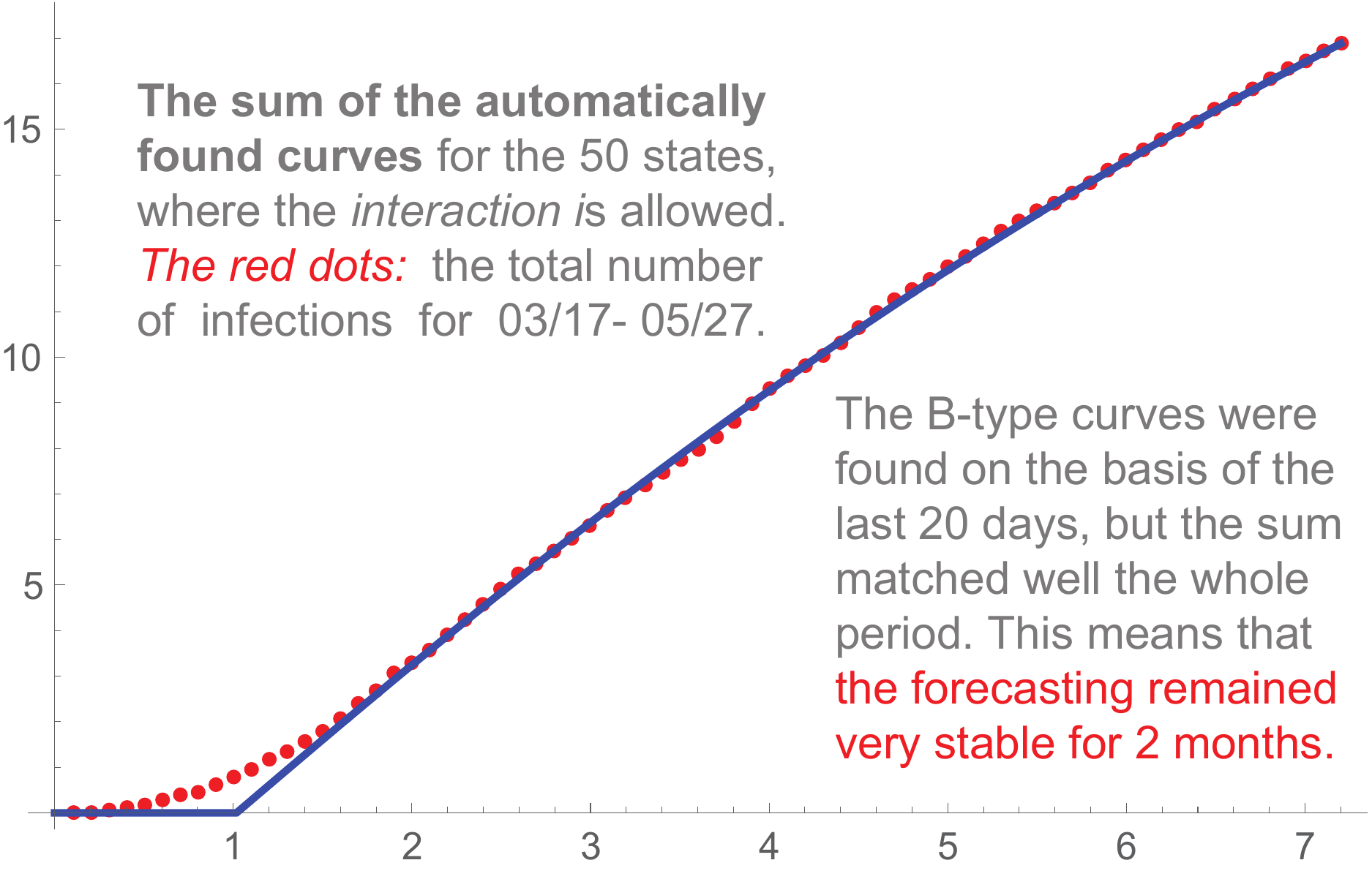}
\vskip -0.2in
\caption{
USA, the sum of the curves for individual states.}
\label{fig:fullusa}
\end{figure*}

The corresponding analysis required
a fully automated system, which was developed
and will be made available.
It produced the
curves and determined the status of every state 
"without human interference". 

The first automated forecast for $50$ states was based on
the period 03/17-05/27; we use for our forecasts for the
USA the data from the site
{\tt https://github.com/nytimes/covid-19-data\,}.
Every state was processed individually. The
{\it red dots} are the total numbers of detected cases, as above.

Our program focuses on the last 20 days; however, the match
with the total number of detected infections 
appeared {\it perfect\,} almost from 03/17 and remained so for
further auto-forecasts for a long period; 
see Figure \ref{fig:fullusa}. Such a high level of stability
is actually rare in any forecasting, which made the
chances to reach the saturation around 9/19 high, However,
the hard measures were significantly reduced at the end of
May practically in all 50 states. 
As a result, the number of states that reached 
phase 2 dropped 
from about 22 (at 5/27) to 8 (at 7/12). Then, in the second half 
of June the USA entered the {\it second wave}. We
provide the automated output of our program from 6/21, one
of the last  before the curve of the total cases in the USA
changed dramatically.

\begin{figure*}[htbp]
\hskip -0.2in
\includegraphics[scale=0.5]{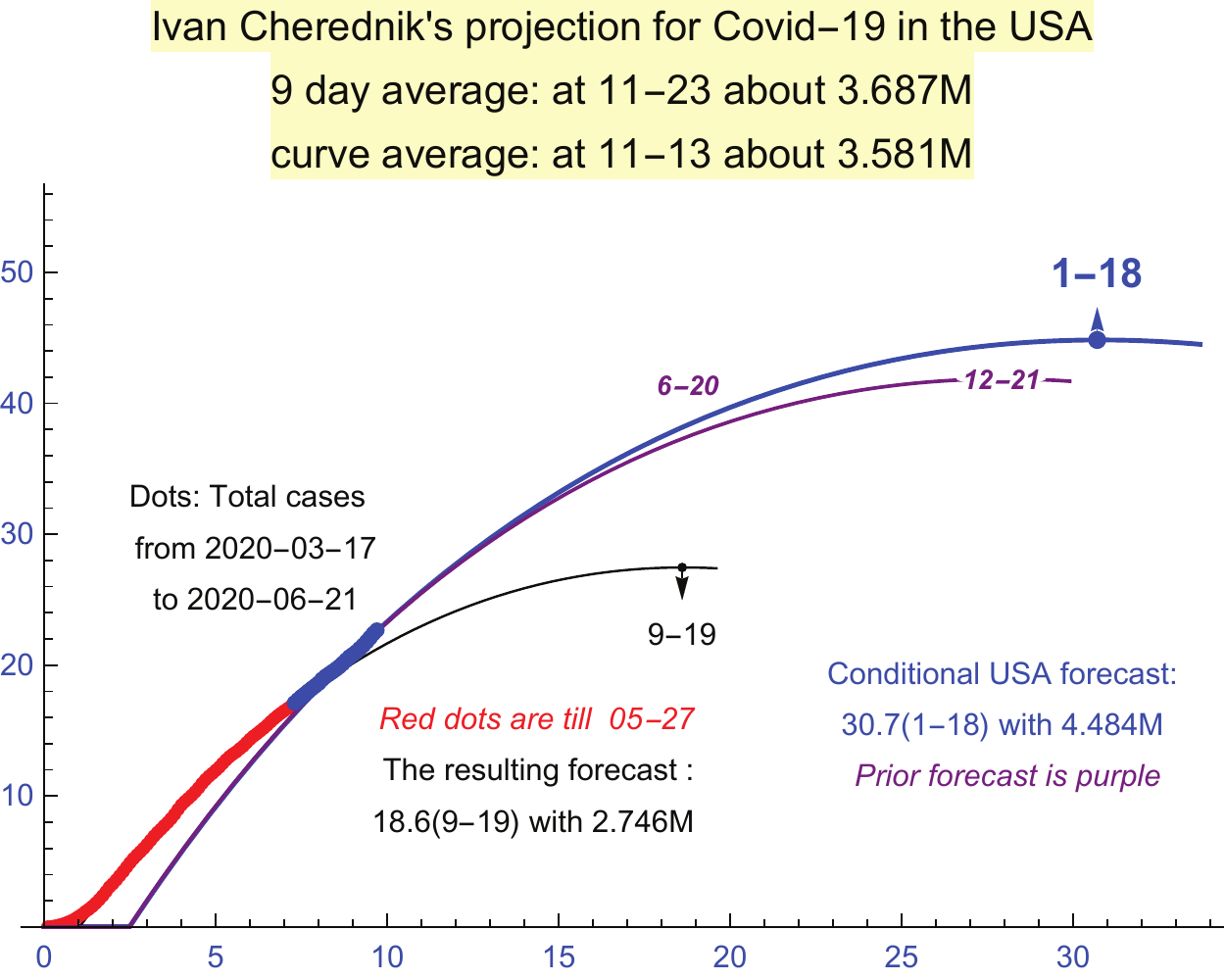}
\vskip -0.2in
\caption{
The forecast for the USA as of 6/21.}
\label{fig:usa-6-21}
\end{figure*}

The "curve average" in this output
is the maximum and the corresponding value of the average
of 9 last curves $u_B(t)$ for the USA, i.e. the moving average. 
The 9-day average is the simple average of the corresponding
maxima; see Figure \ref{fig:usa-6-21}.

\begin{figure*}[htbp]
\hskip -0.2in
\includegraphics[scale=0.5]{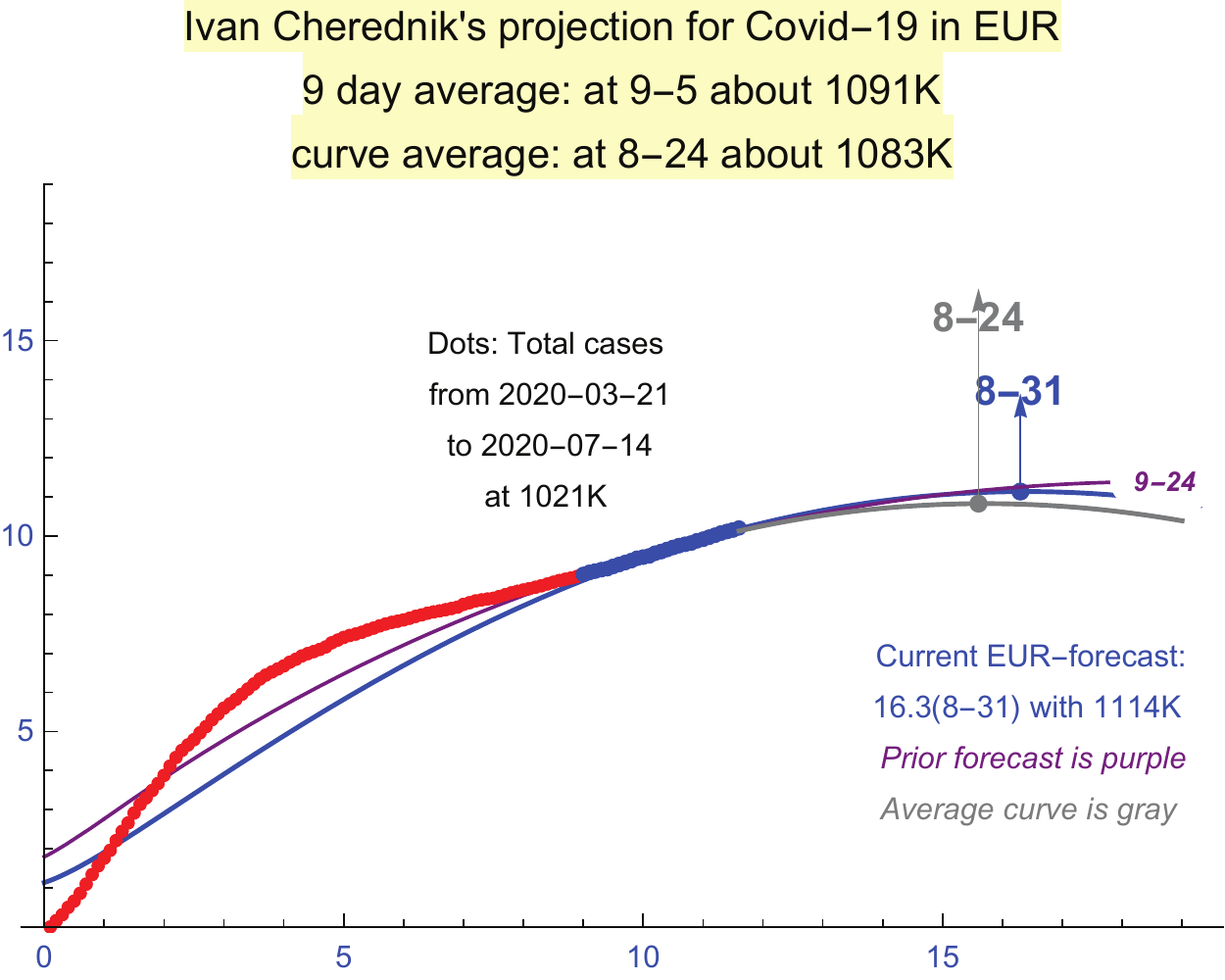}
\vskip -0.2in
\caption{
The forecast for Europe as of 7/14.}
\label{fig:eu-7-14}
\end{figure*}

This is quite different from what our program gave for
Europe {\it till the end of July},
 to be exact for the following 45 countries:
{\footnotesize 
Albania, Andorra, Austria, Belgium, Bosnia and Herzegovina, 
Bulgaria, Croatia, Cyprus, Czech Republic, Denmark, Estonia, 
Faeroe Islands, Finland, France, Germany, Gibraltar, Greece, 
Guernsey, Hungary, Iceland, Ireland, Isle of Man, Italy, Jersey, 
Kosovo, Latvia, Liechtenstein, Lithuania, Luxembourg, Macedonia, 
Malta, Monaco, Montenegro, Netherlands, Norway, Poland, Portugal, 
Romania, San Marino, Serbia, Slovakia, Slovenia, Sweden, 
Switzerland, Vatican}. See Figure \ref{fig:eu-7-14}.

As of July 8, the following
countries had clear second waves:
{\footnotesize Albania, Bosnia and Herzegovina, Bulgaria, Croatia,
Czech Republic, Greece, Kosovo, Luxembourg, Macedonia, Montenegro,
Romania, Serbia, Slovakia, Slovenia.}  Sweden, Poland, 
Portugal and some other countries did not reach phase 2
at that time. 
Nevertheless, the forecasts were sufficiently stable, though with 
a tendency to increase. 
Such stability changed this fall due to the end of the
vacation periods and the beginning of the school year.

The saturation projections are of course of
technical nature, certainly {\it not
for the end of the spread\,}. By design,
the "saturation targets" are supposed to increase over time. 
Moreover, a modest linear growth of the total number of detected
infections is expected after the "saturation", if it
was achieved.
\vskip 0.2cm

\begin{figure*}[htbp]
\hskip -0.2in
\includegraphics[scale=0.5]{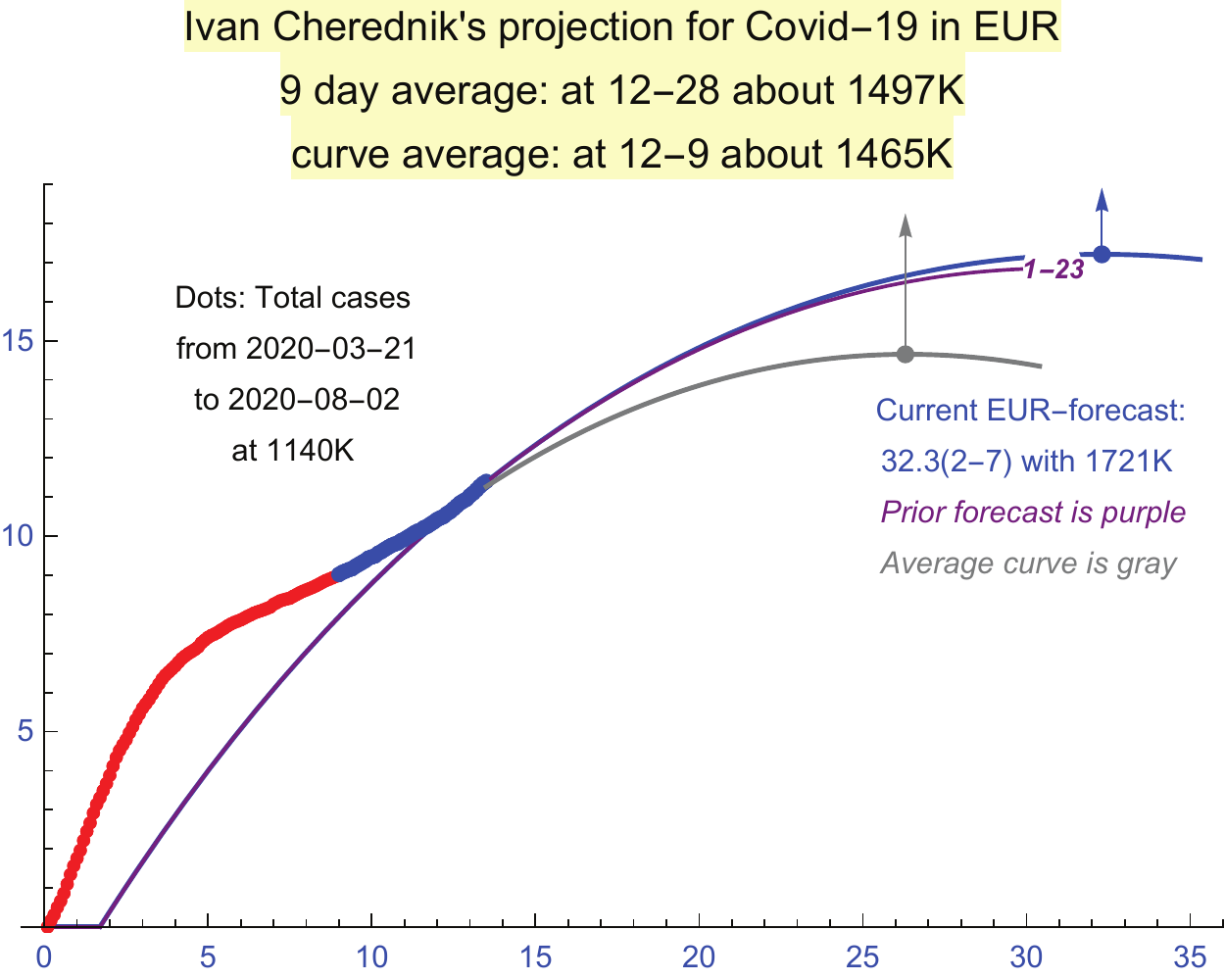}
\vskip -0.2in
\caption{
The forecast for Europe as of 8/02.}
\label{fig:eu-8-2}
\end{figure*}

In Europe, the second phase was 
stable for quite a long period, but then the spread there
changed significantly at the end of July, which is very similar
to the beginning of the "hockey hook"  in the USA shown in
Figure \ref{fig:usa-6-21}. The saturation target in 
Figure \ref{fig:eu-8-2} still exists, but it moved to December and is
only of some symbolic value.
Somewhat later, the 9-day average curves became essentially linear:
the vacation period was over. 

\vskip 0.2cm

{\sf Second waves.}
We will focus on the second wave in Israel, where it appeared
of unexpectedly large magnitude, and in the USA, where it
began with a very large number of new daily 
cases due to the unfinished first wave. The expectations were that
the parameter $c_\circ$ for the 2nd wave could be essentially
the same as $c$ for the 1st. This was based on the
assumption that the termination of the first wave was
mostly due to the "hard measures". So 
if they are reduced or abandoned,  the epidemic can be back 
with about the same parameters.  It appeared actually worse than
expected: the intensity of hard measures during
the 2nd wave dropped significantly in Israel and the USA. 
It is not impossible though that some lower intensity of hard measures 
can have the same impact during the 2nd wave.

With Israel: $c_\circ=2.75, a_\circ=0.1, C_\circ=4.8$, 
and the corresponding
$u_\circ(t)=C_\circ t^{c_\circ+1/2}\,(J_{c_\circ-1/2}
+0.1 J_{1.2-c_\circ}(\sqrt{a_\circ}t)$; the
parameters were determined for the period 6/13- 7/14 and served
well till about 07/30, when the change of the trend was similar
to the transition from phase 1 to phase 2 we observed
in quite a few countries. Then the number of cases began to
increase dramatically.
 
\begin{figure*}[htbp]
\hskip -0.2in
\includegraphics[scale=0.5]{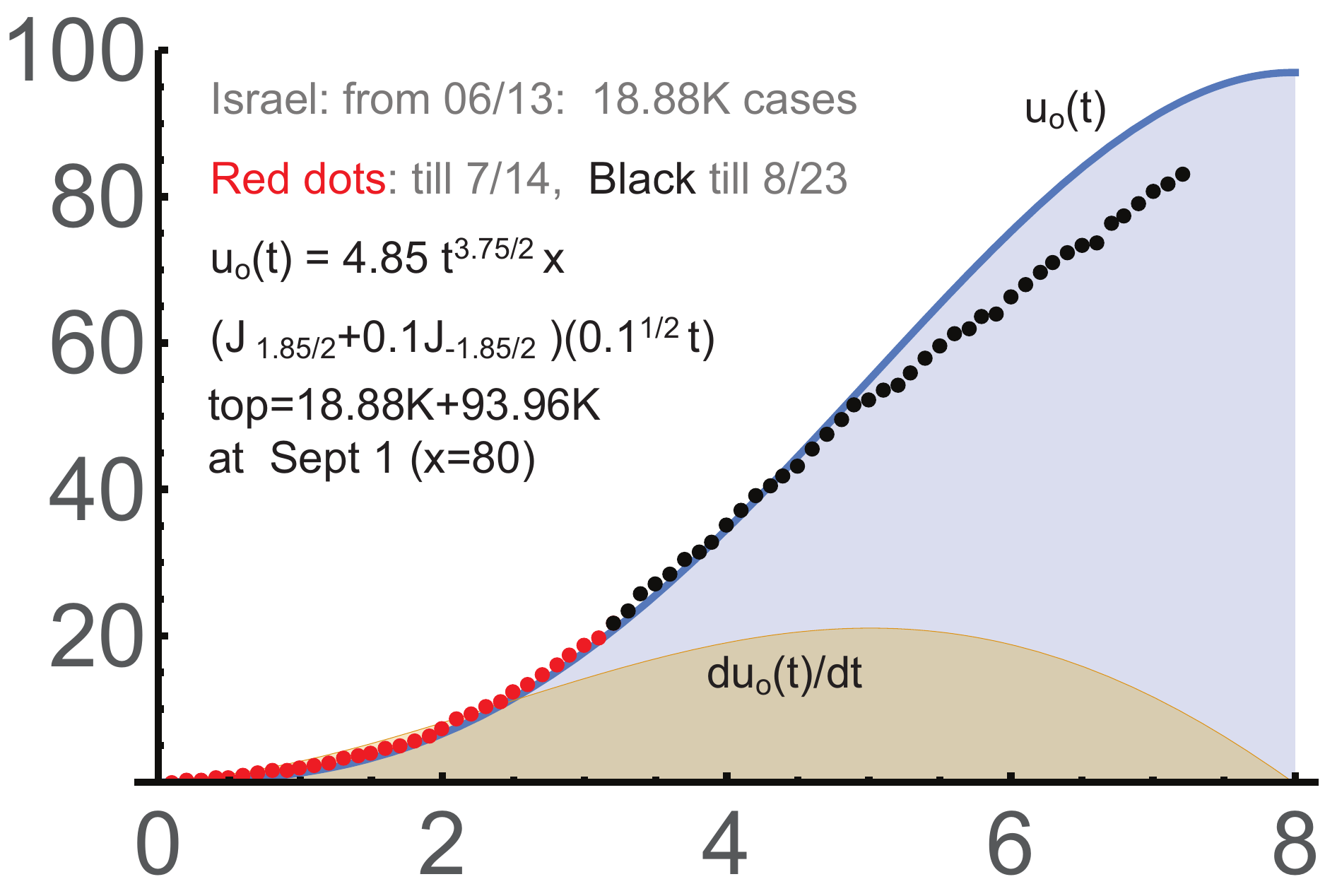}
\vskip -0.2in
\caption{
Second wave in Israel: 6/13-8/23.}
\label{fig:israel-2w}
\end{figure*}

The starting value at 06/13 was $18875$, the number of cases
at 7/7 was $31K$. The black (control) dots were added at 7/14.
See Figure \ref{fig:israel-2w}; $y=$ days$/1K$. 
The parameter $c_\circ$ is somewhat greater that $c=2.6$
for the first wave in Israel, and the intensity of hard
measures, $a_\circ$, is much smaller than $a$.
The significant drop of $a$ can be possibly expected;
$1/\sqrt{a}$ is directly related to the duration of
the wave.

We recall that the first wave in Israel matched well
our $u$-curve, and this wave was managed efficiently.
This was under strict lockdowns. 
The management of the first wave in the USA was not that steady.
Opening schools in 
August-September is one of the key unknown factors, but not the 
only one of course. Also, a significant increase of the number
of new daily cases in Europe at the end of vacation periods was 
expected; this really happened.  
 
\vskip 0.2cm

\begin{figure*}[htbp]
\hskip -0.2in
\includegraphics[scale=0.5]{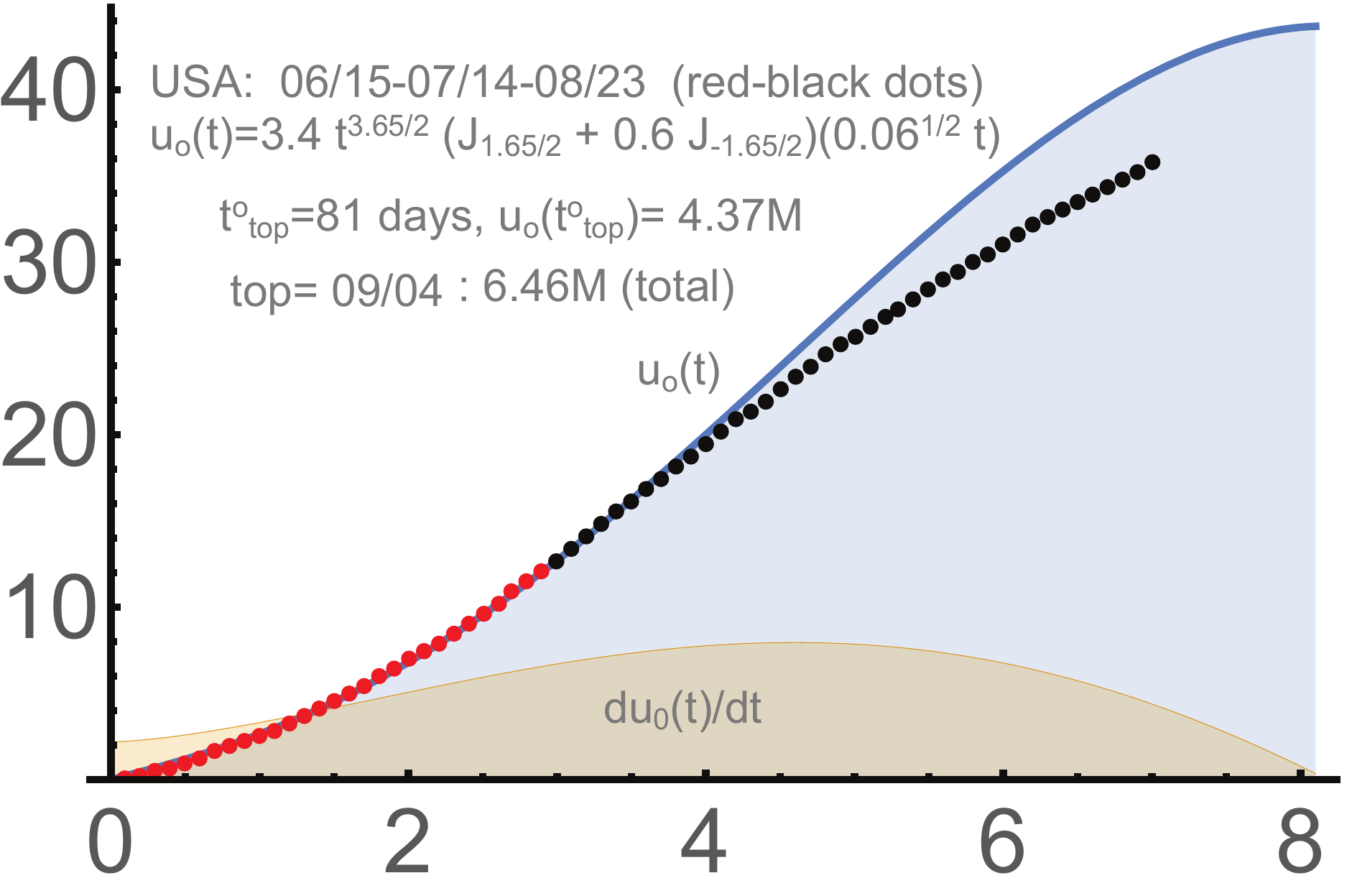}
\vskip -0.2in
\caption{
Second wave in the USA: 6/15-8/23.}
\label{fig:usa-2w}
\end{figure*}

The starting number of the total cases for the second wave
in the USA at 6/15 was $2.1M$; the red dots 
were till 7/14, and the control black dots closely
follow the $u$-curve till about 7/27. 
Upon subtracting $2.1M$, one 
has:  \ 
$a_\circ=0.06,\, c_\circ=2.65,\, C_\circ=3.4$,
$$
u^\circ_{1,2}(t)=\!3.4\,t^{(c_\circ+1)/2}J_{\pm \frac{c_\circ-1}{2}}
(\sqrt{a_\circ}t),\  u_\circ (t)=u^\circ_1(t)+0.6 u^\circ_2(t).
$$ 
Here we use our second Bessel-type
solution $u^2_\circ$,
so the connection of $c_\circ$ with
 $c=2.2$ for the first wave (obtained without $u^2$) is not that 
direct. We used $u^2$ for Israel too, but its coefficient was 
small: $0.1$. 

Generally, finding $a,c,C$ in the {\it early stages\,}
and using them for some {\it preliminary\,}
forecasting seems  more reasonable {\it without} 
$u^2$, even if the latter gives a somewhat better match 
with the "red dots". The usage of $u^2$ mainly addresses the
presence of some 
bulges, 
i.e. some effects of "second order".

Generally, finding the optimal parameters
which are stable enough, 
requires at least passing the "turning points". This
is fully applicable to $a$, but 
the parameter $c$, the transmission
coefficient in our theory, can be seen well in the early stages.

We note  that the presence of wave 1 can generally make
modeling  wave 2 challenging mathematically.
According to \cite{ChAI}, hypergeometric functions may be 
necessary.
However, it appeared that
simple subtracting "the pedestal" from the first
wave made the Bessel functions sufficient to model the
second wave in the USA: Figure \ref{fig:usa-2w}.

The similarity of
the curves of total cases in Israel and the USA
considered above is interesting.  This is
for the second wave and the situations are very different in these
countries. Nevertheless, the curves
almost coincide upon some rescaling.
The second interesting feature is that we see at the end
of July a 
pattern in these curves similar to the switch from phase 1
to phase 2 we observed in many countries during
the first wave. For instance,
see Figure \ref{fig:germ-2}. This resulted in some passage to
phase 2 in the USA, but not in Israel.
\vskip 0.2cm

The situation is of course
very fluid, the levels of new cases are high in both countries, 
and there is uncertainty with the beginning of the Fall there
and everywhere. 
Then the developments were different in both countries.
The USA basically reached phase $2$ as of 09/13; many schools
and business were closed. Israel's curve went up significantly
and eventually this country enforced a total lockdown.

\vskip 0.2cm

\section{\bf Conclusion}
The main result of this paper is that the 
curves of the total numbers of detected infections of {\it Covid-19} 
can be described
with high accuracy by our "two-phase solution". Let us provide 
one more confirmation. It appeared that Figure 24
extended to September 13 matches very well our formula
for proper parameters. As for 9/20, the trend was toward some
linear growth of the total number of cases, but
the applicability of our two-phase solution to the 
second waves of this epidemic is of course an important
argument in favor of the validity of our approach. 

Let us provide some account of  the
projections for the USA
we obtained during several stages:

(1) in contrast to almost all Europe,
the initial $u$\~curve presented in Figure \ref{fig:Bessel} 
with the saturation $t_{top}$ at 5/5 did not
work for the USA;

(2) the switch to mode $(AB)$ and the  
corresponding $w$\~curve with the saturation at 5/30 in Figure 
\ref{fig:Bessel-usa-new} did not work too, in contrast to UK;

(3) the projection "9/19" from
 Figure \ref{fig:usa-6-21}, based on the fact
that $\sim$22 states reached phase 2 in June,
did not materialized as well.
\vskip 0.1cm
\vfil

A very good match  with our Bessel-type solution for the
total number of detected cases in the USA in the middle 
of phase 1 of the first wave was the 
reason to expect (1) or (2), by analogy with
many counties in Europe and some in Asia. The match
was good for quite some time,\, ... before the hard measures were
significantly relaxed.
The projection in (3) was quite stable, but the next
{\it dramatic\,} reduction of hard measures in almost all 50 states
made the saturation at 9/19 unlikely; the number of states in
phase 2 quickly dropped from 22 to 8, and the USA entered the second
wave. It was supposed to be similar to Figure \ref{fig:eu-7-14} 
for Europe, but this did not happen.

With these 3 instances, the "pattern" is basically the same:
{\it hard measures were relaxed upon the
first signs of improvements, well before sufficiently 
small numbers of new daily cases were reached}. There are of
course serious reasons for such a reduction, and
this is applicable not only to the USA. 

We mention that
this kind of "early response" sometimes 
works well, for instance with {\it contrarian 
investing strategies}. However, professional
investing is based on the earliest possible access
to the news and predicting "mass behavior". Here we
face something different. 

Having the 2nd wave on top of 
the "unfinished" first is of course a negative
development. Monitoring new infections, treating those 
infected, and general control of the situation   
become significantly less doable. 
Obviously, the likelihood of such a pattern exists in 
quite a few countries, Brazil and India included.  
\vskip 0.2cm
\vfil

Clear second waves were present in the middle of July 
in 15+ countries and now (the middle of September) are quite
common in Europe. It is likely that  our differential equations 
and Bessel-type solutions
describe them with the same kind of accuracy, as for the first waves.
  As it was predicted,
they appeared right after the end of the
first ones and  their shape and the
strength were quite comparable with those of the first waves. 

Regardless of our theory, this confirms that "hard measures" 
were the 
main reason  for the saturation of {\it Covid-19} in many
countries. The second waves are quite common in
epidemics, but they cannot be so soon if the process is "natural",
i.e. not under intensive management. 

\vskip 0.2cm

Mathematical understanding 
the epidemic
spread and forecasting the results of
their management obviously deserve
a systematic theory.
In this paper, we mostly focus on the uniformity of the
spread of {\it Covid-19} in many countries, namely, on
the uniformity of the curves of total numbers of detected 
infections,  which appeared 
very surprising. We think that our approach explains this.

There are not many different modes of epidemic
management and the ways  to "measure" the
epidemic dynamics. Similar management, if it is the
key factor, can be expected to result in similar features
of the epidemic spread, including the curves of total
cases, the saturation, the passage from phase one to phase two,
and the second waves. 
\vfil

Every country has of course its own path 
during {\it Covid-19}. There can be {\it several\,} different 
patterns here (not just one), so a final theory will be ramified.
The results of our theoretical and numerical  analysis suggest
that by now we have the following major ones: 
\vskip 0.2cm

$(a)$\, The {\it West-European pattern}, with
a successful path to the saturation of both phases and 
modeled very well 
by our "2-phase solution"; 

$(b)$\, the {\it USA pattern}, similar to $(a)$, but with 
as early reductions of hard measures as possible, which 
significantly delays the saturation; 

$(c)$\, {\it "Brazil-India"}, with high $c$, replacing
$R_0$ in our theory,  and with long stages of strong polynomial 
growth of the curves of total cases. 
\vskip 0.2cm

\begin{figure*}[htbp]
\hskip -0.2in
\includegraphics[scale=0.5]{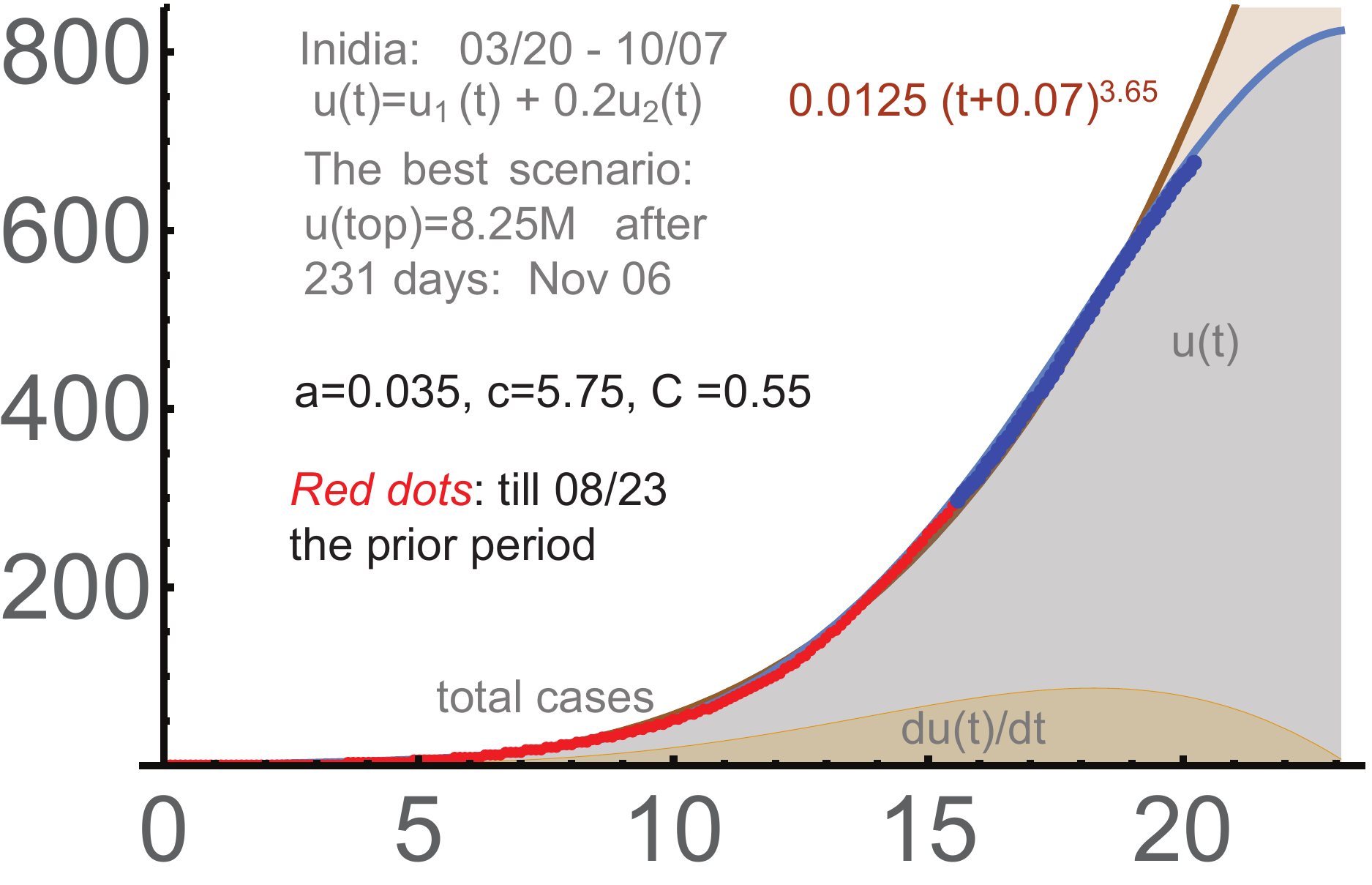}
\vskip -0.2in
\caption{
$India:3/20\!-\!10/7, c\!=\!5.75, a\!=\!0.035, C\!=\!0.55$.}
\label{fig:india-5-1}
\end{figure*}

Let us update Figure \ref{fig:india-fin} till 10/07/2020.
The parameters there were determined {\it before} the
turning point, so they were not that reliable. As of 
the beginning of November, India presumably reached 
this point. Interestingly, only $c$ required some adjustment; 
it is now $c=5.75$ vs $c=5.2$ determined for the period till 8/23.
{\it All other coefficients, including the exponent
 are unchanged}. The same (brown)
power function $0.0125(t+0.07)^{3.65}$
provided a good approximation till the turning point.
The match with our Bessel-type solution was almost perfect, 
though of course much will depend on the further developments
in India. Recall that  $y=$cases$/10K$ in Fig. 
\ref{fig:india-5-1}. The red dots are those used in
Figure \ref{fig:india-fin} (till 8/23). The clear polynomiality
of the curve of total cases in India and the match with
our $u$-functions are of course important confirmations of
our theory. We have:
$$u(t)=0.55\,t^{c/2+0.5}\bigl(J_{c/2-0.5}(\sqrt{a} t)\!+\!
0.2 J_{0.5-c/2}(\sqrt{a} t)\bigr), c\!=5.75,\, a=0.035.
$$

Automated 
modeling epidemics based on our "2-phase solution"
followed by their automated forecasting is quite a challenge. This
seems doable by analogy with the
algorithms described in \cite{ChAI} for trading stocks, which
are working programs. For phase 2, our software can
manage automatically (one-click) any groups of countries. 
This is not an {\it AI-type\,} system so far,
with self-learning and so on, but already quite efficient to
monitor daily the {\it later} stages of {\it Covid-19}. 



For the "end-users", our (or any) projections must be
constantly renewed on the basis of the latest data. 
However it is very important to test the forecasts 
without any later adjustments and
during sufficiently long periods;
 we do this systematically 
in this paper.
The testing periods are the black
and blue dots in our graphs. 

It is of course
important that our theory was created in the middle of the 
epidemic; this provides almost a unique chance to test it 
real-time. Real-time experiments are a must for 
forecasting and trading systems in stock markets. No 
models can be accepted there without such runs and 
carefully crafted tests excluding any usage of "future". 
This kind of discipline is not present in forecasting 
epidemics. 
\vfil

It is not unusual when the epidemic  curves 
are approximated piece-wise using 
frivolous choices of formulas and  parameters.
For {\it Covid-19},  the curves for the numbers of total cases 
can be presented as piece-wise polynomial functions
with fluctuating exponents, 
but this only confirms that their growth 
is no greater than polynomial.

Our $u$-curves used for phase 1
depend only on two major parameters, the initial
transmission rate $c$, which can be seen in the early stages,
and the intensity, $a$, of the measures employed.
If $u^2$ is used, then its coefficient is the third one,
but this is mostly needed to address some "effects of 
second order". 

Forecasting epidemics is of course a natural aim, 
but mathematical models are sometimes considered "acceptable" 
even if they are disconnected with real epidemic data. 
For instance, the usage of $R_0$, 
the {\it reproduction number}, is 
common in the literature in spite of the fact that the 
exponential growth of the number of cases
can be observed only during very short periods, if any. 
The usage of the statistical tools and random
processes is important, but only after the
basic differential or difference equations are proved to be
applicable.

To conclude,  any 
forecasting required and requires mathematics. Any 
analysis or discussion of such complex events 
as epidemics must be based on strict definitions.
For instance, comparing different countries and
various phases is impossible without solid mathematical 
methods.
This is fully applicable to understanding the efficiency
of the measures employed, which can be a difficult task
even if the corresponding mathematical tools are adequate. 

Modeling the spread of {\it Covid-19}  appeared closely related
to the Bessel functions, which can be a challenge for 
researches in the field of mathematical epidemiology. 
The theory of these functions, 
introduced by Daniel Bernoulli and generalized by Bessel,
was renewed recently. We actually use 
some recent developments, but the final formulas
require only very basic "one-line" definitions, simple
to use practically. 

Their quasi-periodicity
is a deep classical result, which provides a brand 
new approach to the saturation of epidemics under
active management. In contrast to the saturation 
due to the {\it herd immunity} or biological factors,
this kind of saturation, the main one for {\it Covid-19}, 
is of unstable nature and 
does require thorough mathematical theory. This paper
is the first step in this direction.    

\vskip 0.2cm

{\bf Acknowledgements.}
I thank very
much David Kazhdan for valuable comments and suggestions; 
a good portion of this paper presents my attempts to answer 
his questions. I also thank Eric Opdam and Alexei Borodin
for their kind interest, and the referee for useful suggestions.
I'd like to thank ETH-ITS for outstanding hospitality;
my special thanks are to Giovanni Felder, Rahul Pandharipande.

\section{\bf Appendix: auto-forecasting}

We will discuss the usage of the 
programs designed for forecasting late stages
of the waves of {\it Covid-19} in any countries or groups of
countries. They are available as the supplement
to the published version of this work;
see also \cite{ChMed}. 
The  second wave of {\it Covid-19} in
the USA is
considered in detail; our 2-phase solution appeared
applicable. 

{\sf Forecasting late stages.}
The programs are designed to be used only at later stages of 
(the waves of) {\it Covid-19} in any countries, groups
of countries, and regions. We provide a 
"universal program" (any countries)
in FOREU.zip, and its version in FORUS.zip designed for 
the USA via auto-forecasting the spread in all $50$ states and calculating the 
{\it superposition\,} of the corresponding
projections. These programs  are 
based on the type $B$ formulas for $u_B(t)$, which
were used in our {\it 2-phase solution}.
They describe {\it only\,} later stages: phase two in
our terminology.

The "universal program" gave quite stable
results for Western Europe, but only till the middle 
of August.
Then the second wave began there, presumably
due to the end of 
vacations and the beginning of the school year.
The USA  program (all states separately) provided stable projections 
for the saturation 
till the middle of June, the beginning of the second wave
in the USA. Then a significant reduction of the hard measures
began. It appeared applicable again in the beginning
of September in the USA, with a path to
the saturation of the second wave. 

The fact that 
many schools and businesses were closed in the USA, in contrast
to Europe, contributed to this. However the trend changed
in the middle of September: the number of new
daily cases became essentially constant, which corresponds
to a linear growth of the number of total cases. 
In Europe, a strong growth of the total
number of (detected) cases began in the end of August, 
so the program became essentially a linear approximation
in September. 
\vskip 0.2cm

\begin{figure*}[htbp]
\hskip -0.2in
\includegraphics[scale=0.5]{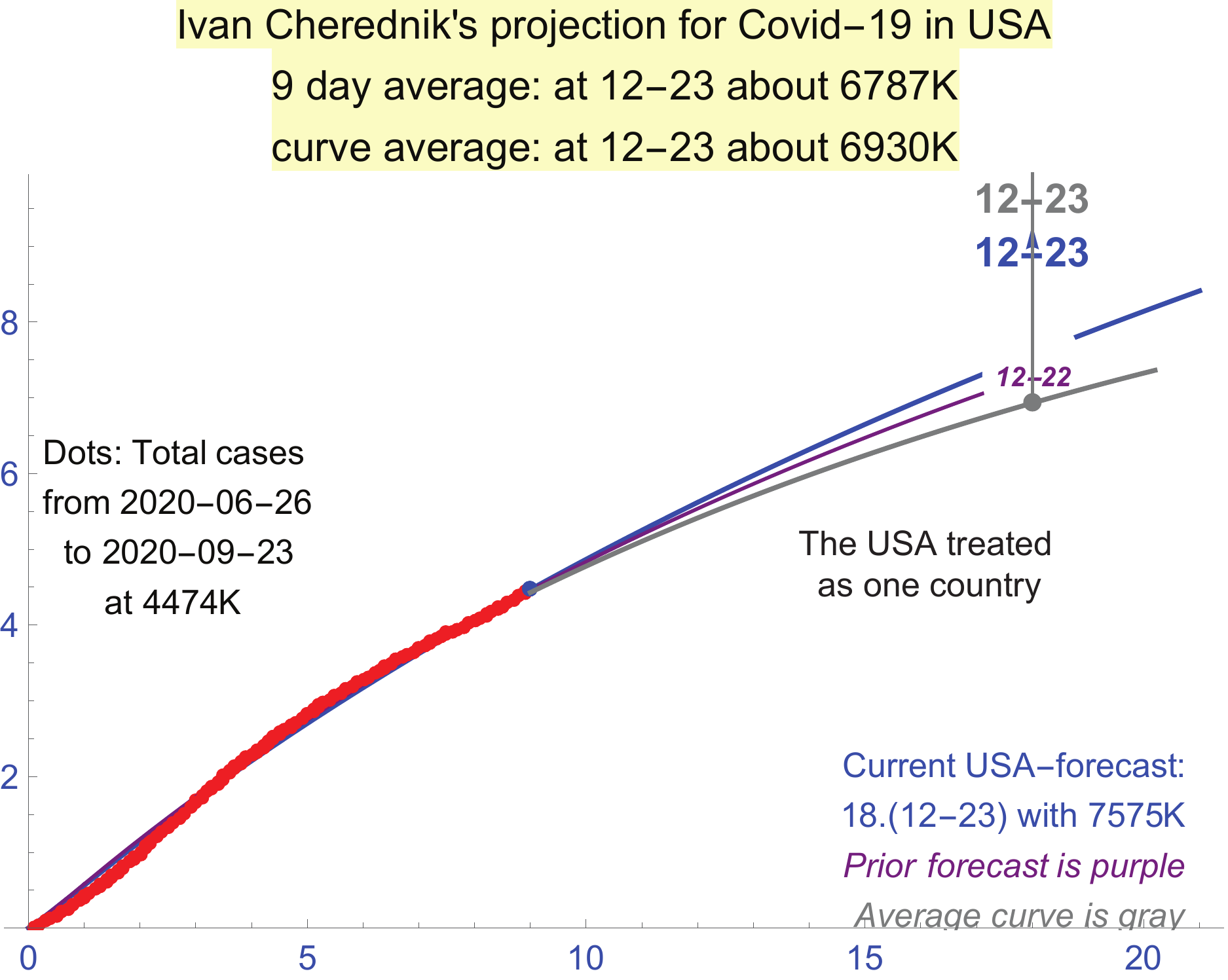}
\vskip -0.2in
\caption{Universal forecast: not via 50 States.}
\label{fig:usaw-9-23}
\end{figure*}

{\sf Two outputs.}
The "universal program" starting with
about 9/20 gave the linear approximation for the total
number of infections. We provide the output
of 9/23 in Figure \ref{fig:usaw-9-23}. The projections are
for 3 months from the date if the saturation cannot be found
during this period. 

The second program, based on the consideration of
all $50$ states, is presumably  more reliable.
The first run we provide is as of 9/12. The projection was
$5.8$M in the first 2 weeks of December  on top
of the initial $2.1$M at 06/16. Actually, only about 17 States
were in phase 2, so this was not really reliable. This
number dropped to about 11 on 9/22. The program 
still found the saturation for the $9$-day average of the 
corresponding curves but it was 12/28. The trend was 
toward the linear growth of the total number of cases.

\begin{figure*}[htbp]
\hskip -0.2in
\includegraphics[scale=0.5]{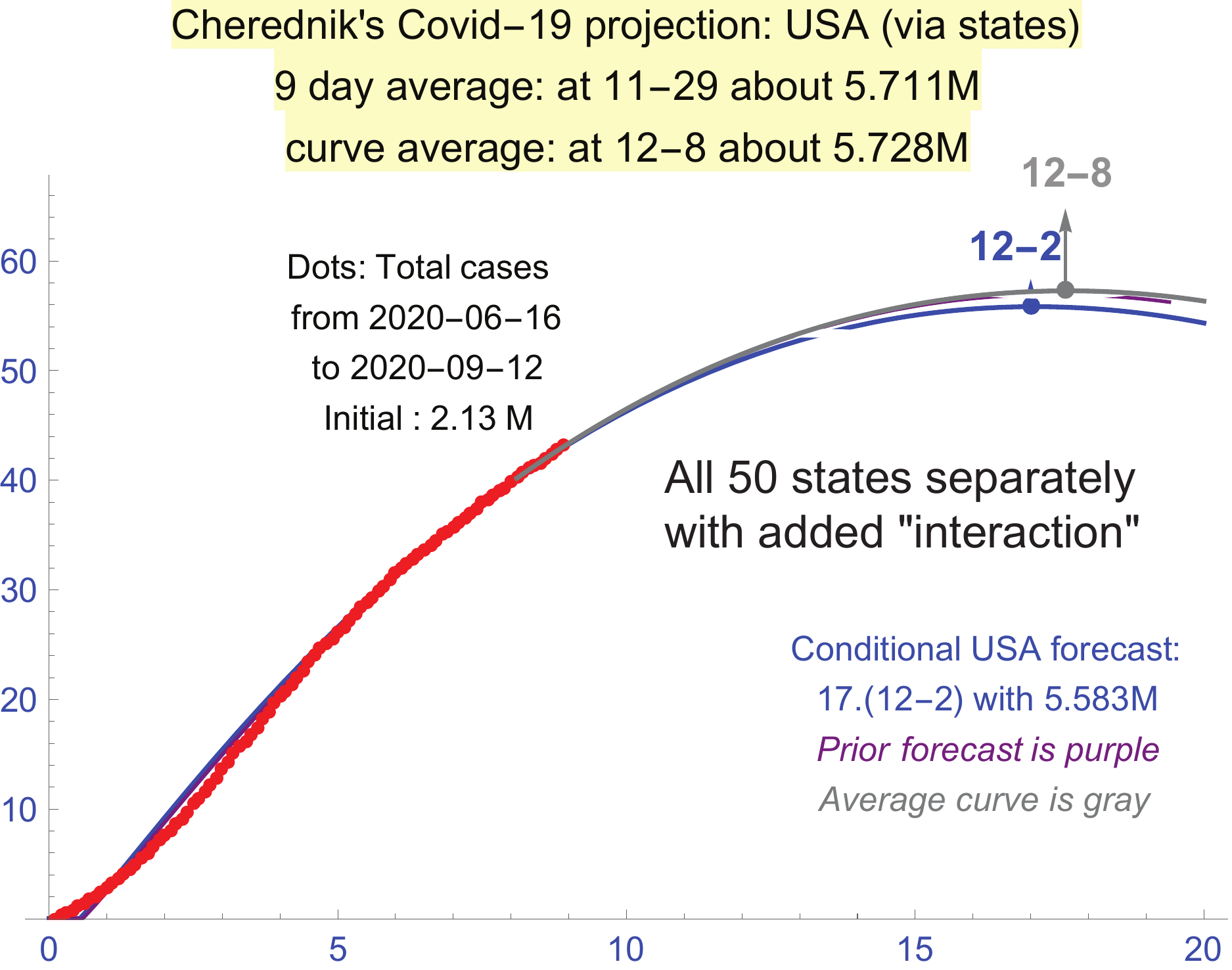}
\vskip -0.2in
\caption{All 50 states separately (9/12).}
\label{fig:usall-9-12}
\end{figure*}

\begin{figure*}[htbp]
\hskip -0.2in
\includegraphics[scale=0.5]{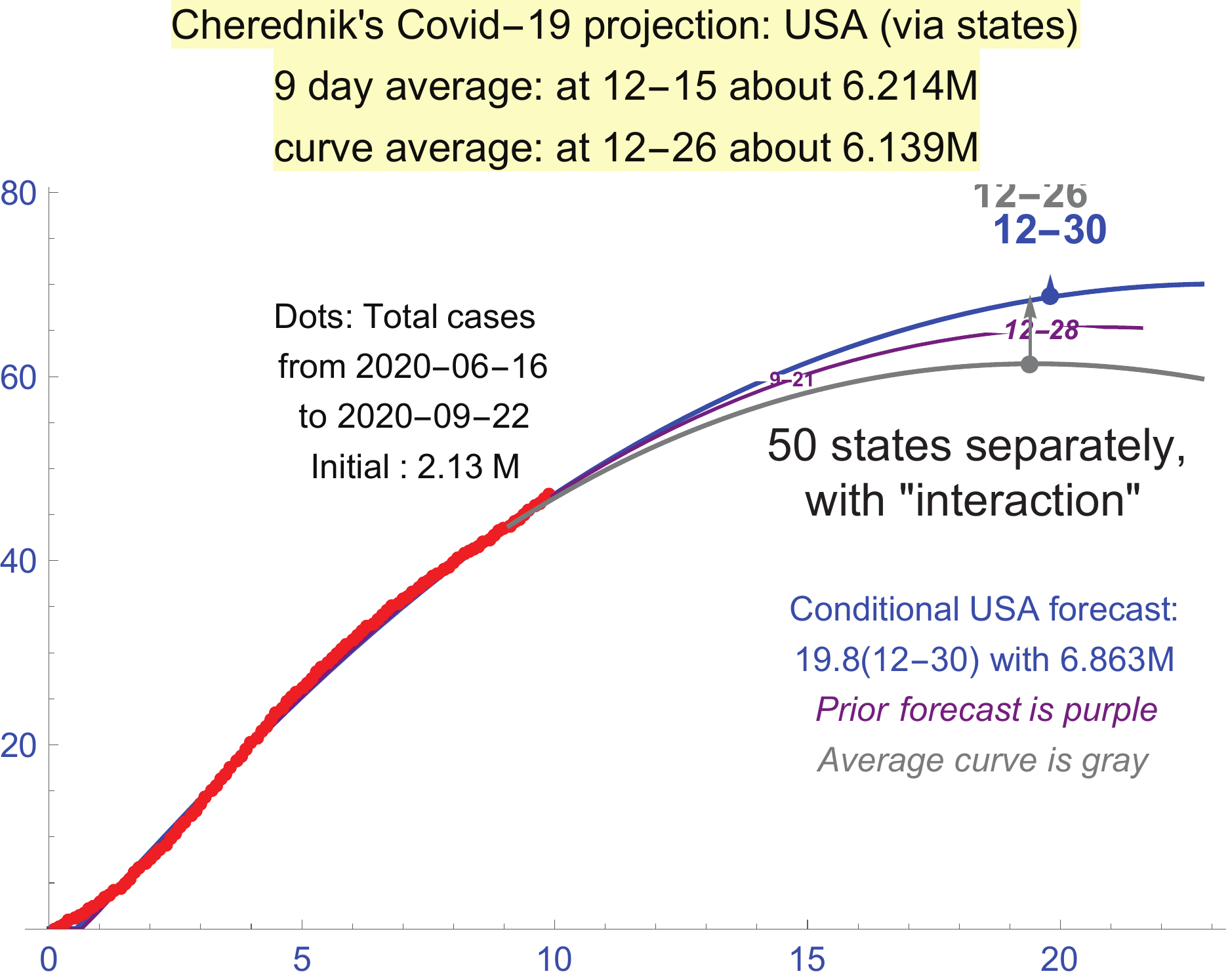}
\vskip -0.2in
\caption{All 50 states separately (9/22).}
\label{fig:usall-9-22}
\end{figure*}

Any reopening of schools and businesses in the
USA can and will
influence these projections, as occurred in Europe. This already 
happened in the middle of June, when the USA approached
the second phase of the 1st wave, after the
hard measures were significantly reduced. Much depends 
on the virus evolution too.
Let us provide two outputs of the program for the
USA: Figures \ref{fig:usall-9-12}, \ref{fig:usall-9-22}.

{\sf Two phase-solution.}
This solution worked well at least till the middle of
September for the
second wave in the USA. The accuracy is comparable
with what we obtained and discussed in this paper for
the first waves in Japan, Israel,
Italy, Germany, UK and the Netherlands. 
Upon subtracting $2.1M$, the parameters we obtained
for the early-middle stage of the 2nd wave in the USA were: 
 $a_\circ=0.06,\, c_\circ=2.65$,
$$
u^\circ_{1,2}(t)=\!3.4\,t^{(c_\circ+1)/2}J_{\pm \frac{c_\circ-1}{2}}
(\sqrt{a_\circ}t),\  u_\circ (t)=u^\circ_1(t)+0.6 u^\circ_2(t).
$$ 

The second phase matched well the following function:
\begin{align*}
u_{B}(t)=\!4.1\, t^{c/2}\cos(d\log(Max(1,t))),
c\!=\!2.65, 
d\!=\!0.435.
\end{align*}

\begin{figure*}[htbp]
\hskip -0.2in
\includegraphics[scale=0.5]{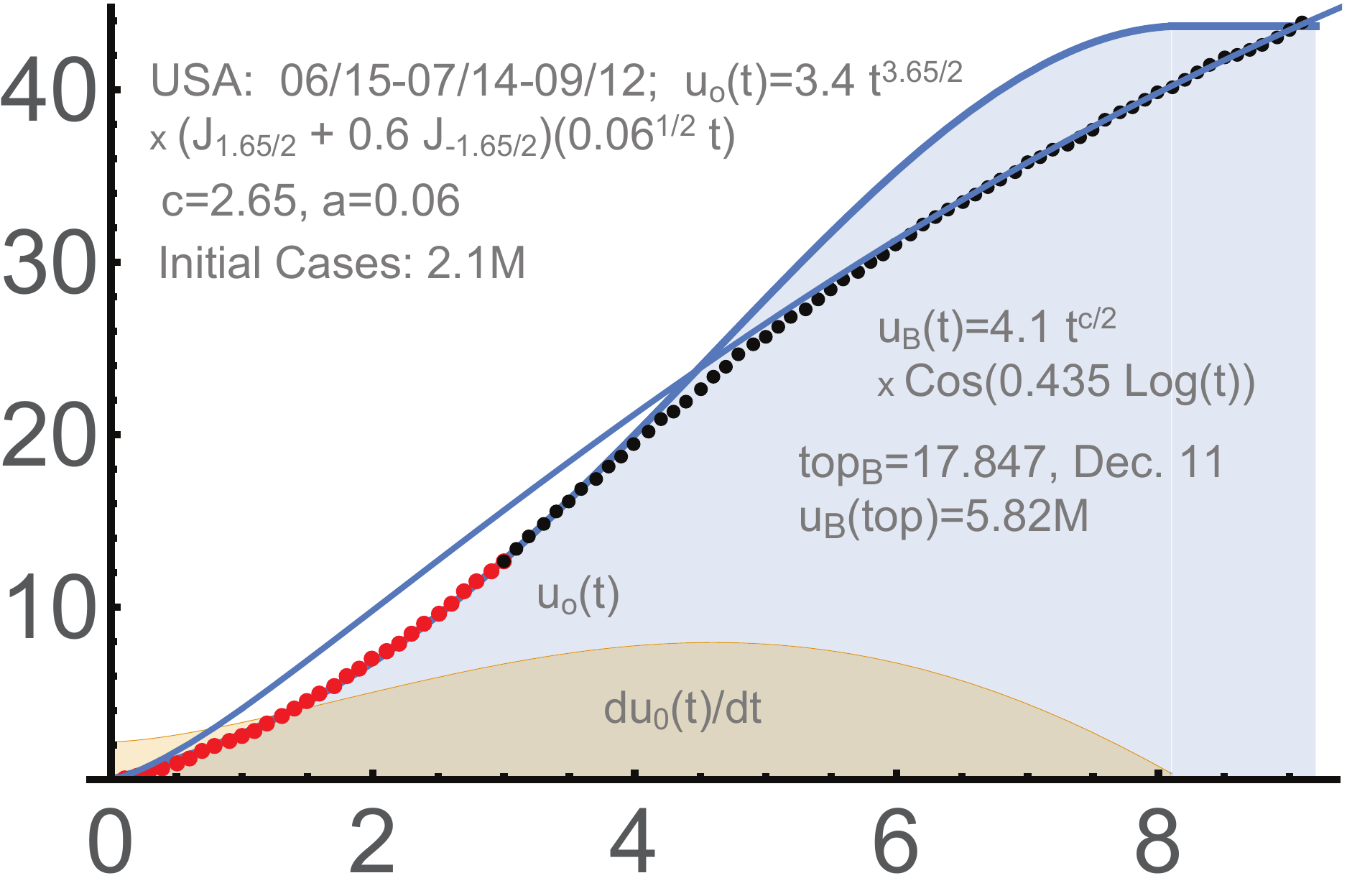}
\vskip -0.2in
\caption{2-phase solution for the 2nd wave in the USA.}
\label{fig:usa-2w-2p}
\end{figure*}

The projected saturation for $u_B$ was given by
the formula
$t_{end}=\exp\bigl(\frac{1}{d}\,tan^{-1}(\frac{c}{2d})\bigr)$.
Numerically, $t_{end}=17.8463$, which is 
178 days from 06/16:  December 11, 2020.
 See Figure \ref{fig:usa-2w-2p}.
This date basically matches 
the auto-projection based on considering all $50$ states
separately in  Figures \ref{fig:usall-9-12},
\ref{fig:usall-9-22}. However, the latter are
actually the last ones before the growth of the number of total
cases became linear in the USA. 
 
{\sf Practical matters.}
The main programs in $FOREU.zip$ are  
"forwor.txt" and "forworstat.txt". In Mathematica,
run "forwor.txt"; it is set for the United States. Upon making
$grp=0$ in "worldfile.txt", it will work for Europe. 
Some countries were excluded in Europe, mostly 
due to problems with data. To change this you need
to open "forwor.txt" and "forworstat.txt" and find the
corresponding places. Mathematica programs are readable. 

Generally, the program is 
"one-click": run the exe-file, after you adjust the path to 
"math.exe" in your computer in the file "mathpath.txt",
and set the countries/regions in
"worldfile.txt" following the "README" file. 

This is similar for the program for 50 states in the USA.
There is no file "worldfile.txt", and currently all states
are included. The initial date is set to 06/16/2020. This
can be adjusted directly in "forusa.txt" and "forustat.txt":
change "datein" (all instances). 

The initial date is 
automatically set to 3 months before "today" in the universal 
program (for any countries), but not in this one. Here it
is the beginning of the wave.  

Mathematica 11 is necessary to use these programs.
For the universal program, 
\noindent
{\tt
raw.githubusercontent.com/owid/covid-19-data/}
is used: the online access to this site is needed.
The names of the countries and regions in
"worworld.txt" must be as at this site. 

\noindent
The site {\tt
raw.githubusercontent.com/nytimes/covid-19-data/} is used
for the program managing the $50$ states in the USA. 
 
The {\it saturation\,} in these programs 
is technical; it is supposed to move over time. 
If the program detects no saturation, then the projections will 
be provided for a 4 month period; in this case, the program 
mostly works as a linear extrapolation. 

The programs are not for any commercial use; the name of their 
creator, Ivan Cherednik, and a link to the Journal or
\cite{ChMed} must be always provided. This is a research tool 
only. The source file
in Mathematica is readable, so you can understand what the
programs really do. Please see the README-files.

\vskip 0.2cm
\bibliographystyle{unsrt}

\end{document}